\documentclass{aa}  
\pdfoutput=1
\usepackage{graphicx}
\usepackage{txfonts}
\usepackage{xcolor}
\usepackage{colortbl}
\usepackage{tabularx}
\usepackage[hyphens]{url}
\usepackage{ulem}
\usepackage{natbib}
\bibpunct{(}{)}{;}{a}{}{,} % to follow the A&A style
\def\ltapprox{\raise 2pt \hbox {$<$} \kern-0.9 em \lower 3pt \hbox
  {$\approx \,$}}
\def\gtapprox{\raise 2pt \hbox {$>$} \kern-0.9 em \lower 3pt \hbox
  {$\approx \,$}}
\def\ltsim{\raise 1pt \hbox {$<$} \kern-0.9em \lower 3pt \hbox {$\sim \,$}}
\def\gtsim{\raise 1pt \hbox {$>$} \kern-0.9em \lower 3pt \hbox {$\sim \,$}}

\DeclareGraphicsExtensions{.pdf,.png}
\begin{document} 

\title{Deep {\it Chandra} observations of PLCKG287.0$+$32.9 : \\
  a clear detection of a shock front {\color{black} in a} 
  %overheating 
  {\color{black} heated} former cool core}

\author{M. Gitti\inst{1,2}, A. Bonafede\inst{1,2},
  F. Brighenti\inst{1,3}, F. Ubertosi\inst{1,2}, M. Balboni\inst{1,4},
  F. Gastaldello\inst{4}, A. Botteon\inst{2}, W. Forman\inst{5}, \\ R.~J. van Weeren\inst{6},
  M. Br{\"u}ggen\inst{7}, K. Rajpurohit\inst{5}, C. Jones\inst{5}}

\institute{ 
Dipartimento di Fisica e Astronomia (DIFA), Universit{\`a} di Bologna, via Gobetti 93/2, 40129 Bologna, Italy \\ \email{myriam.gitti@unibo.it}
\and
Istituto Nazionale di Astrofisica (INAF) -- Istituto di
Radioastronomia (IRA), via Gobetti 101, I-40129
Bologna, Italy
\and
University of California Observatories/Lick Observatory, Department of Astronomy and Astrophysics, Santa Cruz, CA 95064, USA
\and
INAF - Istituto di Astrofisica Spaziale e Fisica Cosmica di Milano,
via A. Corti 12, 20133 Milano, Italy
\and
Smithsonian Astrophysical Observatory, Harvard-Smithsonian Center for Astrophysics, 60 Garden St., Cambridge, MA 02138, USA
\and
Leiden Observatory, Leiden University, PO Box 9513, 2300 RA Leiden,
The Netherlands
\and
University of Hamburg, Hamburger Sternwarte, Gojenbergsweg 112, 21029
Hamburg, Germany
}

\authorrunning{Gitti et al.} 
\titlerunning{{\it Chandra} observations of PLCKG287.0$+$32.9: detection of a shock {\color{black} in a heated} a former cool core}

\date{Accepted: 17 March 2025 }

\abstract 
{%Context:
  The massive, hot galaxy cluster PSZ2 G286.98+32.90 (hereafter PLCKG287,
  $z=0.383$) hosts a giant radio halo and two prominent radio relics which are 
  signs of a disturbed dynamical state. 
  Gravitational lensing analysis shows a complex cluster core with multiple components.
  However, despite optical and radio observations indicate a clear multiple merger, 
  the X-ray emission of the cluster, derived from $XMM$-$Newton$ observations,
  shows only moderate disturbance.
}
{%Aims:
  The aim of this work is to study the X-ray properties and investigate the core heating of such a massive cluster by searching for surface brightness discontinuities and associated temperature jumps that would indicate the presence of shock waves.
  }
{%Methods:
  We present new 200 ks $Chandra$ $X$-$ray$ $Observatory$  ACIS-I observations of 
  PLCKG287 and perform  a detailed analysis to investigate the morphological and thermodynamical properties of the region inside $R_{\rm 500} \, (\sim$1.5 Mpc).}
{%Results:
The global X-ray morphology of the cluster has a comet-like shape, 
oriented in the NW-SE direction, with an $\sim$80 kpc offset between 
the X-ray peak and the Brightest Cluster Galaxy (BCG). 
We detect a shock front to the NW direction at a distance of $\sim$390
kpc from the X-ray peak, characterized by a Mach number $\mathcal{M} \sim$1.3,
as well as a cold front at a distance of $\sim$300 kpc from the X-ray peak, 
nested in the same direction of the shock in a typical configuration expected by a merger.
We also find evidence for X-ray depressions to the E and W, 
that could be the signature of feedback from the active galactic nucleus (AGN).
The radial profile of the thermodynamic quantities show
a temperature and abundance peak in the cluster center, where also the
pressure and entropy have a rapid increase. 
}
{%Conclusions:
Based on these properties,
we argue that PLCKG287 is what remains of a cool core after a heating event. 
{{\color{black} We estimate that both the shock energy and the AGN feedback energy, implied by the analysis of the X-ray cavities, are} sufficient to heat the core to the observed temperature of $\sim$17 keV in the central
$\approx$160 kpc. We discuss the possible origin of the {{\color{black} detected} shock by investigating alternative scenarios of merger and AGN {\color{black} outburst}, finding that they are both energetically viable. {\color{black} However, no single model seems able to explain all the X-ray features detected in this system.} This suggests that the combined action of merger and {\color{black} central AGN feedback is likely necessary to explain the reheated cool core, the large-scale shock and the cold front. The synergy of these two processes may act in shaping
the distribution of cool core and non cool core clusters.}}
}}


\maketitle

%%%%%%%%%%%%%%%%%%%%%%%%%%%%%%%%%%%%%%%%%%%%%%%%%%%%%%%%%%%%%%%%%%%%%%%%%%%%%%%
%%%%%%%%%%%%%%%%%%%%%%%%%%%%%%%%%%%%%%%%%%%%%%%%%%%%%%%%%%%%%%%%%%%%%%%%%%%%%%%
%%%%%%%%%%%%%%%%%%%%%%%%%%%%%%%%%%%%%%%%%%%%%%%%%%%%%%%%%%%%%%%%%%%%%%%%%%%%%%%

\section{Introduction} 
\label{intro.sec}

The hierarchical model of structure formation predicts that clusters
of galaxies grow by successive accretion of smaller sub-units through mergers \citep[e.g.,][for reviews]{Kravtsov_2012,Vikhlinin_2014}. 
A large amount of energy is dissipated in the intra-cluster medium (ICM), and
possibly channeled into the amplification of the magnetic fields
\citep[e.g.,][for a review]{Dolag_2008} and into the acceleration of
high-energy cosmic rays \citep[e.g.,][for reviews]{Petrosian_2008,Brunetti-Jones_2014} that in turn may produce observable radio emission. 
Merger events may also shape the thermodynamical properties of the ICM and thus affect the distribution of cool core and non cool core clusters. However, theory and simulations studies suggest there is ongoing debate on whether mergers are solely responsible for disrupting cool cores, or other processes, as feedback from active galactic nuclei (AGN), are at play \citep[e.g.,][]{McCarthy_2008, Guo_2010, Rasia_2015, Barnes_2018}. Therefore, detailed observations of disturbed systems are key to investigating this problem \citep[e.g.,][]{Molendi_2023}. 

\par
The cluster PSZ2 G286.98$+$32.90, also known in the literature  as PLCKG287.0$+$32.9 (hereafter PLCKG287) is a massive cluster, with $M_{500}\sim 13.7 \cdot 10^{14}$ M$_{\odot}$ \citep{Planck_2016} located at redshift $z = 0.383$ \citep{Daddona_2024}. 
It was the second most significant detection of the Planck Sunyaev-Zel'dovic (S-Z) survey  (Planck collaboration, 2010), and a 10 {\color{black} ks} validation observation performed with {\it XMM-Newton} revealed that it is extremely hot ($kT_{\rm 500} \sim 13$ keV) and luminous ($L_{\rm 0.1-2.4\,keV}^{\rm 500}\sim1.7 \times 10^{45}$ erg s$^{-1}$) within $R_{\rm  500} \, (\sim 1.5$ Mpc). 
PLCKG287 belongs to the CHEX-MATE cluster sample \citep[Cluster HEritage project
with XMM-Newton: Mass Assembly and Thermodynamics at the Endpoint of
structure formation,][]{CHEX-MATE_2021}. Follow-up {\it XMM-Newton} observations have been carried out by the CHEX-MATE collaboration to achieve a robust determination of the thermal properties up to $R_{500}$, confirming that it is a very luminous and hot cluster \citep{Riva_2024}.
\par

The dynamical and thermodynamical properties of the cluster have been analyzed by several authors \citep{Bagchi_2011,Bonafede_2014,Finner_2017,Campitiello_2022,Golovich_2019,Riva_2024,Finner_2024}.
\citet{Bagchi_2011} first analyzed the cluster, concluding that it is undergoing one
or multiple mergers with the main axis oriented in the direction from
North-West (NW) to South-East (SE). 
The merger orientation is confirmed by optical ESO 2.2 m WFI data, which
indicate that PLCKG287 is located within a 6-Mpc long intergalactic
filament and it has undergone a major merger,
slightly misaligned with respect to the main direction of
accretion \citep{Bonafede_2014}. In addition, two more sub-clumps are detected along the filament.
\par
GMRT and JVLA radio observations from 150\,MHz to 3\,GHz show that the cluster hosts diffuse radio emission \citep{Bagchi_2011,Bonafede_2014,Stuardi_2022}, as expected for massive merging clusters. In particular, the cluster hosts two symmetrically radio relics, which are supposed to be generated by merger driven shock fronts, and a radio halo, which is connected to (re)-acceleration of cosmic ray electrons by turbulence \citep[e.g.,][and references therein]{Brunetti-Jones_2014,vanWeeren_2019}.
The radio halo appears roundish in shape at 325 MHz, with a total extent of $\sim$ 250$''$ (1.3 Mpc), and co-spatial with the cluster X-ray emission. At 150 MHz, the halo is reported to be more elongated toward the SE, reaching a largest angular size of $\sim$ 450$''$, corresponding to $\sim$2.4 Mpc \citep{Bonafede_2014}. The relics are found along the merger axis, as usually found for this kind of sources.
Other peculiar features of filamentary radio emission 
have also been detected, whose origin is unknown \citep{Bonafede_2014}. 
\par
More recent Subaru optical observations confirm that the galaxy
distribution is elongated in the SE to NW direction along the merger
axis and that the mass distribution on large scales also follows the
same merger axis \citep{Finner_2017,Finner_2024}. The higher-resolution Hubble
Space Telescope ({\it HST}) mass
map shows that two mass peaks exist within the
central mass clump, and that they also lie along the
merger axis.  The cluster mass is dominated by the primary cluster of mass $M
\sim 1.6 \times 10^{15}$  M$_{\odot}$ centered on
the brightest cluster galaxy (BCG), with three substructures being
$\sim$10\% of the mass of the primary cluster.
Based on the spatial distribution of the substructures, \citet{Finner_2017}
concluded that they are consistent a with merger scenario that may have
created the radio relics.
\par
Despite optical and radio analysis indicate a clear massive and strong multiple merger, the X-ray emission of the cluster shows moderate disturbance with only a single peak. Indeed, the X-ray morphological analysis of \citet{Campitiello_2022} classifies
PLCKG287  as ``mixed morphology'', being ranked 87th out of
118 clusters in the CHEX-MATE sample in terms of morphological
disturbance.  Hence, PLCKG287 could be a case where a merger has not completely disrupted the cluster core, 
as observed in several systems \citep[e.g.,][and references therein]{Douglass_2018, Schellenberger_2023, Rossetti_2010}. 
Yet, the presence of the halo, relics, other thin radio filaments, and the mass distribution reconstructed by weak lensing analysis make the cluster an intriguing case that needs to be studied in more detail.
\par
The aim of this work is to present a detailed analysis of deep {\it
  Chandra} observations of the cluster PLCKG287,  in order to study the
X-ray morphological and thermodynamical properties of the ICM.
We focus our analysis and discussion on the X-ray properties of the cluster, whereas the connection with the radio features will be studied in detail in forthcoming papers based on new observations with MeerKAT L band (Balboni et al, in preparation), uGMRT (Band3 and Band4), and MeerKAT (UHF and S-band) observations (Rajpurohit et al., in preparation). 

The paper is organized as follows:  
In Section \ref{data.sec}, we present the {\it Chandra} observations and the basic steps of the data reduction. In Section \ref{properties.sec}, we present the global properties of the cluster derived from the morphological and spectral analyses inside $R_{\rm 500}$. In particular, we discuss the residual images obtained by subtracting the best-fit $\beta$-model (Sect. \ref{morpho.sec}) and the radial profiles and 2D spectral maps of the thermodynamic properties (Sect. \ref{spectral.sec}).  In Section \ref{fronts.sec} we present the detection of a shock and a cold front based on the detailed analysis of the surface brightness and thermodynamic radial profiles. Our interpretation of the observational results is discussed in Section \ref{discussion.sec}, where we first investigate the possibility that PLCKG287 is a former cool core (Sect. \ref{cc.sec}), then compare the energy required to heat the core (Sect. \ref{heat.sec}) to the energy of the shock (Sect. \ref{shock-energy.sec}) {\color{black} and AGN feedback (Sect. \ref{AGN-energy.sec})}, and finally discuss the possible scenarios for the origin of the shock (Sect. \ref{origin.sec}). We then present our conclusions in Section \ref{conclusion.sec}.

With $H_0 = 70$ km s$^{-1}$ Mpc$^{-1}$ and $\Omega_{\rm M} = 1 - \Omega_{\Lambda} = 0.3$, the luminosity distance to PLCKG287 (z = 0.383) is 2063.7 Mpc and 1$''$
corresponds to 5.23 kpc in the rest frame of the cluster.

%%%%%%%%%%%%%%%%%%%%%%%%%%%%%%%%%%%%%%%%%%%%%%%%%%%%%%%%%%%%%%%%%%%%%%%%%%%%%%%
%%%%%%%%%%%%%%%%%%%%%%%%%%%%%%%%%%%%%%%%%%%%%%%%%%%%%%%%%%%%%%%%%%%%%%%%%%%%%%%
%%%%%%%%%%%%%%%%%%%%%%%%%%%%%%%%%%%%%%%%%%%%%%%%%%%%%%%%%%%%%%%%%%%%%%%%%%%%%%%

\section{Chandra observation and data reduction}
\label{data.sec}

%%%%%%%%%%%%%%%%%%%%%%%%%%%%%%%%%%%%%%%%%

\begin{figure*}[t]
\centerline{
  \includegraphics[width=\columnwidth,bb=36 162 577 630]{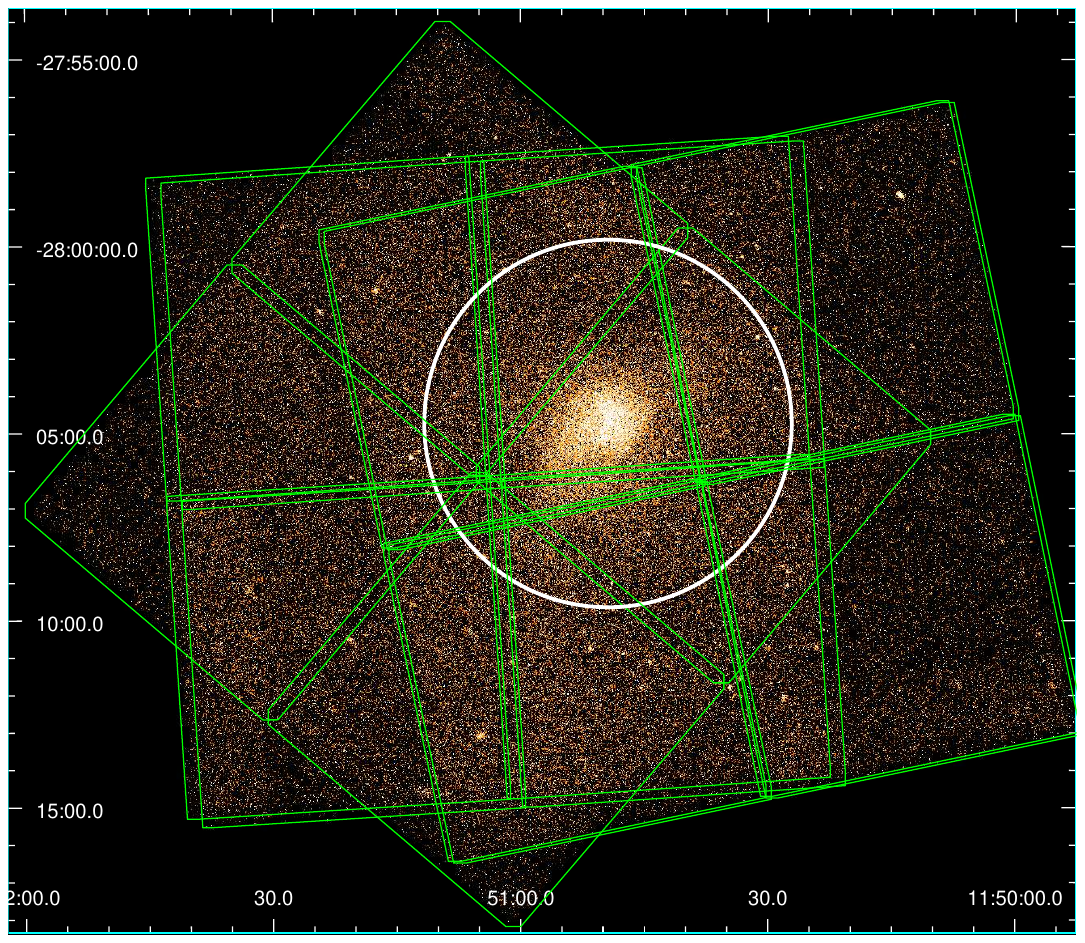}
  \includegraphics[width=\columnwidth,bb=36 165 577 627]{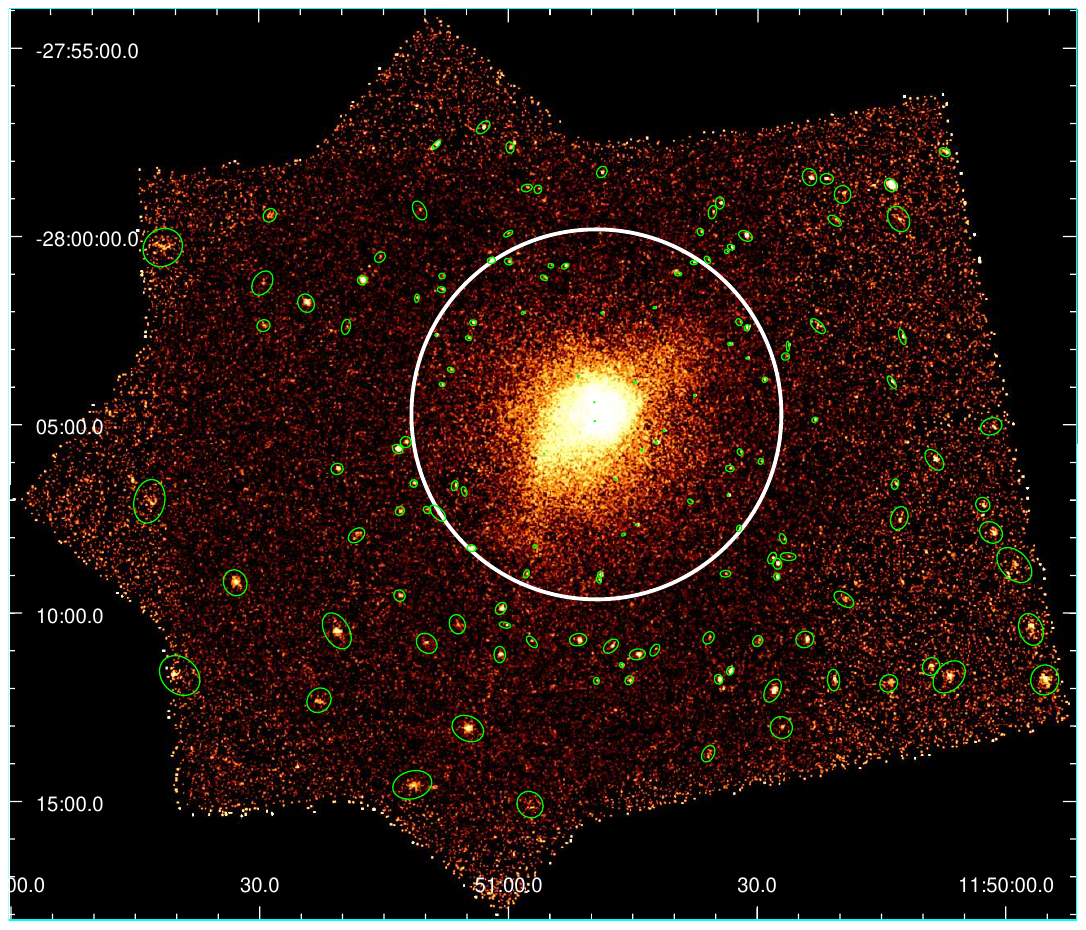}
}
\caption{\label{mosaics.fig} Field of view of the {\it Chandra} observations of PLCKG287 (z=0.383). {\it Left panel:}  Footprint of the five {\it
    Chandra} ObsIDs. {\it Right
    panel:}  Background-subtracted, exposure-corrected mosaic in the
[0.5-7.0] keV band. The detected point sources which were removed from
the analysis are shown with green ellipses. In both panels, the white circle indicates $R_{\rm 500} = 1541$ kpc, which is the region where we focus our analysis. }
\end{figure*}

%%%%%%%%%%%%%%%%%%%%%%%%%%%%%%%%%%%%%%%%%

%%%%%%%%%%%%%%%%%%%%%%%%%%%%%%%%%%%%%%%%%

\begin{figure*}[t]
\includegraphics[width=\textwidth,bb=36 195 577 597]{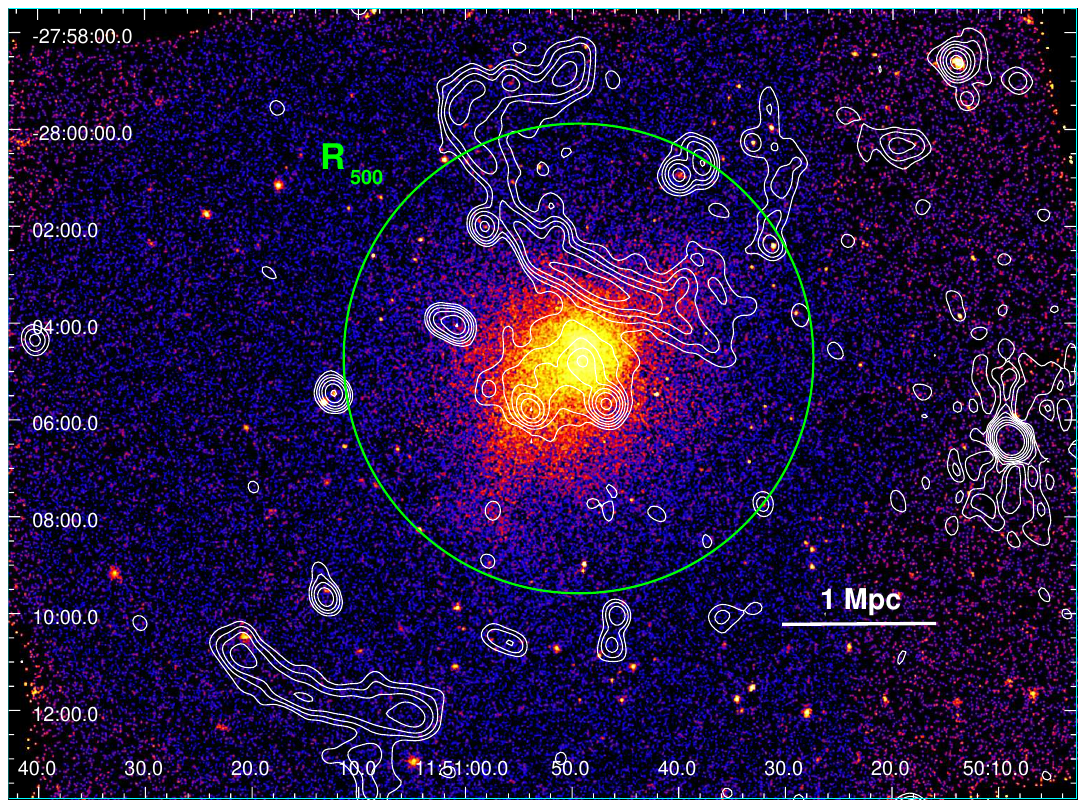}
\caption{\label{mosaic-radio.fig} Background-subtracted,
  exposure-corrected {\it Chandra} mosaic of PLCKG287 in the [0.5-7.0]
  keV band smoothed with a kernel of 4 pixels (1 ACIS pixel =
  $0.492''$). The detected point sources (shown with green ellipses in
  the {\color{black} right} panel of Figure \ref{mosaics.fig}) have been
  subtracted. The green circle indicates $R_{\rm 500} = 1541$ kpc.
  Overlaid in white are the GMRT radio contours at 330 MHz (levels start at $\pm 3 \sigma $ and increase by a factor of 2, where
  $\sigma$ = 0.1 mJy/beam, beam=$12.9'' \times 7.9''$). The radio data
  shows two prominent radio relics to the NW and SE at distances of
  $\sim 500$ kpc and $\sim 2.8$ Mpc kpc from the cluster center. The
  bright X-ray core region is coincident with a giant radio halo.  See
  \citet{Bonafede_2014} for more details on the radio features. }
\end{figure*}

%%%%%%%%%%%%%%%%%%%%%%%%%%%%%%

%%%%%%%%%%%%%%%%%%%%%%%%%%%%%%

\begin{table}
\caption{\label{obs.tab} Summary of the Chandra ACIS-I observations analyzed in
  this work (PI: A. Bonafede).}
% title of Table
\centering                          % used for centering table
\begin{tabular}{c c c c c}        % centered columns (4 columns)
  \hline
  \hline
  ObsID & Obs. Date & $\tau_{\rm exp}$ & $\tau_{\rm clean}$ & chip \\
  \hline
   17165 & 2015-11-17 & 54.4 & 38.2 & I2 \\  
   17166 & 2015-11-24 & 20.8 & 14.5 & I2 \\ 
   17494 & 2015-08-17 & 59.3 & 38.5 & I0 \\ 
   17495 & 2016-03-22 & 32.2 & 22.5 & I2 \\ 
   18807 & 2016-03-23 & 28.7 & 18.2 & I2 \\    
  \hline
  \hline
\end{tabular}
  {\tablefoot{Column 1: ObsID; column 2: Obs. date;
  column 3: exposure time; column 4: net exposure time after
  cleaning; column 5: chip containing the aimpoint.}}
\end{table}  
%%%%%%%%%%%%%%%%%%%%%%%%%%%

We analy{\color{black} z}ed archival {\it Chandra} observations of the cluster PLCKG287
consisting of five different exposures (ObsID 17165, 17166,17494,
17495, 18807, see Table \ref{obs.tab}). Each observation was obtained with
the ACIS-I instrument in  VFAINT mode.  The footprint of the five
available ObsIDs of PLCKG287 is shown in Figure \ref{mosaics.fig} (left
panel), where $R_{\rm 500} = 1541$ kpc \citep{Bagchi_2011} is
indicated with a white circle. The cluster emission extends over all
four ACIS-I chips of each ObsID, therefore we reprocessed
all 20 chips.
Data were downloaded from
the \textit{Chandra Data
  Archive}\footnote{{https://cda.harvard.edu/chaser/}} and were
reprocessed with the software package CIAO (version 4.10) and CALDB
(version 4.7.3) to apply the latest calibration available at the time
of the analysis and to remove bad pixels and flares.  We also improved
the absolute astrometry identifying the point sources of the longest
exposure (ObsID 17494) with the task \texttt{WAVDETECT} and
cross-matching them with the optical catalogue
USNO-A2.0\footnote{{http://tdc-www.harvard.edu/catalogs/ua2.html}}.
The other datasets were then registered to the position of the longest
one.

For the background treatment, we adopted the {\ttfamily blank-sky}
background
files\footnote{https://cxc.cfa.harvard.edu/ciao/threads/acisbackground/}
created by the ACIS calibration team. For each obsID dataset
separately, we first identified and re-projected the ACIS {\ttfamily
  blank-sky} files that match our data. Then we treated each chip
individually, normalizing it to the count rate of the same chip of the
corresponding obsID event file in the 9-12 keV band. The background
event files thus obtained for each chip were then used for the
spectral analysis (Sect. \ref{spectral.sec}).  Since the {\ttfamily blanksky}
files were created for quiescent-background periods, we additionally
filtered our ObsID datasets in the same manner to remove periods of
background
flares\footnote{https://cxc.cfa.harvard.edu/ciao/threads/flare/}.
The net exposure times obtained after this phase of data cleaning are
summarised in Table \ref{obs.tab}, resulting in a final, combined exposure
for the five ObsID datasets of 131.9 ks.

To create a mosaic image, we combined the event files of the five obsID datasets
by running {\ttfamily merge\_obs}, which generates 
an image in units of counts
(\texttt{counts/pixel}), an exposure map
(\texttt{cm}$^2\cdot$\texttt{s}$\cdot$\texttt{counts/ photons}) and an
image in units of flux
(\texttt{photons/(cm}$^2\cdot$\texttt{s}$\cdot$\texttt{pixel}))
resulting from the division between the count image and
the exposure map. The  {\ttfamily merge\_obs} script generates also
the PSF map\footnote{We assumed psfecf=0.9} of the mosaic used to detect the
point sources with \texttt{WAVDETECT}.
To create the associated mosaic background image we proceeded as follows: for
each obsID, we created a background event file by combining with
{\ttfamily dmmerge} the reprojected
{\ttfamily blank-sky} files of each chip. This combined {\ttfamily
  blanksky} event file was then
reprojected to match the corresponding ObsID event file and normalized to
the 9-12 keV band count rate of the corresponding ObsID main chip.
From the combined {\ttfamily blank-sky} event file of each ObsID we
then created a background image and scaled
it to the exposure time of the corresponding obsID.
Finally, we combined the five background images with {\ttfamily reproject\_image}.
The resulting background-subtracted, exposure-corrected mosaic in the
0.5-7.0 keV band is shown in Figure \ref{mosaics.fig} (right panel), where the
detected point sources (then removed from the analysis) are also shown.

Throughout the analysis, error bars are at the 1$\sigma$ confidence levels on a single parameter of interest, unless otherwise stated.

%%%%%%%%%%%%%%%%%%%%%%%%%%%%%%%%%%%%%%%%%%%%%%%%%%%%%%%%%%%%%%%%%%%%%%%%%%%%%%%
%%%%%%%%%%%%%%%%%%%%%%%%%%%%%%%%%%%%%%%%%%%%%%%%%%%%%%%%%%%%%%%%%%%%%%%%%%%%%%%
%%%%%%%%%%%%%%%%%%%%%%%%%%%%%%%%%%%%%%%%%%%%%%%%%%%%%%%%%%%%%%%%%%%%%%%%%%%%%%%

\section{Global cluster properties}
\label{properties.sec}

%%%%%%%%%%%%%%%%%%%%%%%%%%%%%%%%%%%%%%%%%%%%%%%%%%%%%%%%%%%%%%%%%%%%%%%%%%%%%%%
%%%%%%%%%%%%%%%%%%%%%%%%%%%%%%%%%%%%%%%%%%%%%%%%%%%%%%%%%%%%%%%%%%%%%%%%%%%%%%%

\subsection{Morphological analysis}
\label{morpho.sec}

In Figure \ref{mosaic-radio.fig} we show the {\it Chandra} image with
the superimposition of the GMRT radio contours at 330 MHz from
\citet{Bonafede_2014}.  The global X-ray morphology of the cluster has
a comet-like shape oriented in the NE-SW direction.  The central
cluster region encompassed by the radio halo is very bright, with an
X-ray peak (located at RA 10:50:49.4, Dec $-$28:04:44.3) which is
offset by $15.2''$ ($\sim$80 kpc) from the position of the main BCG
\citep[BCG-273, RA 10:50:50.2, Dec $-$28:04:55.7,][]{Daddona_2024}.
This separation, corresponding to roughly 0.05 $R_{\rm 500}$, is
indicative of a disturbed dynamical state
\citep[e.g.,][]{Rossetti_2016}.
We adopt the X-ray peak as the center of the X-ray analyses presented
in this paper. 

\subsubsection{Surface brightess profile}

%%%%%%%%%%%%%%%%%%%

\begin{figure*}
\centerline{
  \includegraphics[width=\columnwidth,bb=0 0 936 720]{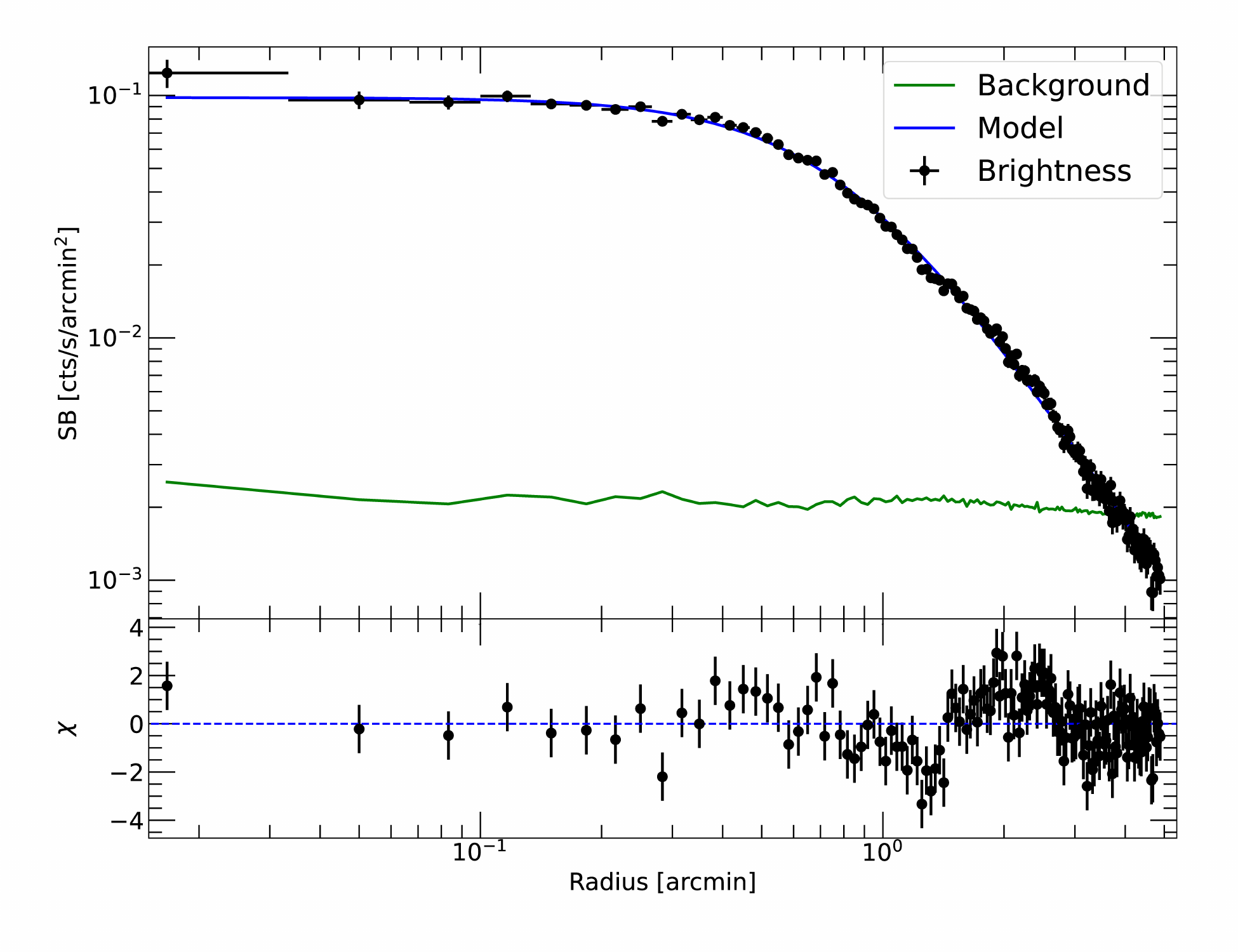}
  \includegraphics[width=\columnwidth,bb=0 0 936 720]{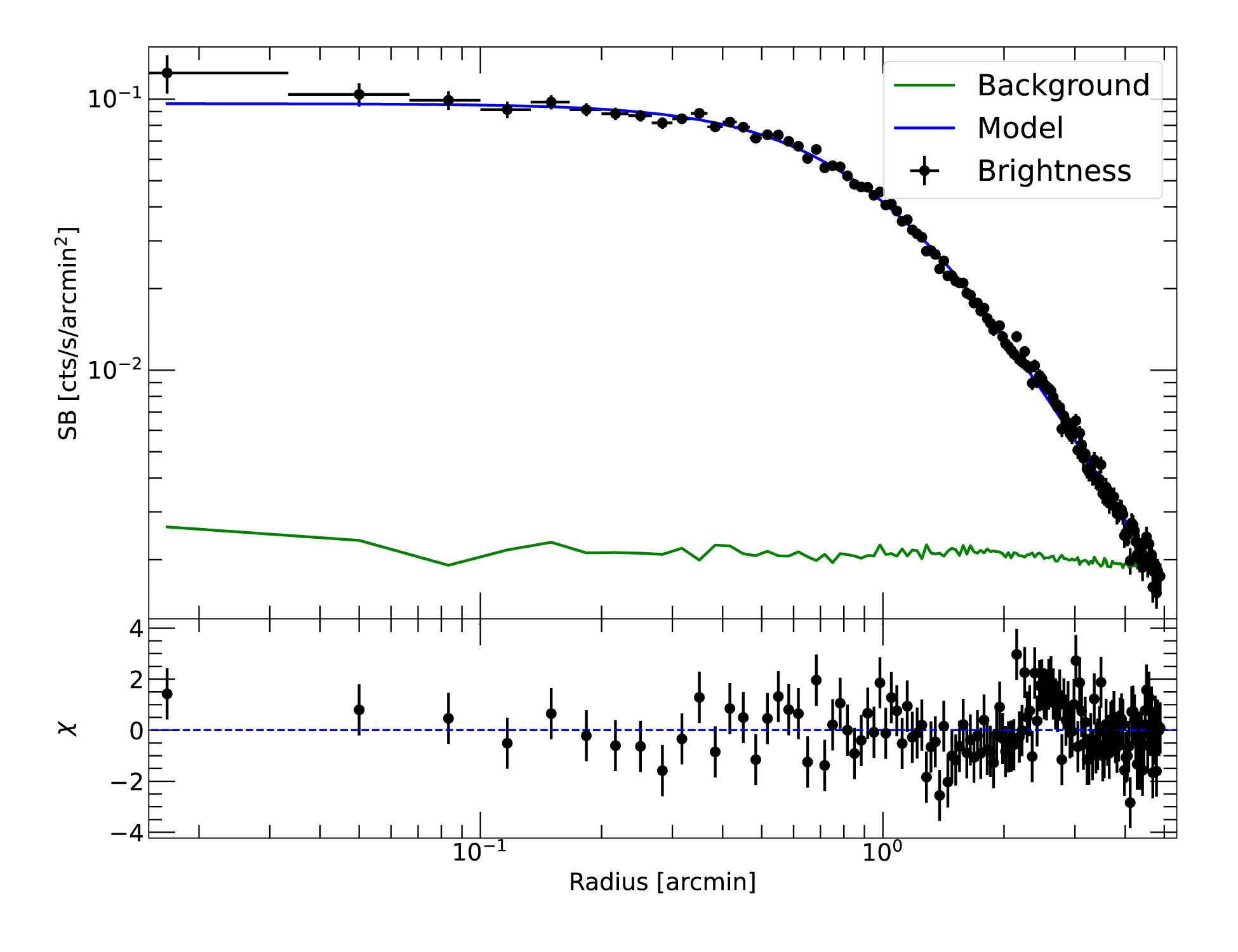}
}
\caption{\label{beta.fig} Background-subtracted, azimuthally-averaged
  radial surface brightness profile in the [0.5-7.0] keV energy band
  extracted in circular ({\it left panel}) and elliptical ({\it right
    panel}) annuli. The best-fit to a single $\beta$-model is shown
  with a blue line in both panels. The best-fit parameters are
  reported in Table \ref{beta.tab}. The radial distances are in units
  of arcmin, where 1 arcmin = 313.8 kpc.}
\end{figure*}

%%%%%%%%%%%%%%%%%%%

\begin{table*}
\caption{\label{beta.tab} Results of the fit to the surface brightness
  profile extracted up to $R_{\rm 500}$.}
% title of Table
\centering                          % used for centering table
\begin{tabular}{c c c c c c c c c}        % centered columns (4 columns)
  \hline
  \hline
  $\beta$-model & $\beta$ & $r_{\rm c1}$ & $r_{\rm c1}$ & $r_{\rm c2}$ & $r_{\rm c2}$ & $\log_{10}$ norm & $R$ & $\chi^2/{\rm dof}$
  \\
  ~ & ~ & (arcmin) & (kpc) & (arcmin) & (kpc) & (counts s$^{-1}$ arcmin$^{-2}$) & ~ & ~
  \\
  \hline
single (ellip.) & $0.595 \pm 0.005$ & $1.045 \pm 0.018$ & $328 \pm 6 $ & - & -& $-1.017 \pm 0.006$ & - & $164/143$
  \\[+1mm]
  single & $0.583 \pm 0.004$ & $0.816 \pm 0.013$ & $255 \pm 4 $ & - & -& $-1.008 \pm 0.006$ & - & $217/143$
  \\[+1mm]
  double & $0.583 \pm 0.004$ & $0.818 \pm 0.013$ & $257 \pm 4$ & $0.015 \pm 0.007$ & $5 \pm 2 $ & $-1.010 \pm 0.006$& $0.60 \pm 0.01$& $215/141$
  \\[+1mm]
  \hline
  \hline
\end{tabular}
{\tablefoot{Column 1:  $\beta$-model; column 2:
  $\beta$ parameter (assumed to be linked for the double $\beta$-model) ;
  column 3 (4): core radius of the first component, $r_{\rm c1}$, in arcmin (kpc); column 5 (6):
  core radius of the second component, $r_{\rm c2}$, in arcmin (kpc); column 7: normalization (see Eqs. \ref{1beta.eq} and \ref{2beta.eq}) ; column 8: ratio of the two components (see Eq. \ref{2beta.eq}); column 9: $\chi^2/{\rm dof}$. The elliptical profile was extracted in annuli with P.A. = 33$^\circ$ and axis ratio (major/minor) = 1.5; the best-fit core radius is the value along the major axis.}}
\end{table*}

%%%%%%%%%%%%%%%%%%%%

To study the X-ray morphology of the ICM we made use of the package
\texttt{pyproffit} \citep{Eckert_2020}, which is the python implementation
of the \texttt{proffit C++} software \citep{Eckert_2011} specifically
designed for the analysis of galaxy cluster X-ray surface brightness
profiles.  By providing as input the mosaic count image, background
image and exposure map\footnote{The exposure map was normalized by its
value at the aimpoint, resulting in units of s as requested by \texttt{pyproffit.}} (Sect. \ref{data.sec}), we extracted the
azimuthally-averaged surface brightness (SB) profile in 2$''$-wide annular bins up to $R_{500}$ (1541 kpc $\sim 4.9'$, indicated by the green circle in
Figure \ref{mosaic-radio.fig}).  We fitted the profile with a single
$\beta$-model, defined by the function:
\begin{equation}
{\rm SB}(r) = {\rm norm} \left (1 + \left(r/r_{\rm c} \right)^2 \right)^{0.5 - 3 \beta} \, ,
\label{1beta.eq}
\end{equation}
\noindent
where norm is the normalization expressed in units of $\log_{10}$ of
the central value, $r_{\rm c}$ is the core radius and
$\beta$ is the slope parameter. 
The best-fit 
$(\chi^2/{\rm dof} = 217/143 = 1.52)$ parameters are $\beta=0.583 \pm 0.004$ and core radius
$r_{\rm c} = (0.816 \pm 0.013)$ arcmin $\sim (246 \pm 1)$ kpc.
We also considered a double $\beta$-model, defined by the function
\begin{equation}
{\rm SB}(r) = {\rm norm} \left[ \left(1 + \left(r/r_{\rm c1}
    \right)^2 \right)^{0.5 - 3 \beta} + R \, \left(1 + \left(r/r_{\rm c2} \right)^2
  \right)^{0.5 - 3 \beta} \right] .
\label{2beta.eq}
\end{equation}
\noindent
We found that, although it provides a better fit to the central bin
($\sim 10.5$ kpc), the F-test indicated that the improvement is
not significant (F-stat = 0.66, F-prob = 0.52).

We further extracted the SB profile in elliptical annuli (P.A. = 33$^\circ$, 
axis ratio = 1.5), finding an improvement in
the fit $(\chi^2/{\rm dof} = 164/143 = 1.15)$. The best-fit parameter are
$\beta=0.595 \pm 0.005$ and core radius (measured along the major axis)
$r_{\rm c} = (1.045 \pm 0.018)$ arcmin $\sim (328 \pm 6)$ kpc.
{\color{black} 
On the other hand, we found that restricting the radial range to the central 1.3 arcmin ($\sim 408$ kpc), the fits with a single
$\beta$-model to the SB profiles extracted in circular annuli and elliptical annuli are indistinguishable ($\chi^2/{\rm dof}= 31.4/35 \; {\rm and} \; 33.0/35$, respectively). We verified that the circular and elliptical fits continue to be consistent at $1 \sigma$ up to $\approx 500$ kpc.} 
We finally verified that changing the radial binning of the profile
extraction by adopting different values and spacing provides
consistent best-fit parameters.

The results of the fits to the radial
profile {\color{black} up to $R_{\rm 500}$} are summarized in Table \ref{beta.tab} and the best-fit of the
circular and elliptical profile are shown in the left and right panels
of Figure \ref{beta.fig}, respectively.

\subsubsection{Residual and unsharp images: analysis of X-ray depressions}

%%%%%%%%%%%%%%%%%%%%%%%%%

\begin{figure*}
\centerline{
   \includegraphics[width=\textwidth,bb=36 264 577 529]{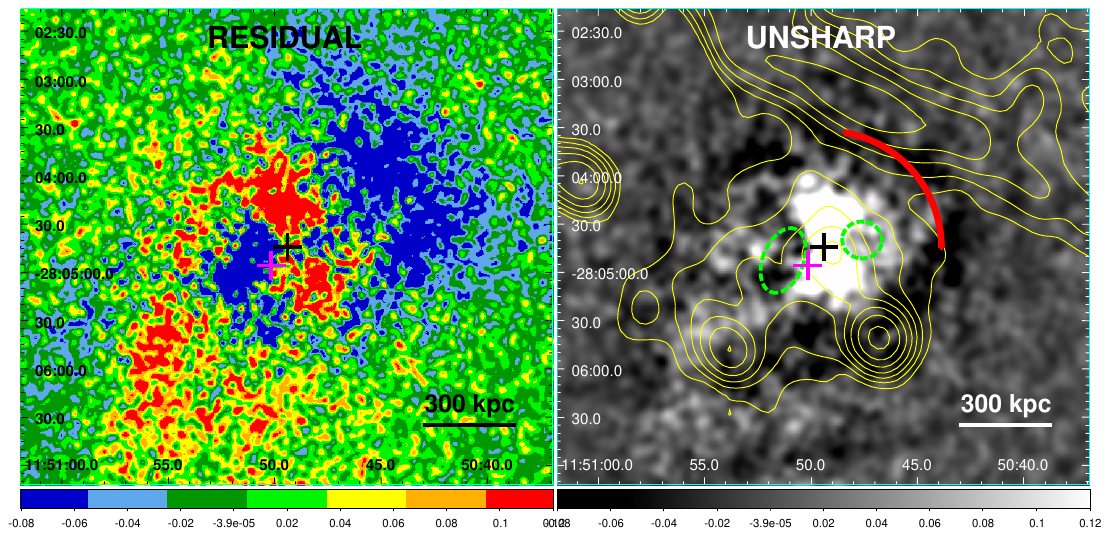}
}
\caption{\label{residual-unsharp.fig} Morphological substructures in PLCKG287 from filtered {\it Chandra} images. {\it Left panel:} Residual {\it
    Chandra} image of the central cluster region obtained by
  subtracting the best-fit elliptical $\beta$-model (see Table
  \ref{beta.tab}).
  As indicated by the color bar, 
  the positive and negative residuals are shown in red and 
  blue, respectively.  {\it Right
    panel:} Unsharp masked image obtained by subtracting a
  40$''$-smoothed image from a 5$''$-smoothed one. The green dashed
  ellipses show the possible X-ray depressions discussed in Section
  {\color{black} \ref{AGN-energy.sec}}, whereas the red arc indicates the position of the
  identified shock front (see Sect. \ref{fronts.sec}). Overlaid are the GMRT radio contours at 330 MHz (same as in Fig.
  \ref{mosaic-radio.fig}). 
  In both panels,
  the black and magenta crosses indicate the location of the X-ray
  peak (RA 10:50:49.4, Dec $-$28:04:44.3) and main BCG (RA 10:50:50.2,
  Dec $-$28:04:55.7), respectively.}
\end{figure*}

%%%%%%%%%%%%%%%%%%%%%%%%

%%%%%%%%%%%%%%%%%%%%%%%%%%%%%%%%%%%%%%%

\begin{figure*}[h]
  \hspace{0.1in}
  \includegraphics[width=0.3\textwidth,bb=18 144 592 718]{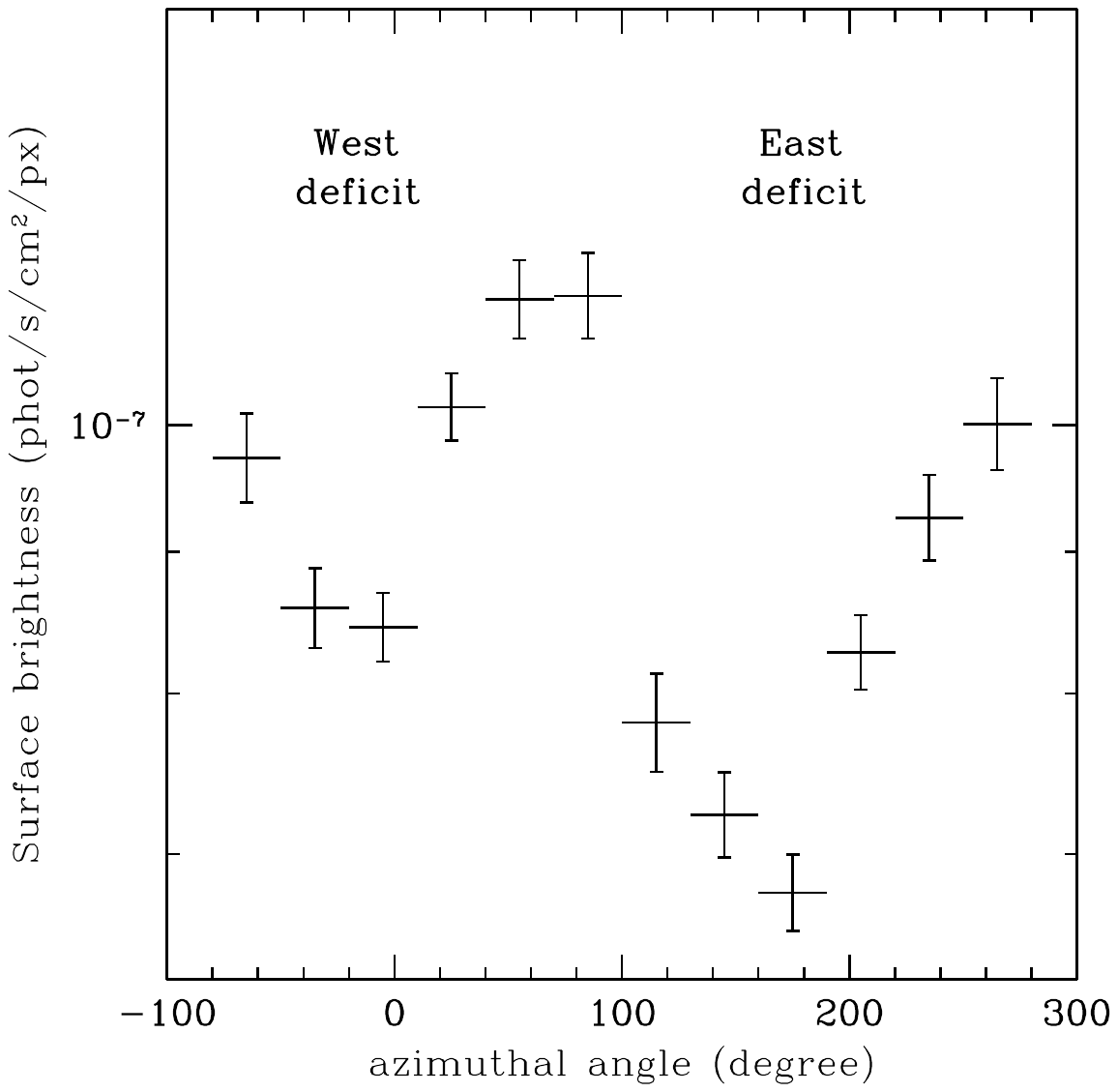}
  \includegraphics[width=0.33\textwidth,bb=36 144 577 648]{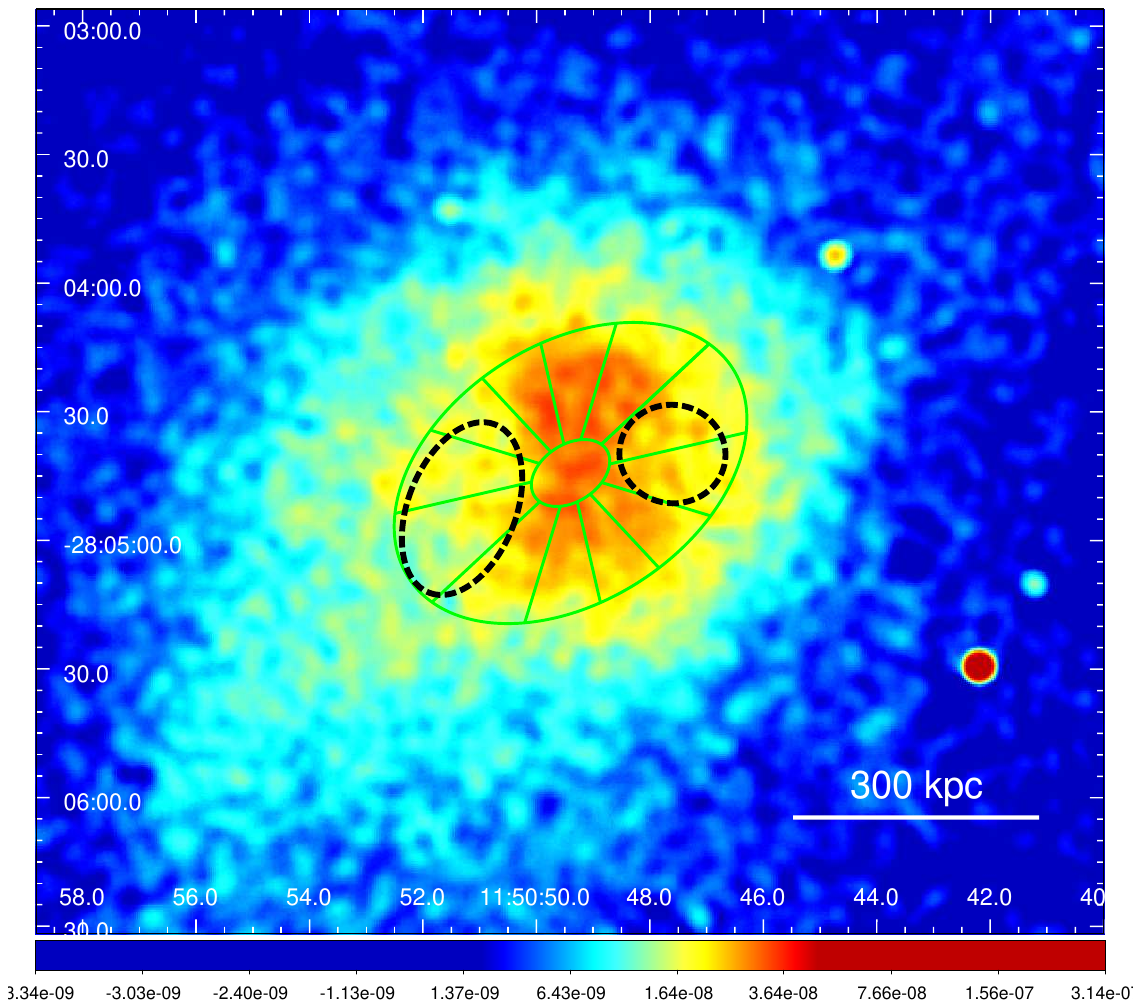}
  \includegraphics[width=0.33\textwidth,bb=0 0 720 576]{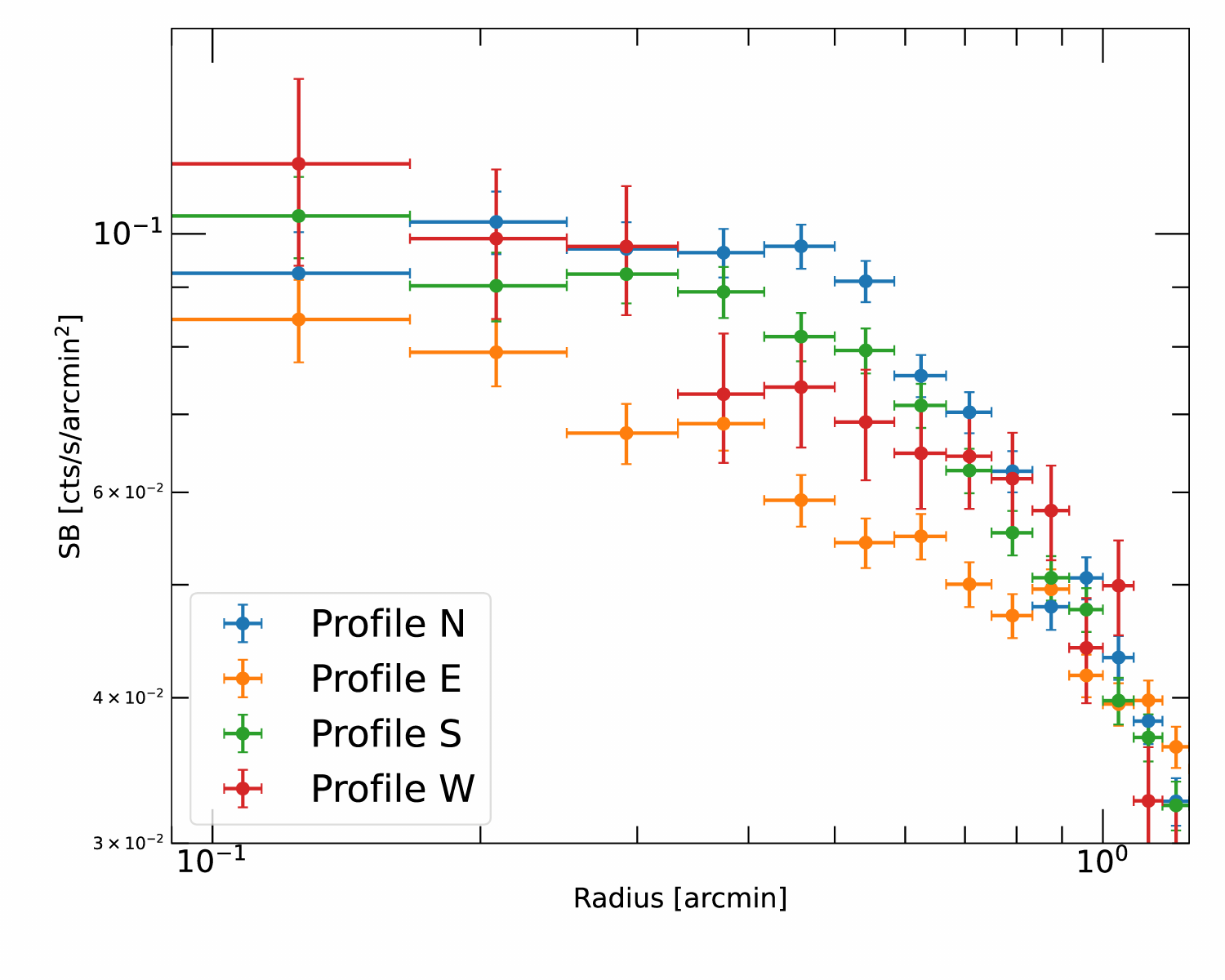}
\vspace{0.0cm}
\caption{\label{cavity.fig} Detection of E-W depressions in PLCKG287.  {\it Left panel:} Azimuthal SB profile
  around the cluster center extracted along the wedges shown in the
  middle panel.  The largest deficit is in the direction of the E
  depression seen in the residual and unsharp images (see
  Fig. \ref{residual-unsharp.fig}). A second dip is visible in the
  azimuthal profile along the direction of the W depression. Angles
  are measured counterclockwise, with $0^{\circ}$ to the West
  direction, and are shown starting from -90$^{\circ}$ to improve the
  visibility of the West deficit in the plot.  {\it Middle panel:}
  Background-subtracted, exposure corrected mosaic [0.5-7.0] keV {\it
    Chandra} image, smoothed with a kernel of 8 pixels, of the central
  region of PLCKG287.  The elliptical wedges used to evaluate the
  azimuthal SB profile around the center were chosen so as to mostly
  encompass the region of each depression (elliptical annulus between
  10$''$ and 45$''$ along the major axis, divided in 12 sectors of
  30$^\circ$ each) and are shown in green panda regions.  The black
  dashed ellipses show the identified X-ray depressions {\color{black} (see also Fig. \ref{cavity-appendix.fig})}. 
  {\it Right panel:} SB profiles extracted in
  5$''$-wide annular bins (with the same geometry as that used for the
  elliptical $\beta$-model, see Fig. \ref{cavity-appendix.fig}) in
  four sectors to the N (blue), E (yellow), S (green) and
  W (red) directions. With respect to the
  SB profiles in orthogonal directions, a clear depression is visible in the E
  profile in the radial range $\approx 0.25'$ - 0.75$'$, 
and a smaller deficit is noticeable in the W profile in the radial 
range $\approx 0.4'$ - 0.6$'$ {\color{black}.}
  The radial distances are indicated along the major axis, in units
  of arcmin (1 arcmin = 313.8 kpc).}
\vspace{-0.1in}
\end{figure*}

%%%%%%%%%%%%%%%%%%%%%%%%%%%%%%%

The residual image of the central region obtained by subtracting the
best-fit elliptical $\beta$-model is shown in Figure
\ref{residual-unsharp.fig} (left panel).
This image illustrates the deviations of the data from the best-fit
$\beta$-model.  Another method typically used in X-ray imaging to
highlight substructures in the ICM is that to produce unsharp masked
images.  In particular, we smoothed the {\it Chandra} image with two
Gaussian functions of small (3$''$-5$''$) and large (10$''$-40$''$)
kernel size; then, we derived an unsharp masked image by subtracting
the large-scale smoothed image from the small-scale smoothed one. We
tried various combinations which image features on different scales,
finding generally consistent results.  In the right panel of Figure
\ref{residual-unsharp.fig}, we show the unsharp image obtained by
subtracting the 40$''$ smoothed image from the 5$''$ smoothed one.
Notable structures such as edges and discontinuities identified in the
residual and unsharp images are analyzed and discussed in
Sect. \ref{fronts.sec} and {\color{black} \ref{shock-energy.sec}}. 
We further note a remarkable spatial correspondence between 
the X-ray feature surrounding the cluster central region 
(best visible in the unsharp mask image, 
right panel of Fig. \ref{residual-unsharp.fig}) and
the outer shape of the giant radio halo. This will be discussed in forthcoming papers (Rajpurohit et al. in preparation, Balboni et al. in preparation).

The visual inspection of both the residual and unsharp mask images also
reveals the presence of possible X-ray depressions in the E-W
direction with respect to the center.  The analysis of the azimuthal
variation of the background-subtracted, exposure corrected SB around
the cluster center, performed in wedges encompassing the X-ray
depressions (see Figure \ref{cavity.fig}, middle panel), indicates
that they are significant, particularly the E one (see Figure
\ref{cavity.fig}, left panel). We also investigated the radial
variation by extracting the SB profile in four sectors (see
Fig. \ref{cavity-appendix.fig}), finding confirmation for a large (size $\approx 80$ kpc)
depression in the E direction and hints of a smaller (size $\approx 60$ kpc) one in the
W direction (see Figure \ref{cavity.fig}, right panel). The
interpretation of these features will be discussed in
Sect. {\color{black} \ref{AGN-energy.sec} and} \ref{feedback.sec}.

%%%%%%%%%%%%%%%%%%%%%%%%%%%%%%%%%%%%%%%%%%%%%%%%%%%%%%%%%%%%%%%%%%%%%%%%%%%%%%%
%%%%%%%%%%%%%%%%%%%%%%%%%%%%%%%%%%%%%%%%%%%%%%%%%%%%%%%%%%%%%%%%%%%%%%%%%%%%%%%

\subsection{Spectral analysis}
\label{spectral.sec}

For each region of interest, we used the command {\ttfamily
  specextract} to extract a spectrum, along with its corresponding
background spectrum and event-weighted response matrices, for each
ObsID and each chip covered by that region. We then combined the
spectra with the command {\ttfamily combine\_spectra} in order to have
one single spectrum and its associated background spectrum for each
region.  We then grouped the spectrum to 25 counts per bin and
performed a spectral fitting with a single absorbed thermal model
({\ttfamily tbabs*apec}) in the energy range 0.5-7.0 keV with XSPEC
version 12.13.0c. The detected point sources (visible in Figure 
{\color{black} \ref{mosaics.fig}},
right) were subtracted before the spectral fitting. We adopted the
abundance ratios of \citet{Asplund_2009} and fixed the redshift
$z=0.383$.  We also fixed the absorbing column density to the galactic
hydrogen column value, $N_{\rm H}=6.93 \times 10^{20}$ cm$^{-2}$
\citep{HI4PI_2016}, and present in the Appendix \ref{freenh.sec} the
analysis performed by leaving $N_{\rm H}$ as a free parameter.

%%%%%%%%%%%%%%%%%%%%%%%%%%%%%%%%%%%%%%%%%%%%%%%%%%%%%%%%%%%%%%%%%%%%%%%%%%%%%%%

\subsubsection{Global ICM properties inside $R_{500}$}
\label{global.sec}

We first measured the global properties inside $R_{500}$ by
considering a circular region of radius = $294.7''$ (= 1541 kpc, see
the green circle in Figure 3).  The best-fit parameters
($\chi^2/{\rm dof}= 514/441=1.17$) are: temperature
$kT = 12.73^{+0.38}_{-0.40}$ keV \citep[consistent with {\it XMM} measurements,][]{Bagchi_2011}, 
and abundance $Z = 0.26\pm0.05$ Z$_{\odot}$. With the model {\ttfamily
  tbabs*clumin*apec} we estimated the unabsorbed luminosity in the
hard and bolometric (0.01-100 keV) bands to be
%$L_{0.4-2.4\, {\rm keV}}= (5.61 \pm 0.03) \times 10^{44}$ erg s$^{-1}$, 
$L_{2-10 \, {\rm keV}}= (1.09 \pm 0.01) \times 10^{45}$ erg
s$^{-1}$ and $L_{\rm bol}= (2.55 \pm 0.01) \times 10^{45}$ erg
s$^{-1}$, respectively.

%%%%%%%%%%%%%%%%%%%%%%%%%%%%%%%%%%%%%%%%%%%%%%%%%%%%%%%%%%%%%%%%%%%%%%%%%%%%%%%

\subsubsection{1D radial profiles of ICM thermodynamic properties}
\label{radial.sec}

\begin{figure*}[t]
\centerline{
  \includegraphics[width=0.9\columnwidth,bb=0 0 461 346]{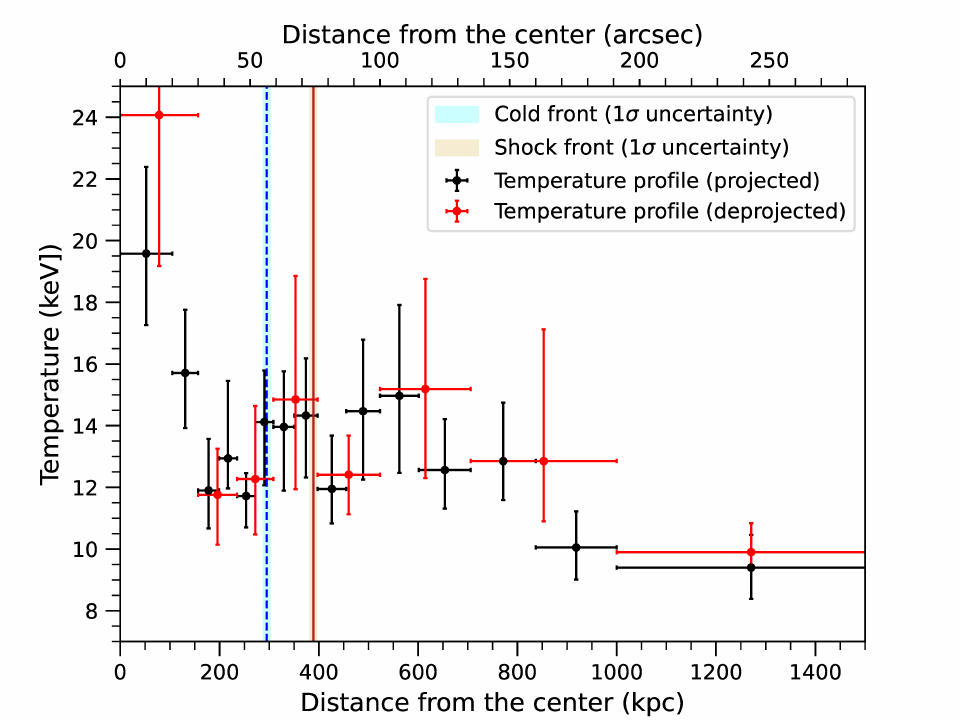}
  \includegraphics[width=0.9\columnwidth,bb=0 0 461 346]{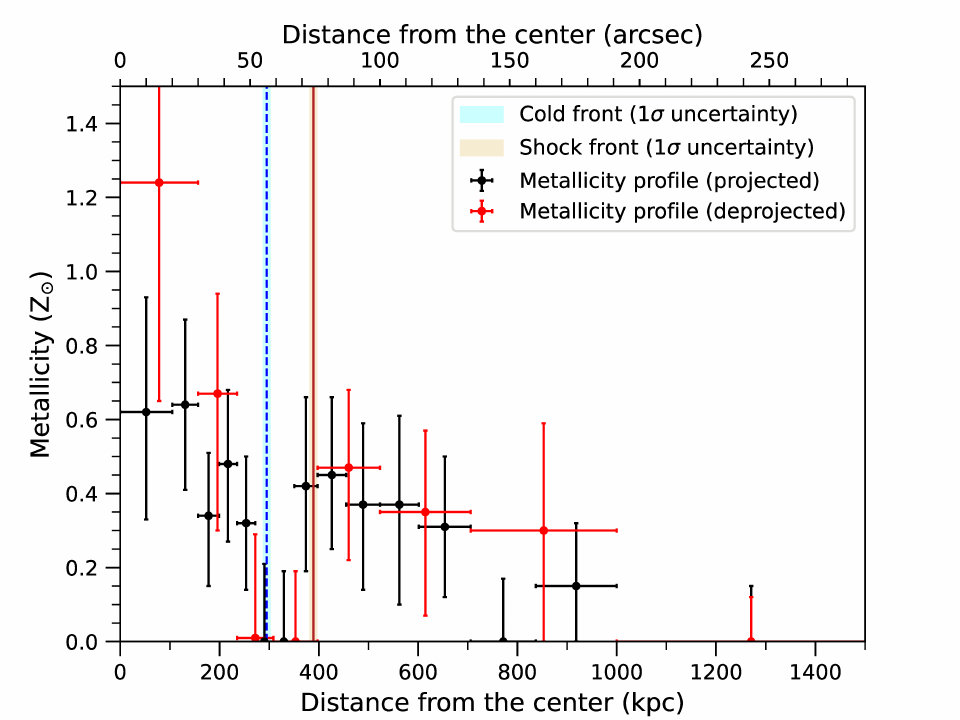}
}
\centerline{
   \includegraphics[width=0.9\columnwidth,bb=0 0 461 346]{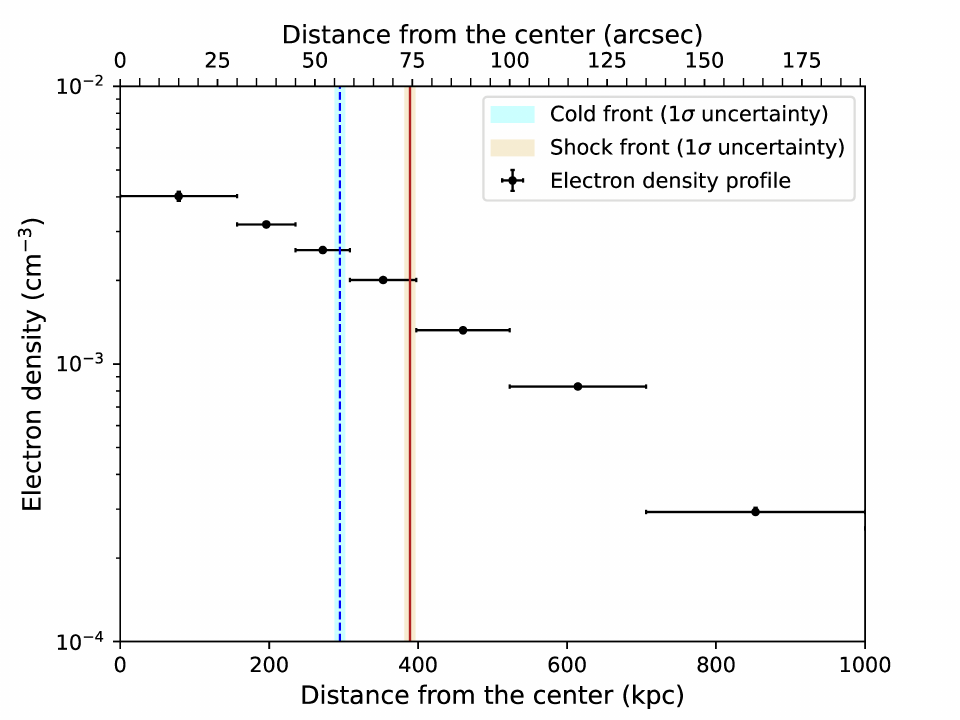}
  \includegraphics[width=0.9\columnwidth,bb=0 0 461 346]{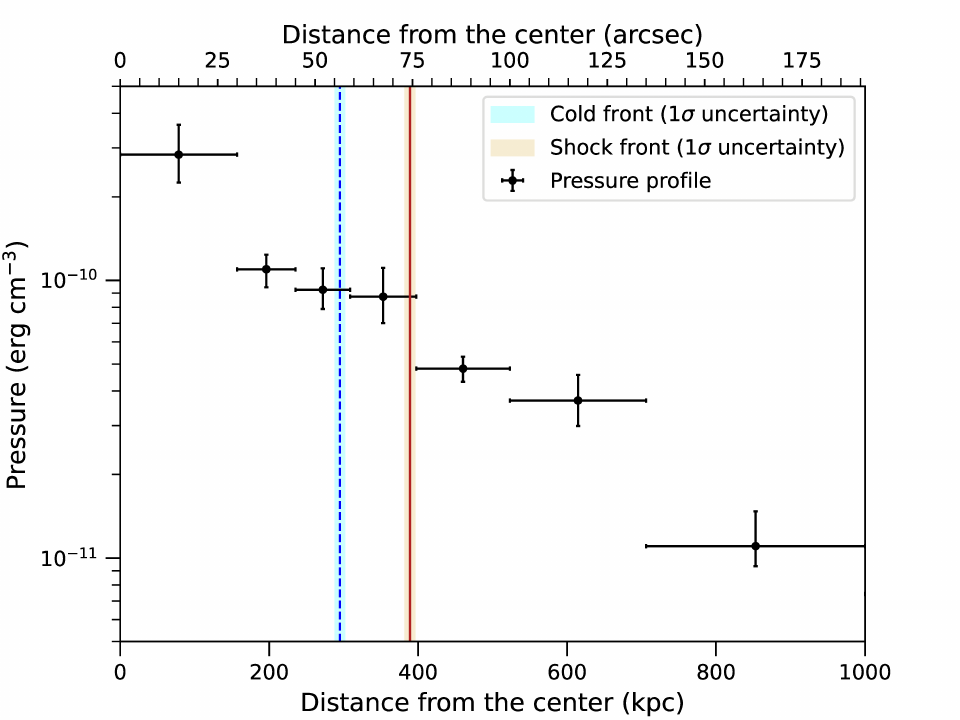}
}
\centerline{
   \includegraphics[width=0.9\columnwidth,bb=0 0 461 346]{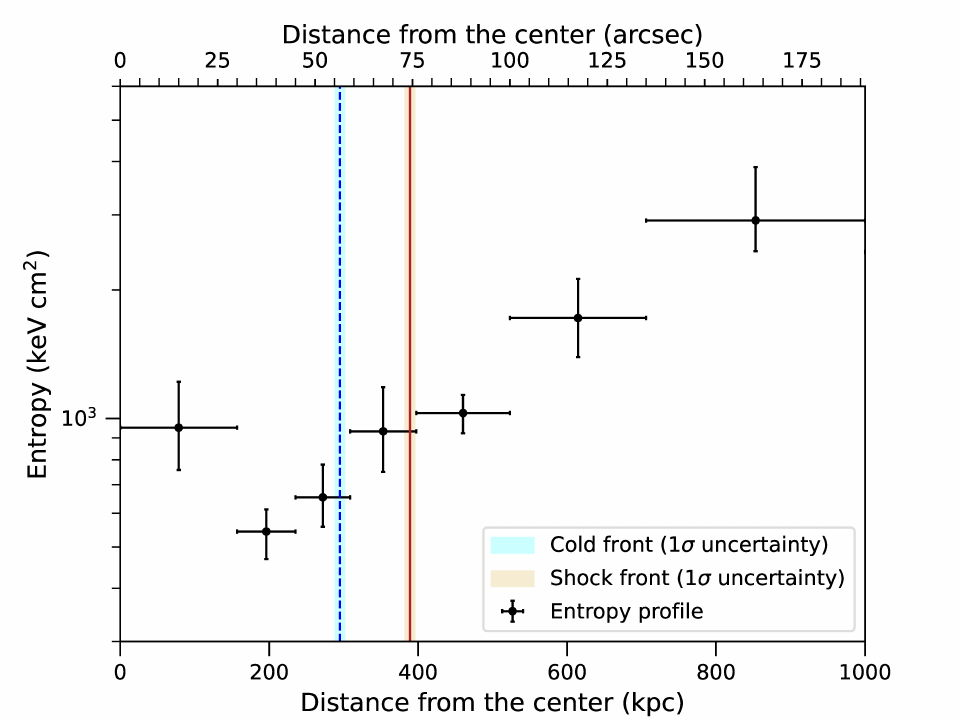}
  \includegraphics[width=0.9\columnwidth,bb=0 0 461 346]{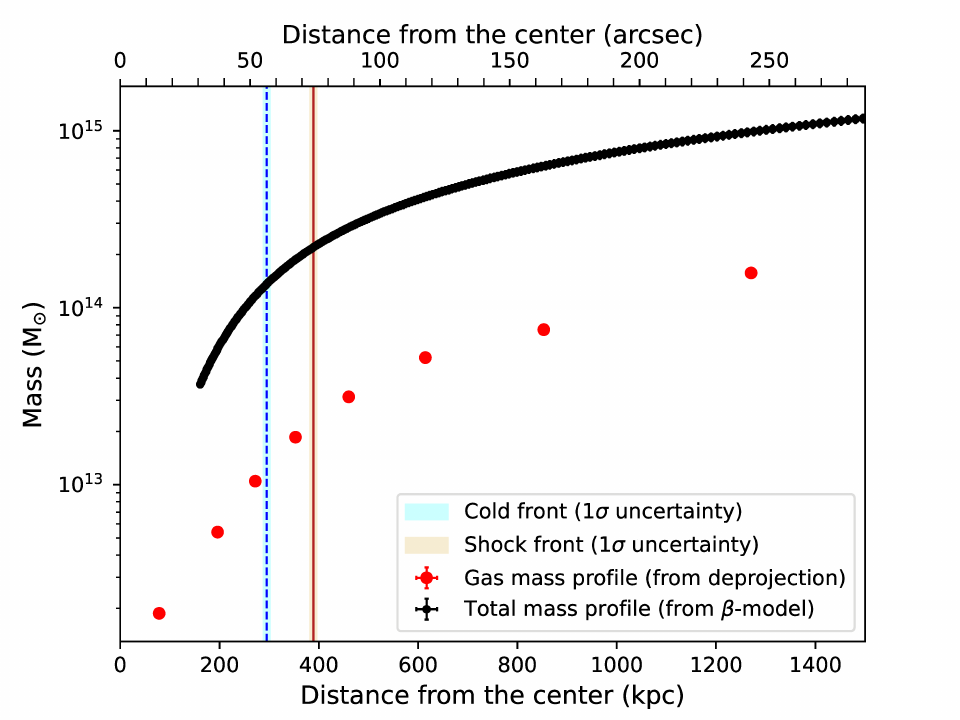}
}
\caption{\label{spectral-profiles.fig} Radial profiles of PLCKG287 obtained from the {\it Chandra} spectral analysis. {\it Top panels:} 
Temperature ({\it left}) and metallicity ({\it right}) 
    from the projected (black) and deprojected (red) spectral analysis (see Tables \ref{annuli.tab} and \ref{profile_deproj.tab}).  {\it Middle and Bottom panels:} 
    Density ({\it middle left}), pressure ({\it middle right}),
  entropy ({\it bottom left}) and integrated gas mass ({\it bottom right}, red points) 
  from the deprojected spectral analysis (see Table \ref{profile_deproj.tab}).
  The total mass profile ({\it bottom right}, black solid line) is estimated from the circular $\beta$-model by assuming an isothermal $kT \sim$ 12.7 keV.
  In all panels, the vertical
  lines show the position of the fronts discussed in Section
  \ref{fronts.sec} (the dashed blue line and solid red line indicate
  the cold front and shock, respectively).
}
\end{figure*}

In order to derive the azimuthally-averaged radial properties of the
ICM, we produced projected temperature and abundance profiles by
extracting spectra in circular annuli
{\color{black} \footnote{\color{black} As derived in Sect. \ref{morpho.sec}, the spherical assumption is a good approximation inside the central $\approx$500 kpc, where our investigation is focused (see Sect. \ref{discussion.sec}).
}
}
centered on the peak of the
X-ray emission (RA 10:50:49.4, Dec $-$28:04:44.3). The annular regions
were chosen in order to collect at least 3,000-3,500 net counts (see
Table \ref{annuli.tab}).  The best-fit parameters obtained from the
fits to the annular spectra are summarized in Table \ref{annuli.tab}
and the derived radial profiles of temperature and abundance are shown
in Figure \ref{spectral-profiles.fig} (top panels).  We note in
  particular that both profiles show an increase in the cluster
  center. This will be discussed in Sect. \ref{discussion.sec}.

%%%%%%%%%%%%%%

\begin{table*}
\caption{\label{annuli.tab} Results of the spectral fit to an absorbed apec model ({\ttfamily tbabs*apec}) performed in the [0.5-7.0] keV energy band in concentric annular regions extracted up to $R_{\rm 500}$.}
% title of Table
\centering                          % used for centering table
\begin{tabular}{c c c c c c c }        % centered columns (4 columns)
  \hline
  \hline
 $R_{\rm in}$ - $R_{\rm out}$ & $R_{\rm in}$ - $R_{\rm out}$ & Exposure & Net counts (\% total) & $kT$ & $Z$ & $\chi^2/$dof ($\chi^2_{\rm red})$  
\\[+1mm]
($''$) & (kpc) & (ksec) & ~ & (keV) & (Z$_{\odot}$) & ~ 
\\[+1mm]
  \hline
0-20 & 0-105 & 131.9 & 4333 (97.7) & $19.58_{-2.32}^{+2.81}$ & $0.62_{-0.29}^{+0.31}$ & 139.00/134 (1.04)
\\[+1mm]
20-30 & 105-157 & 131.9 & 4612 (97.4) & $15.71_{-1.78}^{+2.05}$ & $0.64_{-0.23}^{+0.23}$ & 138.93/140 (0.99)
\\[+1mm]
30-38 & 157-199 & 131.9 & 4039 (96.8) & $11.90_{-1.23}^{+1.67}$ & $0.34_{-0.19}^{+0.17}$ & 145.12/128 (1.13)
\\[+1mm]
38-45 & 199-235 & 131.9 & 3705 (96.2) & $12.94_{-0.97}^{+2.51}$ & $0.48_{-0.21}^{+0.20}$ & 111.63/117 (0.95)
\\[+1mm]
45-52 & 235-272 & 131.9 & 3402 (95.2) & $11.72_{-1.02}^{+0.74}$ & $0.32_{-0.18}^{+0.18}$ & 98.36/108 (0.91)
\\[+1mm]
52-59 & 272-309 & 131.9 & 3276 (94.5) & $14.12_{-2.05}^{+1.67}$ & $<0.21$ & 131.24/106 (1.24)
\\[+1mm]
59-67 & 309-350 & 131.9 & 3373 (93.3) & $13.96_{-2.07}^{+1.80}$ & $<0.19$ & 118.59/110 (1.08)
\\[+1mm]
67-76 & 350-398 & 131.9 & 3308 (91.4) & $14.33_{-2.01}^{+1.85}$ & $0.42_{-0.23}^{+0.24}$ & 108.91/113 (0.96)
\\[+1mm]
76-87 & 398-455 & 131.9 & 3489 (89.0) & $11.95_{-1.12}^{+1.73}$ & $0.45_{-0.20}^{+0.21}$ & 109.89/122 (0.90)
\\[+1mm]
87-100 & 455-523 & 175.1 & 4122 (87.6) & $14.47_{-2.22}^{+2.32}$ & $0.37_{-0.23}^{+0.22}$ & 144.75/143 (1.01)
\\[+1mm]
100-115 & 523-602 & 193.9 & 4165 (84.5) & $14.97_{-2.50}^{+2.94}$ & $0.37_{-0.27}^{+0.24}$ & 157.42/150 (1.05)
\\[+1mm]
115-135 & 602-706 & 244.3 & 4698 (80.2) & $12.56_{-1.25}^{+1.65}$ & $0.31_{-0.19}^{+0.19}$ & 167.54/170 (0.99)
\\[+1mm]
135-160 & 706-837 & 350.7 & 5067 (73.1) & $12.85_{-1.26}^{+1.90}$ & $<0.17$ & 201.42/194 (1.04)
\\[+1mm]
160-191.2 & 837-1000 & 350.7 & 4556 (61.5) & $10.05_{-1.04}^{+1.17}$ & $0.15_{-0.15}^{+0.17}$ & 216.94/206 (1.05)
\\[+1mm]
191.2-294.7 & 1000-1541 & 350.7 & 7289 (39.1) & $9.40_{-1.02}^{+1.06}$ & $<0.15$ & 400.60/385 (1.04)
\\[+1mm]
  \hline
  \hline
\end{tabular}
{\tablefoot{The temperature (in keV) and
metallicity \citep[as a fraction of the solar value,][]{Asplund_2009} were left as free
parameters. The absorbing column density was fixed at the Galactic value $N_{\rm H} = 6.93 \times 10^{20}$ cm$^{-2}$ \citep{HI4PI_2016} and linked among the different annuli. Column 1: inner and outer radius in arcsec; column 2: inner and outer radius in kpc ;
  column 3: combined exposure time; column 4: net counts (and percentage of total counts); column 5: temperature (keV); column 6: abundance (solar value); column 7:  $\chi^2/{\rm dof}$ . The global fit gives $\chi^2/{\rm dof}$= 2390.32/2326 = 1.028.}}
\end{table*}

%%%%%%%%%%%%%%%%%

%%%%%%%%%%%%%%%%%%%%%%%%

\begin{table*}
\caption{\label{profile_deproj.tab} Results of the deprojected spectral fit to an absorbed apec model ({\ttfamily projct*tbabs*apec}) performed in the [0.5-7.0] keV energy band in concentric annular regions extracted up to $R_{\rm 500}$. }
% title of Table
\centering                          % used for centering table
\begin{tabular}{c c c c c c c c}        % centered columns (4 columns)
  \hline
  \hline
 $R_{\rm in}$ - $R_{\rm out}$ & $R_{\rm in}$ - $R_{\rm out}$ &  $kT$ & $Z$ & $n_{\rm e}$ & $p$ & $S$  & $M_{\rm gas}(<R_{\rm out})$
\\[+1mm]
($''$) & (kpc)  & (keV) &  (Z$_{\odot}$) &  ($10^{-3}$ cm$^{-3}$) & ($10^{-11}$ erg cm$^{-3}$) & (keV cm$^2$)  & ($10^{13} M_{\odot}$)
\\[+1mm]
  \hline
0-30 & 0-157 & $24.07_{-4.89}^{+6.72}$ & $1.24_{-0.59}^{+0.88}$ & $4.02_{-0.15}^{+0.15}$ & $28.38_{-5.87}^{+8.01}$ & $952_{-195}^{+267}$ & $0.19 \pm 0.01$
\\[+1mm]
30-45 & 157-235   & $11.76_{-1.62}^{+1.49}$ & $0.67_{-0.37}^{+0.27}$ & $3.18_{-0.04}^{+0.06}$ & $10.97_{-1.52}^{+1.40}$ & $543_{-75}^{+69}$ & $0.54 \pm 0.01$
\\[+1mm]
45-59 & 235-309 & $12.27_{-1.80}^{+2.37}$ & $0.01_{-0.01}^{+0.28}$ & $2.57_{-0.06}^{+0.05}$ & $9.26_{-1.37}^{+1.80}$ & $654_{-96}^{+127}$ & $1.05 \pm 0.01$
\\[+1mm]
59-76 & 309-398  & $14.85_{-2.91}^{+4.00}$ & $<0.19$ & $2.01_{-0.04}^{+0.02}$& $8.74_{-1.72}^{+2.36}$ & $933_{-183}^{+251}$ & $1.86 \pm 0.01$
\\[+1mm]
76-100 & 398-523 & $12.41_{-1.28}^{+1.27}$ & $0.47_{-0.25}^{+0.21}$ & $1.32_{-0.01}^{+0.03}$  & $4.81_{-0.50}^{+0.50}$ & $1030_{-106}^{+106}$ & $3.13 \pm 0.02$
\\[+1mm]
100-135 & 523-706 & $15.19_{-2.89}^{+3.57}$ & $0.35_{-0.28}^{+0.22}$ & $0.83_{-0.01}^{+0.02}$ & $3.69_{-0.70}^{+0.90}$ & $1720_{-328}^{+404}$ & $5.23 \pm 0.04$
\\[+1mm]
135-191.2 & 706-1000 & $12.85_{-1.95}^{+4.27}$ & $0.30_{-0.30}^{+0.29}$ & $0.29_{-0.01}^{+0.01}$ & $1.10_{-0.17}^{+0.37}$ & $2912_{-445}^{+971}$ & $7.52 \pm 0.07$ 
\\[+1mm]
191.2-294.7 & 1000-1541 & $9.90_{-0.44}^{+0.94}$ & $<0.12$ & $0.26_{-0.01}^{+0.01}$ & $0.74_{-0.03}^{+0.07}$ & $2456_{-112}^{+233}$ & $15.73 \pm 0.08$ 
\\[+1mm]
  \hline
  \hline
\end{tabular}
{\tablefoot{
Column 1: inner and outer radius in arcsec; column 2: inner and outer radius in kpc ;
column 3: deprojected temperature (keV); column 4: abundance ($Z_{\odot}$); column 5: electron density (cm$^{-3}$); column 6: pressure ($10^{-11}$ erg cm$^{-3}$);
column 7: entropy (keV cm$^2$); column 8: integrated gas mass ($10^{13} M_{\odot}$). The global fit gives $\chi^2/{\rm dof}$= 1972.66/1920 = 1.027.}}
\end{table*}  

%%%%%%%%%%%%%%%%%%%%

%%%%%%%%%%%%%%%%%

%%%%%%%%%%%%%%%%%%%%

The spectral properties thus derived are the emission-weighted
superposition of radiation originating at all points along the line of
sight through the cluster atmosphere.  In order to account for
projection effects, we further performed a deprojection analysis by considering 
the model {\ttfamily projct*tbabs*apec}, where the first component performs a 
3D to 2D projection of ellipsoidal shells on to elliptical annuli.
The high number of parameters in the {\ttfamily projct} model requires high photon statistics, therefore we binned the annuli two by two (but the last one).
 
We also estimated various quantities derived from the deprojected
spectral fits. The electron density $n_{\rm e}$ is obtained from the
Emission Integral $EI = \int n_{\rm e} n_{\rm p} dV$ given by the
{\ttfamily apec} normalization,
norm=$10^{-14} EI / ( 4 \pi [D_{\rm A} (1+z)]^2 )$. We assumed
$n_{\rm e} \sim 1.2 n_{\rm p}$ in the ionized intra-cluster plasma
\citep[e.g.,][]{Gitti_2012}. By starting from the deprojected density
and temperature values, we then calculated the gas pressure as
$p =n kT$, where we assumed $n = n_{\rm e} + n_{\rm p}$, and the gas
entropy from the commonly adopted definition
$ S = kT \, n_{\rm e}^{-2/3}$.  The results of the deprojection
analysis are reported in Table \ref{profile_deproj.tab} and the
corresponding deprojected temperature, density, pressure and entropy
profiles are shown in Figure \ref{spectral-profiles.fig}.  
We note in particular an entropy and temperature increase in the very center 
($r \lessapprox 160$~kpc). This will be discussed in Sect. \ref{discussion.sec}. 

%%%%%%%%%%%%%%%%%%%%%%%%%%%%%%%%%%%%%%%%%%%%%%%%

\begin{figure*}
 \centerline{
     \includegraphics[width=\textwidth,bb=36 193 577 599]{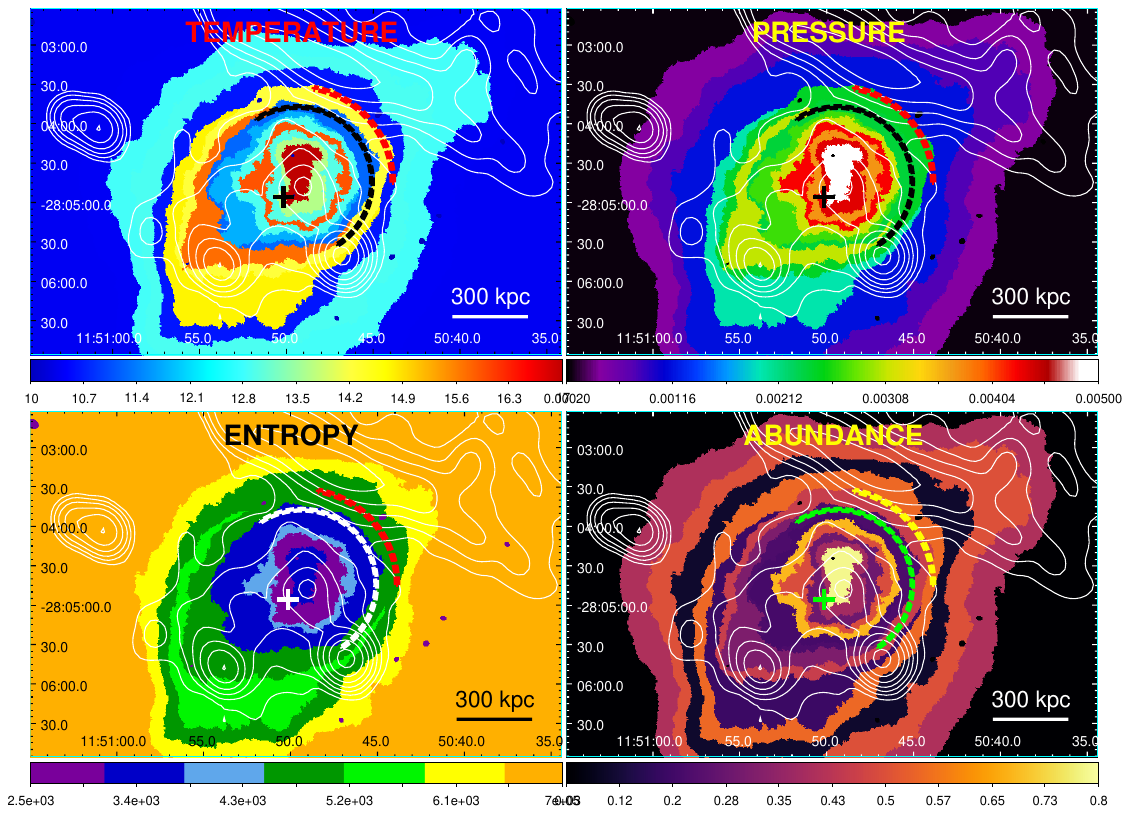}
}
\caption{\label{maps.fig} Spectral maps of the ICM thermodynamical
  properties in PLCKG287 (see Section \ref{maps.sec} for details) with
  overlaid the GMRT radio contours at 330 MHz (same as in Figure
  \ref{mosaic-radio.fig}). {\it Upper panels}: temperature (keV) and
  pseudo-pressure (arbitrary units) maps. {\it Bottom panels}:
  pseudo-entropy (arbitrary units) and abundance (solar units) maps.
  The dashed arcs (indicated in different colors in the
  various panels to enhance visibility depending on the color scale
  adopted) mark the positions of the surface brightness edges detected
  in Section \ref{fronts.sec}. All maps are centered in the X-ray
  peak, whereas the cross indicates the location of the main BCG. }
\end{figure*}

%%%%%%%%%%%%%%%%%%%%%%%%%%%%%%%%%%%%%%%%%%%%%%

We further measured the gas mass profile by integrating in shells the gas
density (estimated from the electron density as
$\rho = 1.92 \, \mu \, n_{\rm e} \, m_{\rm p}$, where $\mu \sim$ 0.6 is the mean
molecular weight and $m_{\rm p}$ is the proton mass), finding
$M_{\rm gas} (< R_{\rm 500}) \sim 1.6 \times 10^{14}$ M$_{\odot}$.  As a
simple estimate, we also derived the analytical profile of the
hydrostatic total mass obtained from the best-fit $\beta$-model
parameters \citep[see e.g., Eq. 20 of][]{Gitti_2012}, by assuming an
isothermal value of $kT = 12.73^{+0.38}_{-0.40}$ keV estimated inside
$R_{\rm 500}$ (see Sect. \ref{global.sec}). We 
consider this a robust approximation of the temperature outside of the cluster core (see
Fig. \ref{spectral-profiles.fig}, top left panel), therefore we show the profile only for
radial distances \gtsim 160 kpc. We estimated
$M_{\rm tot} (<R_{\rm 500}) \approx 1.2 \times 10^{15}$ M$_{\odot}$, in
agreement with the nominal integrated mass estimated from Planck SZ
data \citep[$\sim 1.4 \times 10^{15}$ M$_{\odot}$,][]{Planck_2016}. The mass
profiles are shown in Fig. \ref{spectral-profiles.fig} (bottom right panel).

%%%%%%%%%%%%%%%%%%%%%%%%%%%%%%%%%%%%%%%%%%%%%%%%%%%%%%%%%%%%%%%%%%%%%%%%%%%%%%%

\subsubsection{2D spectral maps of ICM thermodynamic properties}
\label{maps.sec}

To enable high-resolution 2D mapping of the ICM thermodynamic
properties, we built maps by using the contour binning technique
\citep[CONTBIN,][]{Sanders_2006} and setting\footnote{We show the
  spectral maps that we consider the best ones after trying different sets
  of parameters.} a minimum signal-to-noise ratio (S/N) of 55. The
spectrum extracted from each region was fit with a {\ttfamily tbabs *
  apec} model, leaving the temperature, abundance and normalization
free to vary. Consistently with the previous analysis
(Sect. \ref{global.sec} and \ref{radial.sec}), the absorbing column
density was fixed at the value $N_{\rm H}=6.93 \times 10^{20}$
cm$^{-2}$ \citep{HI4PI_2016}.
By combining the temperature and
normalization, we further derived maps of pseudo-pressure and
pseudo-entropy (in arbitrary units) using the method described in
the literature \citep[e.g.,][]{Ubertosi_2023}.

The resulting maps (Figure \ref{maps.fig}) show that the
higher abundance gas is preferentially found inside the central
$\sim 300$ kpc from the X-ray peak and is bounded by a region of cold gas, which
is in turn externally surrounded by a region of hot gas. This
hot ICM region is narrower ($\sim$90 kpc) towards the 
NW direction and corresponds to a region of higher ICM pressure,
possibly indicating shock compression. Across the internal cold ICM region the
pressure is instead nearly constant, suggesting the presence of a
contact discontinuity (cold front).  These features will be further investigated in
Section \ref{fronts.sec} by means of dedicated morphological and
spectral analyses of the radial profiles.

%%%%%%%%%%%%%%%%%%%%%%%%%%%%%%%%%%%%%%%%%%%%%%%%%%%%%%%%%%%%%%%%%%%%%%%%%%%%%%%
%%%%%%%%%%%%%%%%%%%%%%%%%%%%%%%%%%%%%%%%%%%%%%%%%%%%%%%%%%%%%%%%%%%%%%%%%%%%%%%
%%%%%%%%%%%%%%%%%%%%%%%%%%%%%%%%%%%%%%%%%%%%%%%%%%%%%%%%%%%%%%%%%%%%%%%%%%%%%%%

\section{Study of the surface brightness discontinuities}
\label{fronts.sec}

%%%%%%%%%%%%%

\begin{figure*}
  \centerline{
    \hspace{-0.1cm}
    \includegraphics[width=\columnwidth,bb=0 0 576 360]{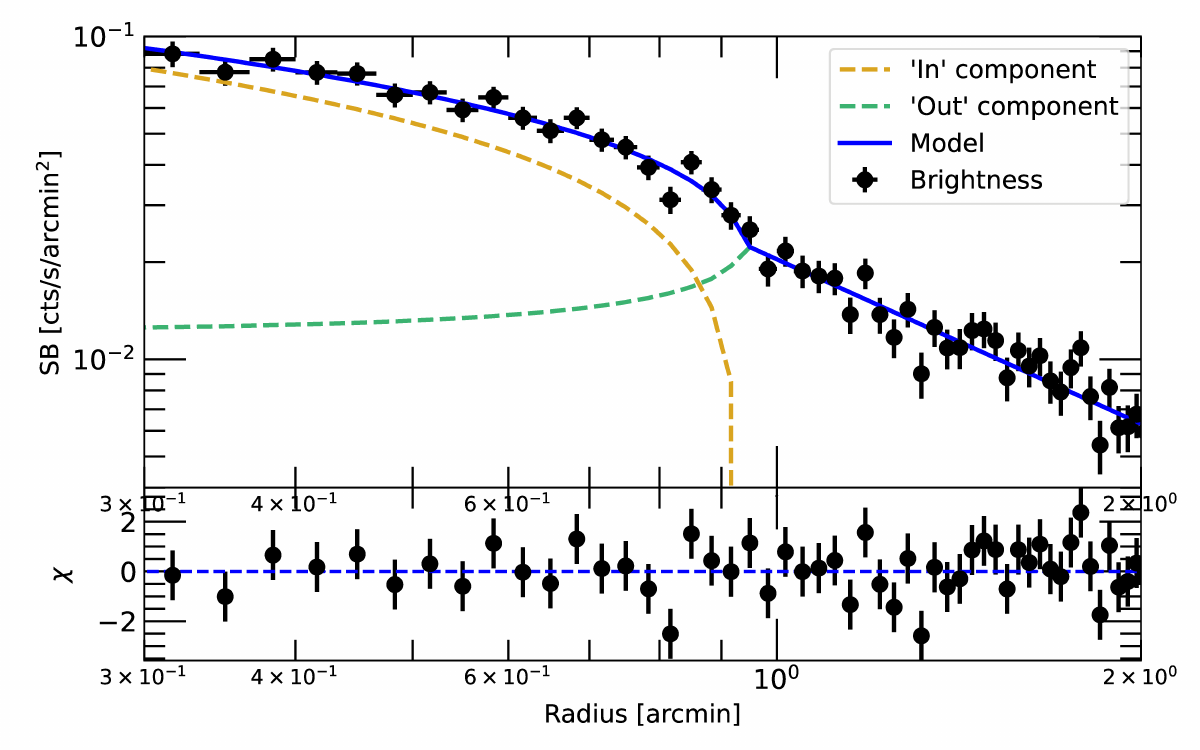}
        \includegraphics[width=\columnwidth,bb=0 0 576 360]{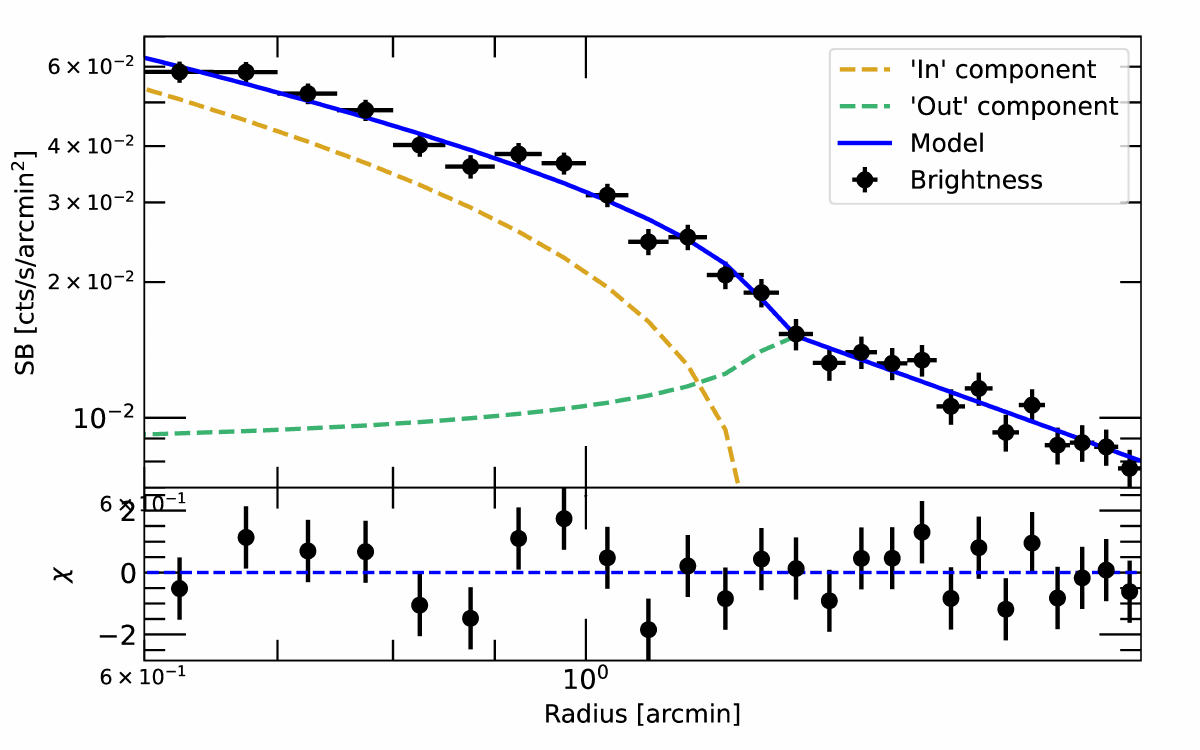}    
}
\caption{\label{sb-edges.fig} Surface brightness profiles across the
  NW direction in a semicircle (from $-57^{\circ}$ to $+123^{\circ}$
  from W and increasing counterclockwise, left panel) and in a
  narrower $80^{\circ}$-wide sector (from $0^{\circ}$ to $80^{\circ}$
  from W and increasing counterclockwise, right panel). In each panel,
  the best-fit broken power-law model is overlaid in blue, while the
  residuals are shown in the lower boxes and the individual components
  of the SB associated to the broken power-law inside (``In'') and
  outside (``Out'') the front are shown with orange and green dashed
  lines, respectively. As discussed in the text, we interpret the
  inner edge (left panel,
  $R_{\rm inner} = 0.94' \pm 0.02' \sim 295 \pm 6$ kpc) as a cold
  front and the outer edge (right panel,
  $R_{\rm outer} = 1.24' \pm 0.02' \sim 389 \pm 6$ kpc) as a shock
  front. The radial distances are in units of arcmin, where 1 arcmin =
  313.8 kpc.}
\end{figure*}

%%%%%%%%%%%%%%%%%%

%%%%%%%%%%%%%%%%%%%%%%%%%%%%%%%%%%%%%%%%%%%%%%%%%%

\begin{figure*}[t]
\hspace{0.in}
  \includegraphics[width=0.33\textwidth,bb=36 172 577 620]{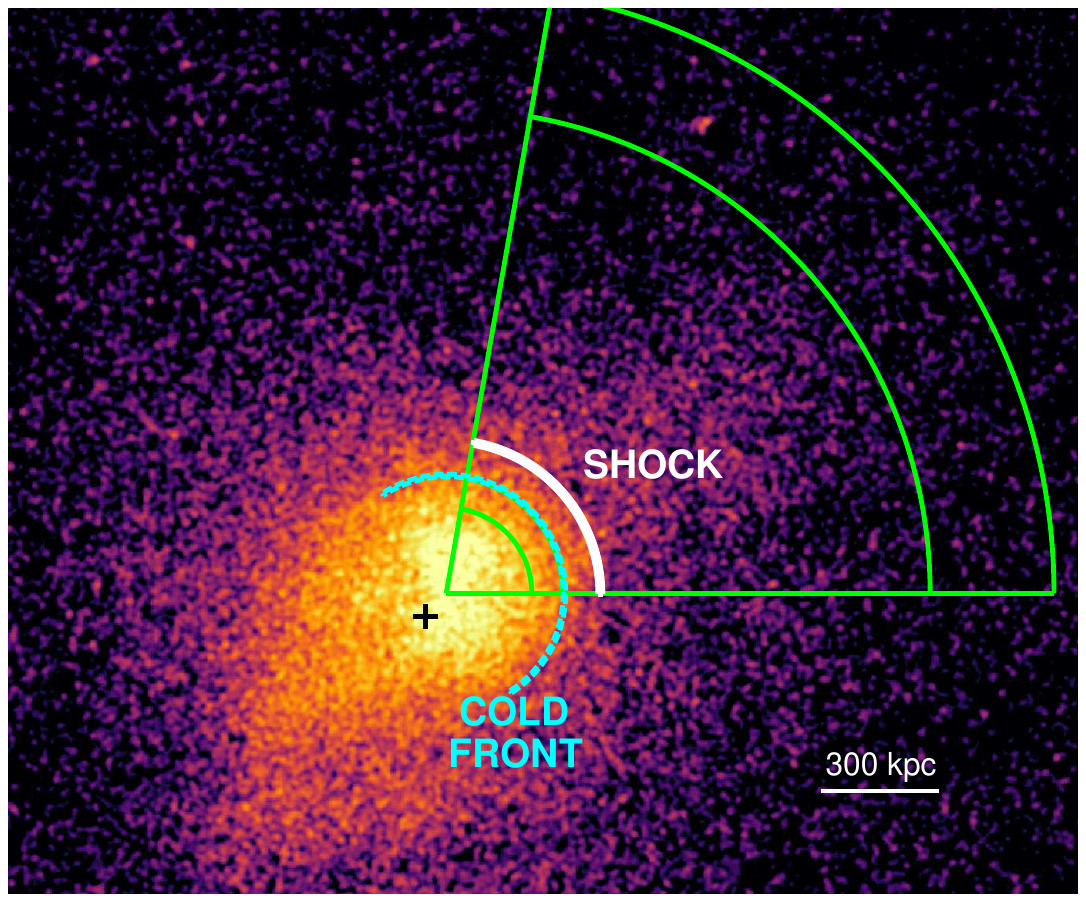}
  \includegraphics[width=0.34\textwidth,bb=0 0 461 346]{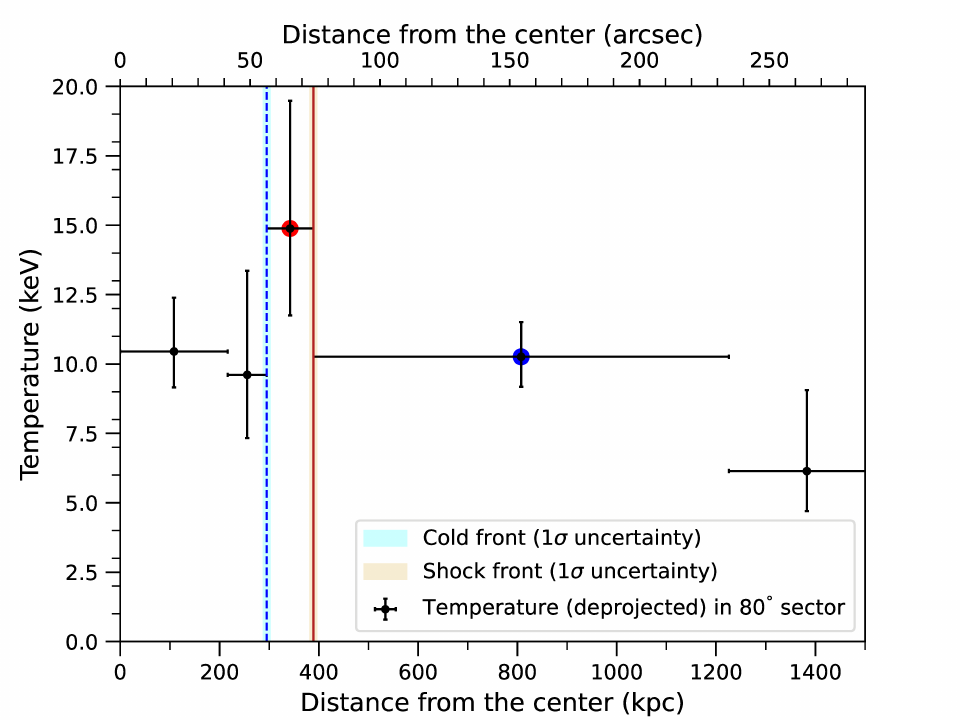}
  \includegraphics[width=0.34\textwidth,bb=0 0 461 346]{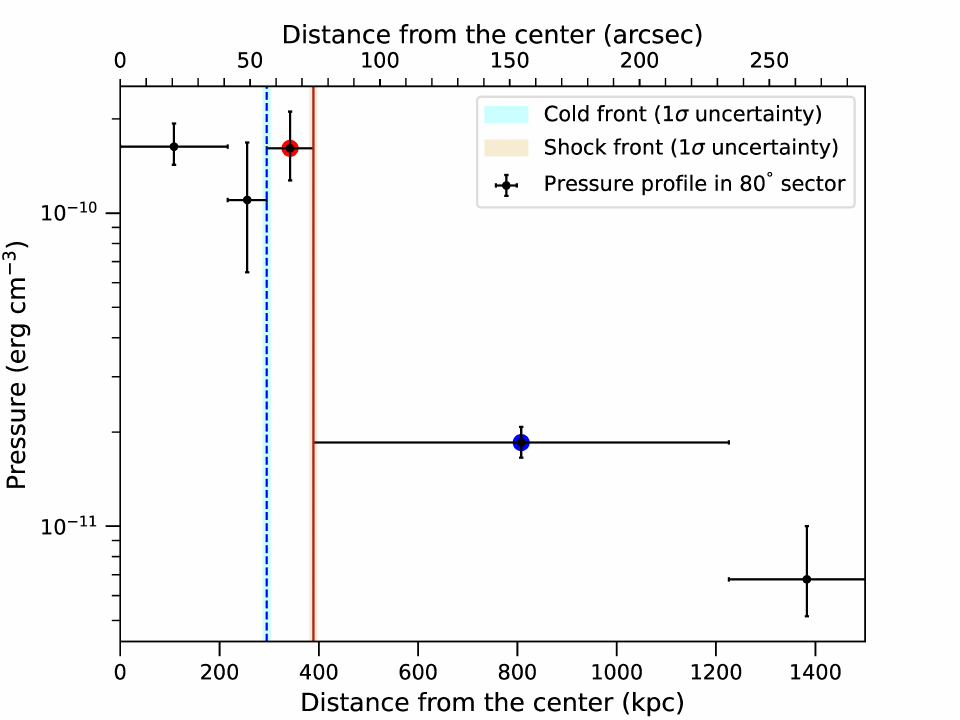}
\vspace{0.0cm}
\caption{\label{sectors.fig} Spectral analysis of the shock and cold fronts in PLCKG287. {\it Left panel:} mosaic [0.5-7.0] keV
  {\it Chandra} image of the central region of PLCKG287 smoothed with a
  kernel of 5 pixels showing the wedges used for the spectral analysis
  of the surface brightness edges. The white arc and the cyan dashed
  semicircle indicates the position of the identified shock and cold
  front, respectively.   {\it Middle and right panels:}
  Deprojected temperature (middle) and pressure (right) profiles
  measured along the wedges shown in the left panel.  The vertical
  lines show the position of the fronts identified in Section
  \ref{fronts.sec}, in particular the dashed blue line and solid red
  line indicate the cold front and shock, respectively.}
\vspace{-0.1in}
\end{figure*}

%%%%%%%%%%%%%%%%%%%%%%%%%%%%%%%%%%%%%%%%%%%%%%%%%%%%%

We investigated the presence and nature of possible surface brightness
discontinuities by performing detailed morphological and spectral
analyses of the radial profiles along different sectors. The strategy
we adopted was to first estimate the exact position and magnitude of
density jumps by studying the surface brightness profile across each
edge and then measure the thermodynamic properties (in particular temperature
and pressure) inside and outside the front to determine its nature.

In particular, we performed a systematic search for edges in the ICM
by extracting surface brightness profiles (centered on the X-ray
peak)\footnote{ We also considered sectors with offset centers in
  order to better represent the shape of the features visually
  identified, finding consistent results.} in circular or elliptical
sectors of varying opening angles (between 30$^{\circ}$ and
180$^{\circ}$) and different binning (between 1$''$ to 3$''$). The
resulting profiles were fit in \texttt{pyproffit} with a single power-law model 
and with a broken power-law model: a surface brightness edge at a distance
$R_{J}$, characterized by a density jump $J$, was considered a
detection if an F-test between the single and the broken power-law
indicated a significant statistical improvement (more than 99\%
confidence).

The best-fit results were obtained in sectors along the NW direction 
{\color{black} (see Fig. \ref{sb-edges.fig})}.
We detected an inner edge in a $180^{\circ}$-wide sector (from
$-57^{\circ}$ to $+123^{\circ}$ from W and increasing counterclockwise) at
$R_{\rm inner} = 0.94' \pm 0.02' = 56.4'' \pm 1.2'' \sim 295 \pm 6$ kpc from the
center, characterized by a density jump $J_{\rm inner} = 1.44 \pm 0.08$
($\chi^2/$dof = 53.0/51).  We also detected an outer edge in a
$80^{\circ}$-wide sector (included in the previous semicircle, from
$0^{\circ}$ to $80^{\circ}$ from W and increasing counterclockwise) at
$R_{\rm outer} = 1.24' \pm 0.02 = 74.4'' \pm 1.2'' \sim 389 \pm 6$ kpc from the
center, characterized by a density jump $J_{\rm outer} = 1.43 \pm 0.10$
($\chi^2/$dof = 24.2/24).  We verified that slightly varying the
radial binning and sector aperture of the profile extraction always
provided consistent results.

Given the geometry of the two nested fronts, which are aligned along
the same direction, we measured their spectral properties along the
common sector having an opening angle of $80^{\circ}$ (from
$0^{\circ}$ to $80^{\circ}$ from W and increasing counterclockwise), 
that is the one in which the outer front is detected. In particular, we
extracted the spectra of the ICM inside and outside each
discontinuity, with the addition of a central region and an outermost
region extending up to $R_{\rm 500}$ to account for deprojection, for a
total of 5 spectral wedges.  Fitting the 0.5-7.0 keV band spectra with
a {\ttfamily projct*tbabs*apec} model returned the deprojected
temperature and electron density, which were combined to derive the
pressure jump across each edge.  A summary of the deprojected
thermodynamic properties along the considered sector is reported in Table
\ref{spectra-front.tab}.

Across the inner edge, located at $\sim 295$ kpc, we measure a
temperature jump $kT_{\rm
  in}/kT_{\rm out} = 0.65^{+0.29}_{-0.25}$ and a pressure jump $p_{\rm
  in}/p_{\rm out} = 0.68^{+0.39}_{-0.35}$. These values are consistent with
the interpretation of this edge as a cold front, that is, a contact
discontinuity characterized by a continuous pressure and a lower
temperature inside the front.

On the other hand, across the outer edge located at $\sim 389$ kpc we measure a temperature jump $kT_{\rm
  in}/kT_{\rm out} = 1.45^{+0.33}_{-0.35}$ and pressure jump $p_{\rm
  in}/p_{\rm out} = 8.7^{+2.8}_{-2.1}$. These values are consistent with
the interpretation of this edge as a shock front which has increased
the temperature\footnote{We verified that the temperature jump is
  detected also in the projected spectral fit: $kT_{\rm
  in}=13.96^{+3.40}_{-2.51}$ keV vs. $kT_{\rm out} =
9.47^{+0.82}_{-0.81}$ keV.} and pressure of the ICM after its passage.
By using the Rankine - Hugoniot (R-H) conditions (e.g.,
\citealt{Landau_1960}),   we estimated the Mach number $\mathcal{M}$ of
the shock front from the best-fit density jump ($J_{\rm outer} = 1.43 \pm 0.10$) 
detected in the surface brightness edge as
\begin{equation}
    \label{machn}
    \mathcal{M} = \left( \frac{3J}{4-J} \right)^{1/2} \, ,
\end{equation}
finding a value $\mathcal{M} = 1.29 \pm 0.07$ (ignoring systematic
errors such as projection and sector choice).
\\

Moreover, the Mach number can be used to predict the expected temperature
jump across the post- and pre-shock region by using again the R-H conditions:

\begin{equation}
    \label{machkt}
    \frac{T_{\rm post}}{T_{\rm pre}} = \frac{5\mathcal{M}^{4} + 14\mathcal{M}^{2} - 3}{16\mathcal{M}^{2}} \, ,
  \end{equation}
finding a value ($1.28 \pm 0.07$) which is consistent with the observed
temperature jump ($ 1.45^{+0.33}_{-0.35}$).

The shock velocity is $ v_{\rm shock}= \mathcal{M} \, c_s$, where
$c_s = \sqrt{\gamma kT / \mu m_{\rm p}}$ is the sound speed in the
upstream (pre-shocked) gas. By considering the upstream temperature
($kT_{\rm in} = 10.26$ keV), we estimated $v_{sh} \approx 2100$ km
s$^{-1}$. The properties of the detected shock front are summarized in Tab. \ref{shock.tab}.

\begin{center}
  \begin{table}
    \caption{\label{spectra-front.tab} Results of the deprojected spectral fit along a  $80^{\circ}$-wide sector enclosing the two fronts identified in Section \ref{fronts.sec} (see Figure \ref{sectors.fig}.)  }
  \begin{tabularx}{\linewidth}{*5{X}}
      \hline
      \hline
  Wedge no. (width) & Net counts & $kT_{\rm}$  (keV)  &  $n_{\rm e}$
  ($10^{-3}$ cm$^{-3}$) &  $p$  ($10^{-11}$ erg cm$^{-3}$)
\\[+1mm]
  \hline
  \#1 $(41'')$& 3551  & $10.45_{-1.29}^{+1.94}$ & $5.32_{-0.10}^{+0.10}$ & $16.30_{-2.03}^{+3.04} $
 \\[+1mm]                           
  \#2 $(15'')$& 1724 & $9.61_{-2.28}^{+3.75}$ & $3.91_{-1.32}^{+1.38}$ & $11.02_{-4.54}^{+5.80}$
  \\[+1mm]
    \end{tabularx}
    
    \nointerlineskip
    \begin{tabularx}{\linewidth}{ *1{X}}
 ---------------------  Cold front  ($R_{\rm cf} = 295 \pm 6$ kpc)    --------------------
 \\[+1mm]
   \end{tabularx}
    
    \nointerlineskip
    \begin{tabularx}{\linewidth}{ *5{X}}
   \#3 $(18'')$& 1620 & $14.88_{-3.13}^{+4.60}$ & $3.69_{-0.06}^{+0.06}$ & $16.11_{-3.40}^{+4.99}$ 
   \\[+1mm]
     \end{tabularx}
    
    \nointerlineskip
    \begin{tabularx}{\linewidth}{ *1{X}}
---------------------  Shock front  ($R_{\rm sh} = 389 \pm 6$ kpc)     --------------------
\\[+1mm]
  \end{tabularx}
    
    \nointerlineskip
    \begin{tabularx}{\linewidth}{ *5{X}}
   \#4 $(160'')$& 6121 & $10.26_{-1.08}^{+1.26}$ & $0.62_{-0.01}^{+0.01}$ & $1.85_{-0.20}^{+0.22}$
  \\[+1mm]
   \#5 $(60'')$& 686 & $6.14_{-1.44}^{+2.92}$ & $0.38_{-0.02}^{+0.02}$ & $0.68_{-0.16}^{+0.32}$ 
  \\[+1mm]
  \hline
  \hline
    \end{tabularx}
    {\tablefoot{Column 1: wedge number (width in arcsec); column 2: net counts in the $[0.5-7.0]$ keV band;
  column 3: deprojected temperature; column 4:
  electron density; column 5: pressure. The derived classification as cold front and shock front is indicated at the boundary of the second and third wedge, respectively. See text for details.}}
\end{table}
\end{center}

%%%%%%%%%%%%%%%%%%%%%%%%%%%%%%%%%%%%%%%%%%%%%%%%%%%%%%%%%%%%%%%%%%%%%%%%%%%%%%%
%%%%%%%%%%%%%%%%%%%%%%%%%%%%%%%%%%%%%%%%%%%%%%%%%%%%%%%%%%%%%%%%%%%%%%%%%%%%%%%
%%%%%%%%%%%%%%%%%%%%%%%%%%%%%%%%%%%%%%%%%%%%%%%%%%%%%%%%%%%%%%%%%%%%%%%%%%%%%%%
      
\section{Discussion}
\label{discussion.sec}
      
The morphological and spectral analyses presented in the previous
sections clearly show the presence of a shock front and hot cluster core with
high metallicity. 
{\color {black} A possible interpretation is that some heating event has increased the temperature of the cluster central region}, which we argue was previously a cool core.

\subsection{A former cool core?}
\label{cc.sec}

%%%%%%%%%%%%%%%%%%%%%%%%

\begin{figure*}[ht]
	\centering
                \includegraphics[width=\columnwidth,bb=18 144 592 718]{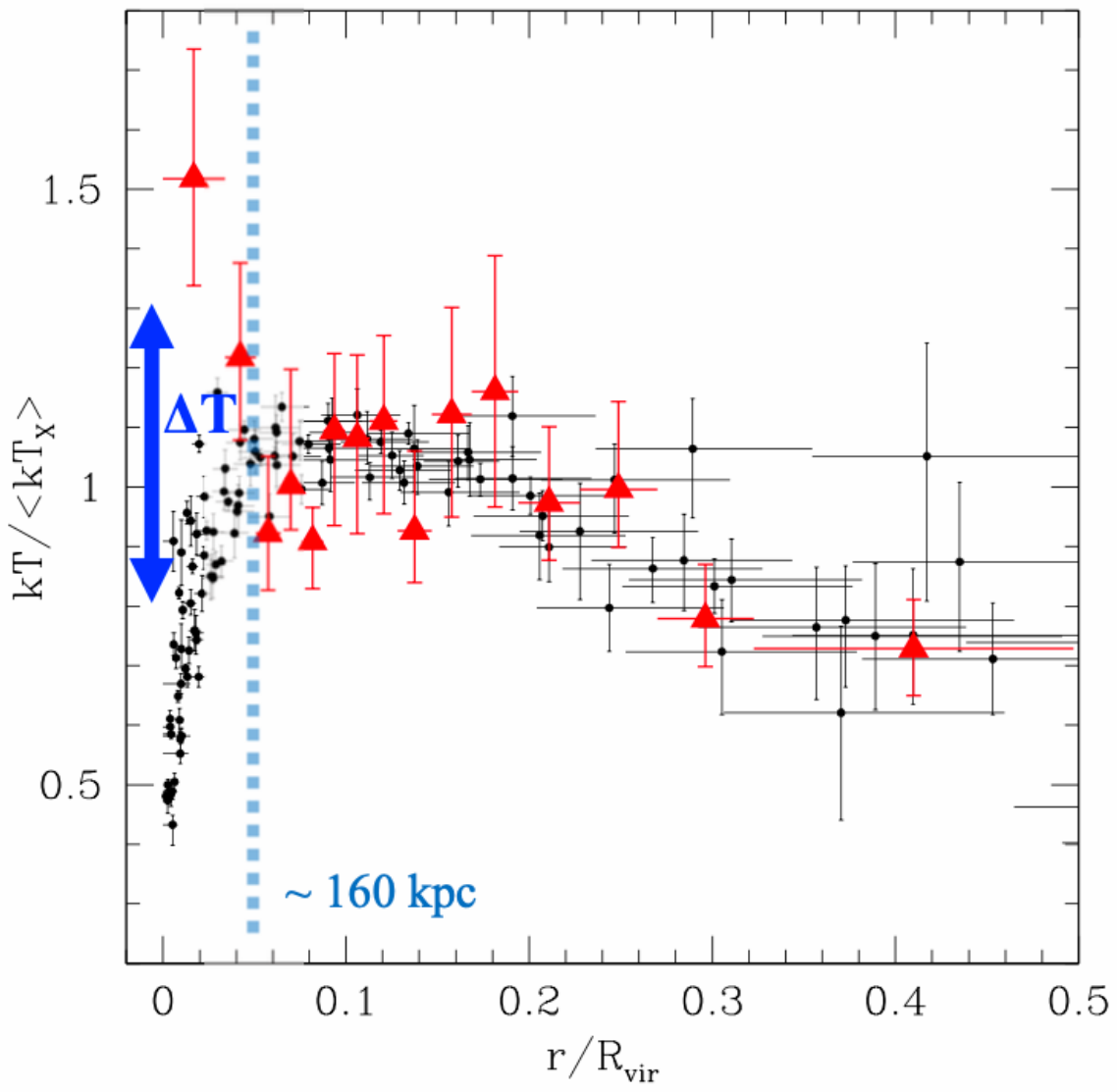}
                \includegraphics[width=0.99\columnwidth,bb=18 144 592 718]{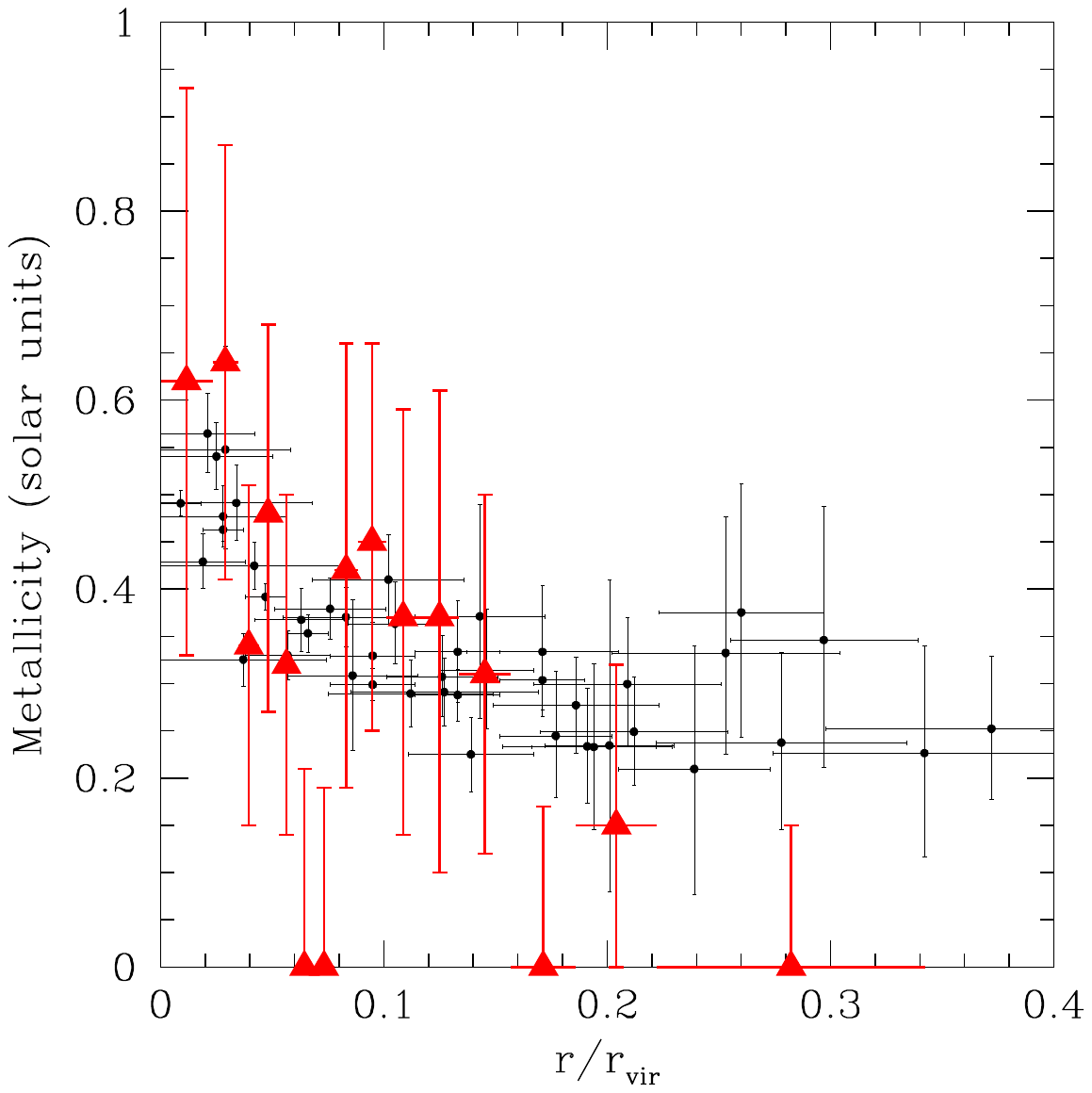}
	\caption{
    Temperature and metallicity profile of PLCKG287 compared to those of typical relaxed clusters. {\it Left panel:} Temperature profiles
          measured for a sample of relaxed clusters presented by
          \citet{Vikhlinin_2005}.  The temperatures are scaled to the
          cluster emission-weighted temperature excluding the central
          70 kpc regions, $<kT_{\rm X}>$.  The profiles for all
          clusters are projected and scaled in radial units of the
          virial radius $R_{vir}$, as estimated from the relation
          $R_{\rm vir} = 2.74 \, {\rm Mpc} \sqrt{<kT_{\rm X}>/10 \,
            {\rm keV}}$ \citep{Evrard_1996}. The values measured for
          PLCKG287 are $<kT_{\rm X}> \sim 13$ keV and
          $R_{\rm vir} \sim 3.1$ Mpc. The
          dotted cyan line indicates the region inside which there is
          a clear departure from the typical profile, with an opposite
          radial trend. We interpret this region as the former cool core, see text for details.  
          {\it Right panel:}
          Metallicity profile measured for PLCKG287 (red triangles)
          overlaid onto the metallicity profile observed for a sample
          of cool core clusters presented by
          \citet{DeGrandi-Molendi_2001}.  The profiles for all
          clusters are projected and scaled in radial units of
          $R_{\rm vir}$.  $H_0 = 50 \mbox{ km
          s}^{-1} \mbox{ Mpc}^{-1}$, $\Omega = 1$, $\Lambda=0$ is
        assumed for this plot and the virial radius is estimated from
        the relation
        $R_{\rm vir} = 3.95 \, {\rm Mpc} \, \sqrt{<kT_{\rm X}>/ \, 10
          \, {\rm keV}}$ \citep{Evrard_1996}, resulting $\sim$ 4.5 Mpc for PLCKG287. }
	\label{scaled.fig}
      \end{figure*}

%%%%%%%%%%%%%%%%%%
      
      The interpretation that PLCKG287 hosts a former cool core is
      mainly supported by the presence of a central peak in the
      metallicity profile, which is consistent with what is typically
      observed in samples of relaxed clusters
      \citep[e.g.,][]{DeGrandi-Molendi_2001, Lovisari_2019,
        Ghizzardi_2021}.  As a comparison, Figure \ref{scaled.fig}
      (right panel) shows the projected metallicity profile of a
      sample of relaxed clusters observed with \textit{BeppoSAX} \citep{DeGrandi-Molendi_2001}, rescaled to the virial radius,
      $R_{\rm vir}$ \citep[estimated from][]{Evrard_1996} {\color{black}.} 
      A strong enhancement in the
      abundance is found in the central regions.  Overlaid is the
      metallicity profile that we measure for PLCKG287 (red
      triangles). Despite the large errorbars, we note that it is
      fully consistent with the strong central enhancement in the
      abundance exhibited by relaxed clusters, thus supporting the
      interpretation of PLCKG287 being a former cool core\footnote{It
        is now well established that dynamically disturbed
        clusters also show an increase of metallicity in the center,
        although not at the level of the peak in relaxed clusters
        \citep[see e.g.,][for a recent measurement on a large sample
        of objects]{Leccardi_2010,Lovisari_2019}. However,
        \citet{Rossetti_2010} argue that this is due to the presence of cool
        core remnants, thus still supporting our interpretation.}.
      In particular, 
       from the spectral fits in the corresponding annular regions we measure 
      $Z = 0.57 \pm 0.17 \, Z_{\odot}$ inside 0.05 $R_{\rm vir}$ ($r< 30''$)
      compared to $Z = 0.27 \pm 0.06 \, Z_{\odot}$ 
      in a region with bounding radii 0.05 $R_{\rm vir}$ and
      $\sim$0.2 $R_{\rm vir}$ ($30'' < r < 120''$), where $R_{\rm vir} \sim 3.1$ Mpc for PLCKG287.
      
      We present below some simple order-of-magnitude {\color{black} energetic} estimates 
      to test the scenario that the inner region of PLCKG287, where we
      measure a high density and an abundance peak, was the cluster cool core
      which has then been heated. 
      {\color{black} In particular, we will first estimate the energy required to heat the core (Sect. \ref{heat.sec}) and compare it with that possibly provided by shock heating (Sect. \ref{shock-energy.sec}) and AGN heating (Sect. \ref{AGN-energy.sec}), and then discuss the possible origin of the detected shock (Sect. \ref{origin.sec}). We will not attempt to develop or discuss detailed hydrodynamic models, which is beyond the aim of this work, but we will limit our discussion to energetic considerations only.}
      
\subsection{Energy required to heat the core}
\label{heat.sec}
      
      To obtain a rough estimate of the temperature of the original
      cool core, which we interpret as the initial temperature before the
      heating event, we adopted a basic approach based on the
      comparison with the typical profile of relaxed clusters.  In
      particular, we show in Figure \ref{scaled.fig} (left panel) the
      projected temperature profile of PLCKG287 (red triangles)
      overlaid to the temperature profiles of a sample of relaxed clusters 
      \citep{Vikhlinin_2005}. 
      The radial distances are scaled to the virial radius
      and the temperatures
      are scaled to the emission-weighted cluster temperature,
      excluding the central 70 kpc region usually affected by
      radiative cooling.
      To be consistent with the widely-adopted scaled profiles
      presented by \citet{Vikhlinin_2005}, we thus estimated the
      emission-weighted temperature used for the scaling of PLCKG287
      by excluding the central bin (see Table \ref{annuli.tab}).
      We note that the overall temperature
      profile measured for PLCKG287 is generally consistent within the
      scatter of the profiles observed in a relaxed cluster for radial
      distances 
      $r$ \gtsim $0.05 \, R_{\rm vir}$, 
      corresponding 
to a physical scale $r$ \gtapprox $160$ kpc. On the
      other hand, in the central region there is a clear departure
      from the typical profile, with an opposite radial
      trend\footnote{A similar result comes from the comparison with
        the temperature profiles derived for the clusters of the
        CHEX-MATE sample \citep{Rossetti_2024}.}.
      In particular, for radial distances $r$ \ltapprox $160$ kpc,
PLCKG287 shows a negative gradient $dT/dr$ with higher temperatures (scaled $kT > 1$) with respect to the typical
central trend, which instead shows a positive gradient $dT/dr$ and lower temperatures (scaled $kT < 1$).  In our interpretation,
      the region inside $\sim$160 kpc thus corresponds to the heated cool
      core, where we estimated an average core temperature of
      $\sim 17$ keV (by fitting the combined spectra of the two
      innermost bins, see Table \ref{annuli.tab}). On the other hand,
      by assuming that the original cool core in PLCKG287 featured the
      typical temperature drop in the central regions observed for
      relaxed clusters, we estimated that the average pre-heated
      temperature of PLCKG287 inside \ltapprox $160$ kpc was initially
      below $\sim$ 12 keV, reaching local values as low as $\approx 5$
      keV in the very center (see Figure \ref{scaled.fig}, left
      panel). Therefore, the heating event raised the average
      ICM temperature in the core by at least $5$ keV, 
      from the initial value \ltsim $12$ keV to the currently observed
      value of $\sim 17$ keV.

{\color{black} We verified that the results are not affected by the spherical assumption: we performed a new spectral analysis by considering elliptical annular bins with the same geometry as that of the elliptical $\beta$-model (see Table \ref{beta.tab}). In particular, we extracted spectra in a central ellipse (with major axis $\sim$160 kpc) and in an adjacent elliptical annulus (with major axis $\sim$160-300 kpc), finding that the measured temperature and metallicity values are consistent within $1 \sigma$ to the ones derived in circular bins. 
We further note that the high temperature values measured by {\it Chandra} are consistent with those measured also by XMM-Newton \citep[PLCKG287 is the most massive and hottest cluster in the CHEX-MATE sample: see the red points in Fig. 3, right panel, of][]{Riva_2024}\footnote{\color{black} 
Even though cross-calibration differences may exist \citep[see for example the level of 10-15\% estimated by][]{Wallbank-2022}, 
the comparison with the temperatures measured in various systems by NuSTAR, with a much broader energy range, confirms the ability of {\it Chandra} to measure high temperature values.}}.

The energy required to increase the ICM temperature by $k \Delta T \approx 5$
keV can be measured as:
\begin{equation}
 E_{\rm heat} = \frac{3}{2} \frac{k \Delta T}{\mu m_{\rm p}} \, M_{\rm
   gas} \, ,
\end{equation}
where $M_{\rm gas}$ is the gas mass derived by integrating the
electron density (see Table \ref{profile_deproj.tab}) in shells.  We
obtained a gas mass $M_{\rm gas} \sim 1.9 \times 10^{12}$ M$_{\odot}$
inside the central $\approx 160$ kpc (which is the region
corresponding to the heated cool core), thus leading to an estimate
of $E_{\rm heat} \approx 4 \times 10^{61}$ erg.  

%%%%%%%%%%%%%%%%%%%%%
      
\subsection{Energy of the shock}
\label{shock-energy.sec}

\begin{table}
  \caption{\label{shock.tab} Summary of the shock properties}
\begin{center}
  \begin{tabular}{@{}lllll}
%~ & \multicolumn{2}{c}{Cavity N} &  \multicolumn{2}{c}{Cavity S}
%\\
\hline
    \hline
%    $\mathcal{M}$ & $v_{\rm shock}$ & $J$ & $T_{\rm jump, exp}$ & $T_{\rm jump, obs}$ & $E_{\rm sh}$ \\
%    ~ & (km/s) & ~ & ~ & ~ & (erg)\\
$\mathcal{M}$ & $v_{\rm sh}$ & $J$ & $T_{\rm post}/T_{\rm pre}$ & $E_{\rm sh}$ \\
    ~ & (km/s) & ~ & ~ & (erg)\\
    \hline
%    $1.29 \pm 0.07$ & $\sim$2100 &  $1.43\pm0.10$  & $1.28\pm0.07$ & $1.45^{+ 0.33}_{-0.35}$ & $8 f \times 10^{62}$
    $1.29 \pm 0.07$ & $\sim$2100 &  $1.43\pm0.10$  & $1.45^{+ 0.33}_{-0.35}$ & $8 f \times 10^{62}$
    \\[+1mm]
 \hline
\hline
  \end{tabular}
\end{center} {\tablefoot{Column 1: Mach number ; column 2: shock velocity; column 3: compression factor;
    column 4: temperature jump measured from a
    deprojected spectral analysis (see Tab. \ref{spectra-front.tab} and Fig. \ref{sectors.fig}); column 5: shock energy,
    where $f$ is the fraction of the volume of
    the shocked gas, ranging from $\sim$0.1 to 1 depending on the
    actual geometry of the shock ($f \sim$0.12 for a shock which
    propagated through the cluster only along the $\sim$1.5 steradian solid
    angle corresponding to the limited projected angle of 80 degrees in
    which we detected it, and $f$ = 1 for an isotropic shock) }}
\end{table}

%%%%%%%%%%%%%%%%

In order to verify whether the {\color{black} detected} shock can be responsible for heating
the cluster core, we need to compare $E_{\rm heat}$ (just derived in
Sect. \ref{heat.sec}) with the energy that has been dumped by the
shock.
{\color{black} A lower limit for the} shock energy, $E_{\rm sh}$, can be estimated as \citep[e.g.,][]{David_2001}:
\begin{equation}
  E_{\rm sh} = \frac{3}{2} \, \Delta p \, V_{\rm sh} =  \frac{3}{2}  (p_{\rm post} -
  p_{\rm pre}) V_{\rm sh} \, ,
\end{equation}
where $p_{\rm post}$ and $p_{\rm pre}$ are the pressures in the
post-shock and pre-shock regions, respectively, and $V_{\rm sh}$ is the volume of shocked gas.
By substituting the expression for the pressure jump expected from the
R-H conditions :
\begin{equation}
\frac{p_{\rm post}}{p_{\rm pre}} = \frac{5 \mathcal{M}^{2} - 1}{4}
\end{equation}
we can express the shock energy as a function of the measured
post-shock pressure and Mach number as:
\begin{equation}
 E_{\rm sh} =  \frac{3}{2} p_{\rm post} V_{\rm sh} \, \left( \frac{5
     \mathcal{M}^2 -  5}{5 \mathcal{M}^2 -1} \right) \, .
  \label{Eshock.eq}
\end{equation}  
As the volume of shocked gas, we considered the volume of a sphere of
radius $R_{\rm sh} = R_{\rm outer} \sim 389$ kpc , reduced by a
fiducial factor $f$ that depends on the fraction of the volume
occupied by the shocked gas. We therefore estimate a volume of shocked gas
of the order of
$V_{\rm sh} = 4/3 \pi f R_{\rm sh}^3 \sim 2.5 \, f \times 10^8 $
kpc$^3$, where $f$ can range from $\sim 0.1$ to 1 depending on the
actual geometry of the shock ($f \sim 0.12 $ for a shock which propagated
through the cluster only along the $\sim 1.5$ steradian solid angle corresponding to
the limited projected angle of $80^{\circ}$ in which we detected it,
and $f=1 $ for an isotropic shock).  Considering the Mach number and
post-shock pressure estimated in Section \ref{fronts.sec} (see Table
\ref{spectra-front.tab}), we finally estimated a shock energy
$E_{\rm sh} \approx 8 \, f \times 10^{62} $ erg, which results of the order of
$E_{\rm sh} \approx 10^{62} $ erg if $f \sim 0.12$.
The shock properties are summarized in Tab. \ref{shock.tab}.

Comparing this with
the energy required to heat the core 
($E_{\rm heat} \approx 4 \times 10^{61} $, see Sect. \ref{heat.sec}), we conclude that even
considering the lower limit of the geometrical factor $f \sim 0.12$
(that is, considering that the shock propagated through the cluster
only along the direction in which we detect it), the amount of
energy injected by the shock is a factor \gtsim 2 higher than that
required to heat the core.  In the case of a semispheric or isotropic
shock ($f = 0.5$ or 1), its energy would still be enough to heat the
whole region inside the shock radius ($R_{\rm sh} \sim$ 390 kpc), where
we estimated a gas mass of $1.9 \times 10^{13}$ M$_{\odot}$ and an
energy required to heat it of $E_{\rm heat} \approx 2 \times 10^{62}$
erg (considering a $k \Delta T \approx 2$ keV estimated from the
comparison with the typical temperature profile, Figure \ref{scaled.fig}).
In summary, despite the uncertainties and approximations of our
estimates, there are clear indications that the shock injects sufficient energy to overheat the core in PLCKG287.

{\color{black} However, this scenario seems viable only if the shock originated in the cluster center where it was stronger in the past, which could be the case if the central AGN had driven it. On the contrary, it seems implausible that a large-scale merger shock would preferentially heat the former dense cool core.
In fact,} from Figure \ref{scaled.fig} (left panel) we see that $\Delta T$ is largest at the center, {\color{black} right} where the gas density is the highest.  In particular, at $r \sim 0$
the observed temperature is a factor of $\sim 3$ higher than that of the typical cool cores. Such a temperature jump would imply a shock with Mach number $\mathcal{M}\sim 2$, significantly
stronger than the shock that is heating gas at larger radii.  {\color{black} We will investigate the possible origin of the detected shock in Section 5.5.
}

%%%%%%%%%%%%%%%%%%%%%%%%%%%%%%%%%%%%%

{\color{black}
\subsection{Energy of AGN feedback}
\label{AGN-energy.sec}
}

\begin{figure}
\centerline{
    \includegraphics[width=\columnwidth,bb=36 180 577 612]{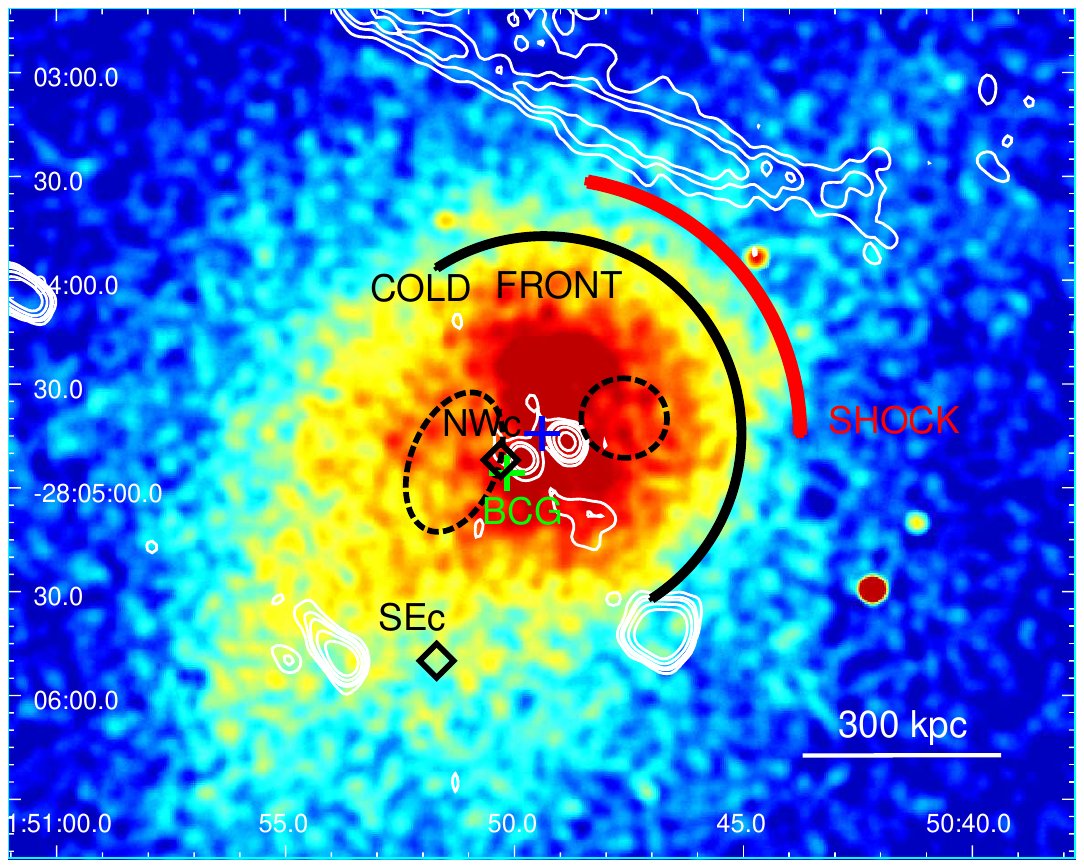}
}
\caption{\label{final-discussion.fig} Background-subtracted, exposure
  corrected mosaic [0.5-7.0] keV {\it Chandra} image, smoothed with a
  kernel of 8 pixels, of the central region of PLCKG287, with
  superimposed in white the GMRT radio contours at 610 MHz
  \citep[levels start at $\pm 3 \sigma $ and increase by a factor of
  2, where $\sigma$ = 0.07 mJy/beam,
  beam=$6.7'' \times 5.3''$,][]{Bonafede_2014}. The black dashed
  ellipses show the possible X-ray cavities whereas the red and black
  arcs indicates the position of the shock and cold front,
  respectively. The locations of the X-ray peak and main BCG are
  indicated by blue and green crosses, respectively, whereas the
  positions of the dark matter substructures identified by weak
  lensing analysis are shown by black diamonds and labeled following
  the same nomenclature as in \citet{Finner_2017}.  }
\end{figure}

As we discussed in Section \ref{cc.sec}, we argue that PLCKG287 is a former cool core.
Furthermore, the visual inspection of both the residual and the
background-subtracted, exposure corrected mosaic images revealed the
presence of X-ray depressions (see Figure \ref{cavity.fig} {\color{black} and \ref{cavity-appendix.fig})}.  We also
note that the locations of the two depressions are opposite with
respect to the central radio source associated with the BCG (white
contours overlapping the green cross indicating the BCG in Figure
\ref{final-discussion.fig}).
From the GMRT radio data \citep{Bonafede_2014}, we estimated that at
610 MHz the BCG has a flux density of $\sim$ 2.4 mJy, with a spectral
index\footnote{We define the spectral index, $\alpha$, as
  $S_{\nu} \propto \nu^{- \alpha}$, where $S_{\nu}$ is the flux
  density at the frequency $\nu$.}  $\alpha \sim$ 0.9 calculated
between 610 MHz and 323 MHz.  This corresponds to a radio
power\footnote{The monochromatic radio power at the frequency $\nu$ is
  estimated from
  $P_{\nu}=4 \pi \, D_{\rm L}^2 \, S_{\nu} \, (1+z)^{(\alpha - 1)}$ 
  \citep[e.g.,][]{Condon_2018}.} of $\sim$1.2 $\times 10^{24}$ W/Hz.
By extrapolating, assuming a straight spectral index of $\sim 0.9$ up
to higher frequencies, we estimated a total flux density of $1.2$ mJy
at 1.3 GHz, corresponding to a radio power of $\sim 6 \times 10^{23}$
W/Hz.  These results, which are confirmed by new wideband radio observations obtained in L
and S bands with MeerKAT in the frequency range $\sim 1$-$3$ GHz 
(Balboni et al. in preparation; Rajpurohit in preparation),
indicate that the BCG hosts a radio AGN, thus corroborating the
interpretation of the X-ray depressions as possible cavities.
If confirmed, this would be the signature of AGN feedback and thus
possibly point to an AGN-driven origin 
{\color{black} of the heating event}.

Assuming that the X-ray depressions are cavities carved by a past AGN
outburst, we calculated the cavity properties 
{\color{black} (summarized in Table \ref{cavity.tab})}
and compared them with
{\color{black} the energy estimated in Sect. \ref{heat.sec}}. In particular, {\color{black} based on the volume $V$ of the cavities estimated from the ellipsoidal shapes indicated in Fig. \ref{cavity.fig}, and on the pressure $p$ of the surrounding hot gas derived from the deprojected spectral analysis (see Table \ref{profile_deproj.tab})}, we estimated that the {\color{black} energy required to inflate them is} $E_{\rm cav}\approx 4\times10^{61}$ erg, {\color{black} where we assumed the typically-adopted value of 4$pV$ of energy per cavity \citep[e.g.,][]{McNamara-Nulsen_2007}.}
This quantity, which is typically considered a proxy {\color{black} (likely a factor of a few lower)} of the total AGN outburst mechanical energy, \citep[e.g.,][]{Birzan_2004,McNamara-Nulsen_2007,Gitti_2012,Fabian_2012}, is comparable to the energy required to heat the former cool core (see Sect. \ref{heat.sec}), thus indicating that
AGN feedback may be {\color{black} energetically} responsible for overheating the core.
{\color{black} 
This is consistent with simulation results indicating that if a substantial portion of energy from the most energetic AGN outbursts ($10^{61} - 10^{62}$ erg, as in PLCK287) is deposited near the centers of cool core clusters, which is a likely scenario given the dense cores, these AGN outbursts can completely remove the cool cores, thereby transforming them into non cool core clusters \citep{Guo_2010}.
}

\subsection{Origin of the shock}
\label{origin.sec}

We discuss below two alternative scenarios for the origin of the {\color{black} detected} shock: the
origin by merger and the origin by the same AGN outburst {\color{black}which generated the X-ray cavities}.

%%%%%%%%%%%%%%%%%%%%%%%%%%

%%%%%%%%%%%%%%%%%%%%%%%%%%%%%%%

\subsubsection{Merger}
\label{merger.sec}

In the X-ray morphological analysis of the CHEX-MATE sample, 
PLCKG287 is classified as ``mixed'' 
in terms of morphological disturbance \citep{Campitiello_2022}.
The dynamically active nature of PLCK287 is probed by the
presence of two radio relics and a radio halo \citep{Bagchi_2011,
  Bonafede_2014}, but the merger scenario is still not clear \citep[e.g.,][]{Golovich_2019}.
Subaru and {\it HST} weak-lensing analysis detected
multiple peaks in the mass distribution of PLCKG287, revealing the
complexity of this system \citep{Finner_2017}.  In particular, the
{\it HST} mass map shows two mass peaks within the central mass clump, with
one primary cluster of mass $M_{\rm NWc} = 1.59 \times 10^{15}$
M$_{\odot}$ (labeled NWc in Figure \ref{final-discussion.fig}) which
is likely to be undergoing a merger with one subcluster of mass
$M_{\rm SEc} = 1.16 \times 10^{14}$ M$_{\odot}$ located at a projected
distance $d \sim 320$ kpc to the SE (labeled SEc in
Fig. \ref{final-discussion.fig}).  
The observation of surface brightness discontinuities in the ICM of merging clusters is particular useful to determine whether a significant component of the merger is occurring on the plane of the sky, as edges are easily hindered by projection effects \citep[e.g.,][]{Botteon_2018}. The detection of a cold front nested within a shock front in PLCKG287 (see Sect. \ref{fronts.sec} and Fig. \ref{final-discussion.fig}) may thus suggest the presence of a collision component along the SE-NW axis where the shock is driven by the merger and is located ahead of the infalling subcluster core, in a similar fashion as the Bullet Cluster \citep[e.g.,][]{Markevitch_2002, Clowe_2004}.
\\

To investigate the consistency of the merger origin scenario we
estimated 
the energy available during the merger event for heating, that is the energy that can be dissipated in
the ICM, and make a comparison with the shock properties.
We estimated the merger velocity predicted by free-fall from the
turn-around radius of the two subclusters, $v_{\rm merger,ff}$, as
\citep[e.g.,][]{Ricker_2001, Sarazin_2001}:

\begin{equation}
  v_{\rm merger,ff} \approx 2930 \, {\rm km/s} \left(
    \frac{M_{\rm NWc} + M_{\rm SEc}}{10^{15} M_{\odot}} \right)^{\frac{1}{2}}
  \left(\frac{d}{\rm 1 \, Mpc}\right)^{-\frac{1}{2}} \, ,
  \label{vmerge.eq}
 \end{equation}
 finding an estimate of $v_{\rm merger} \approx 6800$ km/s, which must
 be considered as an upper limit as $d$ is a projected distance, hence a lower limit to the true separation of the two subclumps.
 This leads to an estimate of the maximum kinetic energy that can be
 dissipated in the ICM inside the volume corresponding to the shock distance
 ($\sim 390$ kpc) of
\begin{equation}
%E_{\rm kin} = \frac{1}{2} \left(  M_{\rm NWc} + M_{\rm SEc} \right) v^2_{\rm merger} {\rm ,}
E_{\rm kin,gas} = \frac{1}{2} M_{\rm gas} v^2_{\rm
  merger,ff}  \, {\rm .}
\end{equation}
With a gas mass of $\sim 1.9 \times 10^{13} M_{\odot}$ inside $\sim 390$
kpc (see Table \ref{profile_deproj.tab} and
Fig. \ref{spectral-profiles.fig}), we thus estimate
$E_{\rm kin,gas} \approx 9 \times 10^{63}$ erg.
This largely exceeds the energy we
derived for an isotropic shock
($E_{\rm sh} \approx 8 \times 10^{62}$ erg, see Table \ref{shock.tab}), thus the merger scenario is energetically viable.
Although these are only
order-of-magnitude estimates, based on many assumptions and affected
by
projection effects,
this comparison
may further suggest that the shock energy is not fully
thermalized, and that
some fraction may go into turbulence,
magnetic fields, or cosmic rays \citep[e.g.,][]{Sarazin_2001}, thus
powering the diffuse radio sources observed in this cluster
\citep{Bagchi_2011, Bonafede_2014}.  
~\\

{\color{black} 
On the other hand, the geometry of the shock and cold front system might challenge the merger scenario.}
The ratio between the stand-off distance $\Delta = R_{\rm sh} - R_{\rm cf} \sim 94$ kpc 
to the cold front radius $R_{\rm cf} = 295$ kpc is $\sim 0.3$, which implies a 
shock Mach number $\mathcal{M} \sim 3$, according to the standard theory of shocks generated 
by a blunt body moving in a uniform medium \citep[e.g.,][]{Vikhlinin_2001, Sarazin_2001, Zhang_2019}.
This finding is 
at odds with the observed $\mathcal{M} \sim 1.3$. 
More recent merging simulations further exacerbate this disagreement, as they show that
the ratio $\Delta/R_{\rm cf}$ increases in the post-merger phase
reaching typical values $\Delta/R_{\rm cf} \sim 1 - 2$ for $\mathcal{M}\sim 1.5-2$ \citep{Zhang_2019}.
Lower $\Delta/R_{\rm cf}$ values are characteristic only of the pre-merger phase and cannot be applied to PLCKG287 because the observed geometry of the cold front-shock configuration 
suggests that this is a post-merger system.  
We therefore investigate the possibility that the shock may have
a different origin.

 \subsubsection{AGN {\color{black} outburst}}
 \label{feedback.sec}

%%%%%%%%%%%%%%%%%%%%%%%%%%%

\begin{table}
  \caption{\label{cavity.tab} Properties of the cavities indicated in Figure \ref{cavity.fig}. }
\begin{center}
  \begin{tabular}{@{}lccc}
%~ & \multicolumn{2}{c}{Cavity N} &  \multicolumn{2}{c}{Cavity S}
%\\
\hline
\hline
Quantity & Cavity East &  Cavity West & Total
  \\
\hline
$a$~ (kpc)  & $110$ & $65$ & -
\\[+1mm]
$b=c$~ (kpc)  & $66$ & $60$ & -
\\[+1mm]
$r_{\rm eff}$~ (kpc)  & $78$ & $62$ & - 
\\[+1mm]
$D$~ (kpc)  & $142$ & $127$ & -
\\[+1mm]
$V \, ({\rm cm}^3)$ & $5.9 \times 10^{70}$ & $2.9 \times 10^{70}$ & $8.8 \times 10^{70}$
\\[+1mm]
$E_{\rm cav}$~ (erg)& $2.6 \times 10^{61}$ & $1.3 \times 10^{61}$  & $3.9 \times 10^{61}$
\\[+1mm]
$t_{\rm s}$~ (yr) & $8.0 \times 10^7$ & $7.1 \times 10^7$ & - 
\\[+1mm]
$P_{\rm cav,s} \, ({\rm erg \, s}^{-1})$ & $1.0 \times 10^{46}$ & $5.6 \times 10^{45}$ & $1.6 \times 10^{46}$
\\[+1mm]
$t_{\rm exp}$~ (yr) & $4.4 \times 10^7$ & $3.5 \times 10^7$ & - 
\\[+1mm]
$P_{\rm cav,exp} \, ({\rm erg \, s}^{-1})$ & $1.9 \times 10^{46}$ & $1.2 \times 10^{46}$ & $3.1 \times 10^{46}$
 \\[+1mm]
\hline
\hline
  \end{tabular}
\end{center} {\tablefoot{Cavity volumes, $V$, are calculated assuming
    a prolate ellipsoidal shapes with semimajor axis $a$ and semiminor
    axes $b=c$.  Cavity powers, $P_{\rm cav}$, are calculated as the
    ratio of the cavity energy, $E_{\rm cav}$, by the cavity age,
    assuming $4 pV$ of energy per cavity. The cavity age was estimated
    in two ways: i) as the sound crossing timescale
    $t_{\rm s} = D/c_{\rm s}$, where $D$ is the cavity distance from
    the center and $c_{\rm s}$ is the sound speed, and ii) as the
    expansion time $t_{\rm exp} = r_{\rm eff}/c_{\rm s}$, where
    $r_{\rm eff} = (a b^2)^{1/3}$ is the effective radius.  We
    considered the spectral values of temperature and pressure
    obtained in the second radial bin of the deprojected spectral
    analysis (see Table \ref{profile_deproj.tab}), where we estimated
    a value of the sound speed of $c_{\rm s} = 1740$ km s$^{-1}$. We
    note that due to projection effects the cavity volume and age
    estimates should be considered as lower limits. }}
\end{table}

%%%%%%%%%%%%%%%%%%%%%%%%%%%%%%%%%%

As discussed in Sect. \ref{AGN-energy.sec}, {\color{black} the X-ray depressions can be interpreted as cavities carved by a past AGN outburst} and thus possibly point to an AGN-driven origin of the shock, also supported by the fact
that the shock is detected along the direction of the cavity pair
axis. 
%To investigate the consistency of the AGN shock scenario, we 
{\color{black} 
The AGN outburst origin of the shock would be energetically viable if $E_{\rm cav}$ is of the same order as $E_{\rm sh}$ or a few times lower, as typically found in systems where both cavity systems and associated shocks were detected \citep[e.g.,][]{Randall_2011,Liu_2019,Ubertosi_2023}\footnote{Some exceptions have been observed in which the shock is not the dominant heating source \citep[e.g.,][]{Forman_2020}.}. 
To test the consistency of this scenario, we thus compared the cavity properties with the shock ones.  In particular, we note that} assuming that the shock propagated only along a limited
direction through the cluster (that is the one where we actually
detected it, leading to $f \sim 0.12$), the energy of the shock
($\approx 10^{62}$ erg, see Table \ref{shock.tab}) would be {\color{black} only} a factor of a few higher than that of the cavities ($\approx 4 \times 10^{61}$ erg, see Table \ref{cavity.tab}).
Therefore, the scenario of an AGN origin of the shock is energetically plausible. 

We further estimated the cavity age
from both the sound crossing time (that is the time necessary to reach
the current distance from the center assuming that they are moving at the
sound speed) and the expansion time (that is the time necessary to
expand to the current size, assuming that they are expanding at the sound
speed), resulting of the order of {\color{black} 70-80 Myr} and
$\sim${\color{black} 40 Myr}, respectively. Due to projection effects, these
ages must be considered as lower limits.
The cavity properties are summarized in Table \ref{cavity.tab}.
For comparison, in the scenario of an AGN-driven shock {\color{black} we can estimate
an upper limit for its age} by considering the time it
took to travel from the center to the observed shock radius,
$R_{\rm sh} \sim 390$ kpc, {\color{black} at the current shock speed 
($v_{\rm sh} \approx$ 2100 km s$^{-1}$, see Sect. \ref{fronts.sec} and Table \ref{shock.tab})}.  
This results in an
estimate of the shock age of
$t_{sh} = R_{\rm sh}/v_{sh} \approx$ {\color{black} 180 Myr, which should be
considered as an upper limit because the initial Mach number of the shock
was probably higher. 
Although this upper limit for the shock age might be consistent with 
the lower limit for the cavity age  ($\, \gtapprox$ 40-80 Myr), the large difference
is a warning against the interpretation of a
common origin of both the cavities and shock as due to the central AGN
outburst. }

 We finally
 note that {\color{black} in general} an AGN-driven shock would be able to produce a
  central temperature peak
  \citep[e.g.,][]{Brighenti-Mathews_2003,Gaspari_2011a,Gaspari_2011b}, as the one observed in PLCKG287, whereas a
  merger shock would 
  generate a heated region downstream the front where the gas will eventually cool adiabatically as the shock moves in the outskirts
  \citep[e.g.,][]{Heinz_2003b,Mathis_2005}.
  If the AGN heats the gas in the center, it can generate a negative
  entropy gradient making the ICM convectively unstable. This is
  consistent with the entropy profile derived from the deprojected spectral analysis, where we observe an increase
  of entropy in the central bin. {\color{black} Together with  the high density and abundance peak (Fig. \ref{spectral-profiles.fig}), this suggests}
  that this structure was the cool core before the heating event
  (Sect. \ref{cc.sec}). 
  The entropy inversion and
  pressure drop are indications that the ICM is not relaxed at the
  center, but it is probably expanding {\color{black} after the  AGN heating event. 
  Detailed gas dynamical models would be necessary to validate or disprove this scenario.} 

%%%%%%%%%%%%%%%%%%%%%%%%%%%%%%%%%%%%%%%%%%%%%%%%%%%%%%%%%%%%%%%%%%%%%%%%%%%%%%%
%%%%%%%%%%%%%%%%%%%%%%%%%%%%%%%%%%%%%%%%%%%%%%%%%%%%%%%%%%%%%%%%%%%%%%%%%%%%%%%      
%%%%%%%%%%%%%%%%%%%%%%%%%%%%%%%%%%%%%%%%%%%%%%%%%%%%%%%%%%%%%%%%%%%%%%%%%%%%%%%

\section{Summary and Conclusions}
\label{conclusion.sec}

\begin{table}
  \caption{\label{summary.tab} Summary of the energy estimates presented in this work}
\begin{center}
  \begin{tabular}{@{}lll}
\hline
    \hline
    ~ & Energy (erg) & Section \\
\hline
    Core heating & $\approx 4 \times 10^{61}$ & \ref{heat.sec}
\\[+1mm]
    Shock &  $\approx 8 f^{(*)} \times 10^{62}$ & \ref{shock-energy.sec}
%\\[+1mm]
% Merger & $\approx 3 \times 10^{64}$ & \ref{merger.sec}
\\[+1mm]
Cavities & $\approx 4 \times 10^{61}$ & {\color{black} \ref{AGN-energy.sec}}
\\[+1mm]
Merger heating & $\approx 9 \times 10^{63}$ & \ref{merger.sec}\\
\hline
\hline
  \end{tabular}
\end{center} {\tablefoot{ $(^*) f$ is the fraction of the volume of
    the shocked gas. }}
\end{table}

In this work, we presented the analysis of new $\sim$200 ks {\it Chandra}
observations of the galaxy cluster PLCKG287, which is likely
undergoing a merger with a small ($\sim 10$\% of the mass) subcluster located at a projected
distance $\sim 320$ kpc to the SE \citep{Finner_2017}, and which is  
known to host
a giant radio halo and two prominent radio relics in the NW and SE
directions at distances of $\sim 500$ kpc and $\sim 2.8$ Mpc from the cluster center \citep{Bonafede_2014}.
The main results of our detailed morphological and spectral
investigations, focused on
the region inside $R_{\rm 500} (\sim 1.5$ Mpc), are summarized below:

\begin{itemize}

\item The cluster shows a comet-like X-ray morphology oriented in the
  NW-SE direction, with a bright X-ray core region which is co-spatial
  with the known giant radio halo. The X-ray peak is spatially offset
  by $\sim 80$ kpc from the main BCG, which is indicative of a
  disturbed dynamical state.  The $\beta$-model produces a good fit
  (Table \ref{beta.tab}) to the surface brightness profile centered on
  the X-ray peak,
  % (RA 10:50:49.4, Dec $-$28:04:44.3),
  and the residual and unsharp images (Fig. \ref{residual-unsharp.fig})
  reveal the presence of X-ray depressions to the E and W
  (Fig. \ref{cavity.fig} {\color{black} and \ref{cavity-appendix.fig}}).\\

\item From a spectral analysis inside $R_{\rm 500}$
  we measure a global temperature $kT$=$12.73^{+0.38}_{-0.40}$ keV and
  abundance $Z$=$0.26^{+0.05}_{-0.05}$ Z$_{\odot}$, with a luminosity
  of $L_{2-10 \, {\rm keV}}$=$(1.09 \pm 0.01) \times 10^{45}$ erg/s. 
  %in the 2-10 keV band. 
  The integrated gas mass is
  $M_{\rm gas} (< R_{\rm 500}) \sim 1.6 \times 10^{14}$ M$_{\odot}$
  and the hydrostatic total mass obtained from the best-fit
  $\beta$-model parameters is
  $M_{\rm tot} (< R_{\rm 500}) \approx 1.2 \times 10^{15}$ M$_{\odot}$.\\

\item The radial profile of thermodynamic quantities
  shows a {\color{black} density,} temperature and abundance peak in the cluster center, where
  also the pressure and entropy have a rapid increase (Fig. \ref{spectral-profiles.fig}).  This is an indication that the
  ICM is not relaxed at the center, but is dynamically unstable and
  probably in expansion.  The 2D spectral maps 
  (Fig. \ref{maps.fig}) further show that the
  higher abundance gas is preferentially found inside
  the central $\sim 300$ kpc from the X-ray peak and is bounded by a region of cold gas,
  which is in turn externally surrounded by a region of hotter
  gas. This hot ICM region is narrower ($\sim$90 kpc) towards the
  NW direction and corresponds to a region of
  higher ICM pressure, possibly indicating shock compression. Across
  the internal cold ICM region the pressure is instead nearly
  constant, suggesting the presence of a contact discontinuity.\\

\item By means of accurate morphological and spectral analyses in the
  regions identified in the residual and spectral maps, we detect a
  shock front in a $80^{\circ}$-wide sector to the NW direction at a
  distance $R_{\rm sh} = 389 \pm 6$ kpc from the center, characterized
  by a density jump $= 1.43 \pm 0.10$, Mach number
  $\mathcal{M} = 1.29 \pm 0.07$ and shock velocity
  $v_{sh} \approx 2100$ km s$^{-1}$. Across the shock, we measure a
  temperature jump $= 1.45^{+0.33}_{-0.35}$, which is consistent with
  that expected from the R-H conditions. We estimate a shock energy of
  $E_{\rm sh} \approx 8 \, f \times 10^{62} $ erg, where $f$ is a
  filling factor indicating the fraction of the volume occupied by the
  shocked gas (see Table \ref{shock.tab} for a summary of the shock
  properties).  We also detect a cold front in a semicircle to the
  NW direction at a distance $R_{\rm cf} = 295 \pm 6$ kpc from the center,
  characterized by a density jump $= 1.44 \pm 0.08$ and a temperature
  jump $=0.65^{+0.29}_{-0.25}$.\\

\item {\color{black} 
From GMRT radio data we estimate that the BCG has a 610 MHz radio power of $\sim$1.2 $\times 10^{24}$ W/Hz and a spectral index $\alpha \sim 0.9$, indicating that it hosts a radio AGN.
Therefore, the two large depressions detected in the 
X-ray surface brightness can be most easily explained as the result of an AGN feedback episode. We estimate that the energy associated to the AGN outburst, which occurred $\gtsim$ 40 Myr ago, is  $\; \gtsim \, 4\times 10^{61}$ erg. } \\

\item {\color{black} The observations above point to a scenario in which a heating event has heated a former cool core. }
  The interpretation that PLCKG287 {\color{black} hosted a cool core} is mainly
  supported by the presence of a central peak in the metallicity
  profile, which is typically observed in samples of cool core
  clusters (Fig. \ref{scaled.fig}, right panel).
  By assuming that the original cool core featured the
  typical temperature drop in the central regions observed for relaxed
  clusters (Fig. \ref{scaled.fig}, left panel), we
  estimate that the heating event raised the average ICM temperature
  in the core ($r$ \ltapprox $160$ kpc) by approximately $5$ keV
 (see Sect. \ref{cc.sec}), corresponding to an energy $E_{\rm heat} \approx 4 \times 10^{61}$ erg (see
 Sect. \ref{heat.sec}).
{\color{black} Based on simple energetic considerations only, we explore two scenarios
to explain the heated core. In the first case, the detected shock crossed the core region, raising the core temperature by $\sim 5$ keV. Our approximate estimate for the shock energy  indicates that this picture is plausible.
In the second model, an AGN outburst heats the core from inside out, but no attempt is made to single out the specific heating process (likely still a shock, although not necessarily the one we detect). We estimate that the energy associated with the X-ray cavity formation is again of the same order as the energy required to heat the former cool core. We therefore argue that heating by both the detected shock and by an AGN outburst are energetically viable
scenarios (see Table \ref{summary.tab} for a summary of the energy estimates). However, we briefly discuss that neither of the two scenarios alone seems able to explain all the observed features of the X ray emission in PLCKG287.
} \\

\item We discuss the possible origin of the {\color{black} detected} shock by investigating the
  alternative scenarios of {\it (i)} merger, which is supported by the
  observed configuration of a cold front nested inside a shock along
  the same direction, as typically expected to be produced during
  merging processes (Sect. \ref{merger.sec}); and {\it (ii)} AGN
  feedback, which is supported by the presence of X-ray cavities in
  opposite positions with respect to the central radio BCG along the
  direction of the shock, as expected if both the cavities and shock
  are produced by a central AGN outburst (Sect. \ref{feedback.sec}).
  We find that both scenarios are energetically viable (see Table
  \ref{summary.tab}), {\color{black} although time-scale considerations favor a merger origin of the shock.}
  
\end{itemize}

We finally note that although the merging scenario is more plausible
to explain the origin of the detected shock and particularly the
large-scale structure observed in PLCKG287, a merger event is not
expected to produce an increase of the central temperature and
entropy. This can instead be produced on smaller scales by AGN
feedback. Therefore, the observed temperature and entropy peaks, that
we measure in the core, could indicate the concomitant effect of the
central AGN, thus pointing to a combined action of merging and AGN
feedback in heating the former cool core of this system. 
More generally, this indicates that mergers are not solely responsible for disrupting cool cores, 
but AGN feedback may contribute to shaping the thermodynamical properties of clusters.

\section*{Acknowledgments}

{\color{black} We thank the anonymous referee for insightful comments and constructive suggestions which stimulated a critical discussion of the results}. MG and FU acknowledge the financial contribution from contract PRIN 2022 - CUP J53D23001610006.
RJvW acknowledges support from the ERC Starting Grant ClusterWeb 804208.
MB acknowledges funding by the Deutsche Forschungsgemeinschaft (DFG) under Germany's Excellence Strategy -- EXC 2121 ``Quantum Universe'' --  390833306 and the DFG Research Group "Relativistic Jets".
WF acknowleges support from the Smithsonian Institution, the Chandra
High Resolution Camera Project through NASA contract NAS8-0306, NASA
Grant 80NSSC19K0116 and Chandra Grant GO1-22132X. KR acknowledges the Smithsonian Combined Support for Life on a Sustainable Planet, Science, and Research administered by the Office of the Under Secretary for Science and Research.
This research has made use of data obtained from the Chandra Data
Archive and Chandra Source Catalog and software provided by the
Chandra X-ray Center (CXC) in the application packages CIAO. The
National Radio Astronomy Observatory is a facility of the National
Science Foundation operated under cooperative agreement by Asso-
ciated Universities, Inc.  
\\
{\it Facilities}: CXO 
\\
{\it Software}: astropy \citep{astropy_2013,astropy_2018}, APLpy \citep{aplpy_2012},
Numpy \citep{vanderWalt_2011, Harris_2020}, 
%Scipy (Jones et al. 2001), 
CIAO \citep{Fruscione_2006}, XSPEC \citep{Arnaud_1996}.

\bibliographystyle{aa}
%\bibliography{../../bibliography-gitti}
\bibliography{plckg287-gitti.bbl}

\begin{thebibliography}{69}
\expandafter\ifx\csname natexlab\endcsname\relax\def\natexlab#1{#1}\fi

\bibitem[{{Arnaud}(1996)}]{Arnaud_1996}
{Arnaud}, K.~A. 1996, in Astronomical Society of the Pacific Conference Series,
  Vol. 101, Astronomical Data Analysis Software and Systems V, ed. G.~H.
  {Jacoby} \& J.~{Barnes}, 17

\bibitem[{{Asplund} {et~al.}(2009){Asplund}, {Grevesse}, {Sauval}, \&
  {Scott}}]{Asplund_2009}
{Asplund}, M., {Grevesse}, N., {Sauval}, A.~J., \& {Scott}, P. 2009, \araa, 47,
  481

\bibitem[{{Astropy Collaboration} {et~al.}(2018){Astropy Collaboration},
  {Price-Whelan}, {Sip{\H{o}}cz}, {G{\"u}nther}, {Lim}, {Crawford}, {Conseil},
  {Shupe}, {Craig}, {Dencheva}, {Ginsburg}, {VanderPlas}, {Bradley},
  {P{\'e}rez-Su{\'a}rez}, {de Val-Borro}, {Aldcroft}, {Cruz}, {Robitaille},
  {Tollerud}, {Ardelean}, {Babej}, {Bach}, {Bachetti}, {Bakanov}, {Bamford},
  {Barentsen}, {Barmby}, {Baumbach}, {Berry}, {Biscani}, {Boquien}, {Bostroem},
  {Bouma}, {Brammer}, {Bray}, {Breytenbach}, {Buddelmeijer}, {Burke},
  {Calderone}, {Cano Rodr{\'\i}guez}, {Cara}, {Cardoso}, {Cheedella}, {Copin},
  {Corrales}, {Crichton}, {D'Avella}, {Deil}, {Depagne}, {Dietrich}, {Donath},
  {Droettboom}, {Earl}, {Erben}, {Fabbro}, {Ferreira}, {Finethy}, {Fox},
  {Garrison}, {Gibbons}, {Goldstein}, {Gommers}, {Greco}, {Greenfield},
  {Groener}, {Grollier}, {Hagen}, {Hirst}, {Homeier}, {Horton}, {Hosseinzadeh},
  {Hu}, {Hunkeler}, {Ivezi{\'c}}, {Jain}, {Jenness}, {Kanarek}, {Kendrew},
  {Kern}, {Kerzendorf}, {Khvalko}, {King}, {Kirkby}, {Kulkarni}, {Kumar},
  {Lee}, {Lenz}, {Littlefair}, {Ma}, {Macleod}, {Mastropietro}, {McCully},
  {Montagnac}, {Morris}, {Mueller}, {Mumford}, {Muna}, {Murphy}, {Nelson},
  {Nguyen}, {Ninan}, {N{\"o}the}, {Ogaz}, {Oh}, {Parejko}, {Parley}, {Pascual},
  {Patil}, {Patil}, {Plunkett}, {Prochaska}, {Rastogi}, {Reddy Janga},
  {Sabater}, {Sakurikar}, {Seifert}, {Sherbert}, {Sherwood-Taylor}, {Shih},
  {Sick}, {Silbiger}, {Singanamalla}, {Singer}, {Sladen}, {Sooley},
  {Sornarajah}, {Streicher}, {Teuben}, {Thomas}, {Tremblay}, {Turner},
  {Terr{\'o}n}, {van Kerkwijk}, {de la Vega}, {Watkins}, {Weaver}, {Whitmore},
  {Woillez}, {Zabalza}, \& {Astropy Contributors}}]{astropy_2018}
{Astropy Collaboration}, {Price-Whelan}, A.~M., {Sip{\H{o}}cz}, B.~M., {et~al.}
  2018, \aj, 156, 123

\bibitem[{{Astropy Collaboration} {et~al.}(2013){Astropy Collaboration},
  {Robitaille}, {Tollerud}, {Greenfield}, {Droettboom}, {Bray}, {Aldcroft},
  {Davis}, {Ginsburg}, {Price-Whelan}, {Kerzendorf}, {Conley}, {Crighton},
  {Barbary}, {Muna}, {Ferguson}, {Grollier}, {Parikh}, {Nair}, {Unther},
  {Deil}, {Woillez}, {Conseil}, {Kramer}, {Turner}, {Singer}, {Fox}, {Weaver},
  {Zabalza}, {Edwards}, {Azalee Bostroem}, {Burke}, {Casey}, {Crawford},
  {Dencheva}, {Ely}, {Jenness}, {Labrie}, {Lim}, {Pierfederici}, {Pontzen},
  {Ptak}, {Refsdal}, {Servillat}, \& {Streicher}}]{astropy_2013}
{Astropy Collaboration}, {Robitaille}, T.~P., {Tollerud}, E.~J., {et~al.} 2013,
  \aap, 558, A33

\bibitem[{{Bagchi} {et~al.}(2011){Bagchi}, {Sirothia}, {Werner}, {Pandge},
  {Kantharia}, {Ishwara-Chandra}, {Gopal-Krishna}, {Paul}, \&
  {Joshi}}]{Bagchi_2011}
{Bagchi}, J., {Sirothia}, S.~K., {Werner}, N., {et~al.} 2011, \apjl, 736, L8

\bibitem[{{Barnes} {et~al.}(2018){Barnes}, {Vogelsberger}, {Kannan},
  {Marinacci}, {Weinberger}, {Springel}, {Torrey}, {Pillepich}, {Nelson},
  {Pakmor}, {Naiman}, {Hernquist}, \& {McDonald}}]{Barnes_2018}
{Barnes}, D.~J., {Vogelsberger}, M., {Kannan}, R., {et~al.} 2018, \mnras, 481,
  1809

\bibitem[{{B{\^{\i}}rzan} {et~al.}(2004){B{\^{\i}}rzan}, {Rafferty},
  {McNamara}, {Wise}, \& {Nulsen}}]{Birzan_2004}
{B{\^{\i}}rzan}, L., {Rafferty}, D.~A., {McNamara}, B.~R., {Wise}, M.~W., \&
  {Nulsen}, P.~E.~J. 2004, \apj, 607, 800

\bibitem[{{Bonafede} {et~al.}(2014){Bonafede}, {Intema}, {Br{\"u}ggen},
  {Girardi}, {Nonino}, {Kantharia}, {van Weeren}, \&
  {R{\"o}ttgering}}]{Bonafede_2014}
{Bonafede}, A., {Intema}, H.~T., {Br{\"u}ggen}, M., {et~al.} 2014, \apj, 785, 1

\bibitem[{{Botteon} {et~al.}(2018){Botteon}, {Gastaldello}, \&
  {Brunetti}}]{Botteon_2018}
{Botteon}, A., {Gastaldello}, F., \& {Brunetti}, G. 2018, \mnras, 476, 5591

\bibitem[{{Brighenti} \& {Mathews}(2003)}]{Brighenti-Mathews_2003}
{Brighenti}, F. \& {Mathews}, W.~G. 2003, \apj, 587, 580

\bibitem[{{Brunetti} \& {Jones}(2014)}]{Brunetti-Jones_2014}
{Brunetti}, G. \& {Jones}, T.~W. 2014, International Journal of Modern Physics
  D, 23, 30007

\bibitem[{{Campitiello} {et~al.}(2022){Campitiello}, {Ettori}, {Lovisari},
  {Bartalucci}, {Eckert}, {Rasia}, {Rossetti}, {Gastaldello}, {Pratt},
  {Maughan}, {Pointecouteau}, {Sereno}, {Biffi}, {Borgani}, {De Luca}, {De
  Petris}, {Gaspari}, {Ghizzardi}, {Mazzotta}, \& {Molendi}}]{Campitiello_2022}
{Campitiello}, M.~G., {Ettori}, S., {Lovisari}, L., {et~al.} 2022, \aap, 665,
  A117

\bibitem[{{CHEX-MATE Collaboration} {et~al.}(2021){CHEX-MATE Collaboration},
  {Arnaud}, {Ettori}, {Pratt}, {Rossetti}, {Eckert}, {Gastaldello}, {Gavazzi},
  {Kay}, {Lovisari}, {Maughan}, {Pointecouteau}, {Sereno}, {Bartalucci},
  {Bonafede}, {Bourdin}, {Cassano}, {Duffy}, {Iqbal}, {Maurogordato}, {Rasia},
  {Sayers}, {Andrade-Santos}, {Aussel}, {Barnes}, {Barrena}, {Borgani},
  {Burkutean}, {Clerc}, {Corasaniti}, {Cuillandre}, {De Grandi}, {De Petris},
  {Dolag}, {Donahue}, {Ferragamo}, {Gaspari}, {Ghizzardi}, {Gitti}, {Haines},
  {Jauzac}, {Johnston-Hollitt}, {Jones}, {K{\'e}ruzor{\'e}}, {Le Brun},
  {Mayet}, {Mazzotta}, {Melin}, {Molendi}, {Nonino}, {Okabe}, {Paltani},
  {Perotto}, {Pires}, {Radovich}, {Rubino-Martin}, {Salvati}, {Saro},
  {Sartoris}, {Schellenberger}, {Streblyanska}, {Tarr{\'\i}o}, {Tozzi},
  {Umetsu}, {van der Burg}, {Vazza}, {Venturi}, {Yepes}, \&
  {Zarattini}}]{CHEX-MATE_2021}
{CHEX-MATE Collaboration}, {Arnaud}, M., {Ettori}, S., {et~al.} 2021, \aap,
  650, A104

\bibitem[{{Clowe} {et~al.}(2004){Clowe}, {Gonzalez}, \&
  {Markevitch}}]{Clowe_2004}
{Clowe}, D., {Gonzalez}, A., \& {Markevitch}, M. 2004, \apj, 604, 596

\bibitem[{{Condon} \& {Matthews}(2018)}]{Condon_2018}
{Condon}, J.~J. \& {Matthews}, A.~M. 2018, \pasp, 130, 073001

\bibitem[{{D'Addona} {et~al.}(2024){D'Addona}, {Mercurio}, {Rosati}, {Grillo},
  {Caminha}, {Acebron}, {Angora}, {Bergamini}, {Bozza}, {Granata},
  {Annunziatella}, {Gargiulo}, {Gobat}, {Tozzi}, {Girardi}, {Lombardi},
  {Meneghetti}, {Schipani}, {Tortorelli}, \& {Vanzella}}]{Daddona_2024}
{D'Addona}, M., {Mercurio}, A., {Rosati}, P., {et~al.} 2024, arXiv e-prints,
  arXiv:2401.16473

\bibitem[{{David} {et~al.}(2001){David}, {Nulsen}, {McNamara}, {Forman},
  {Jones}, {Ponman}, {Robertson}, \& {Wise}}]{David_2001}
{David}, L.~P., {Nulsen}, P.~E.~J., {McNamara}, B.~R., {et~al.} 2001, \apj,
  557, 546

\bibitem[{{De Grandi} \& {Molendi}(2001)}]{DeGrandi-Molendi_2001}
{De Grandi}, S. \& {Molendi}, S. 2001, \apj, 551, 153

\bibitem[{{Dolag} {et~al.}(2008){Dolag}, {Bykov}, \& {Diaferio}}]{Dolag_2008}
{Dolag}, K., {Bykov}, A.~M., \& {Diaferio}, A. 2008, \ssr, 134, 311

\bibitem[{{Douglass} {et~al.}(2018){Douglass}, {Blanton}, {Randall}, {Clarke},
  {Edwards}, {Sabry}, \& {ZuHone}}]{Douglass_2018}
{Douglass}, E.~M., {Blanton}, E.~L., {Randall}, S.~W., {et~al.} 2018, \apj,
  868, 121

\bibitem[{{Eckert} {et~al.}(2020){Eckert}, {Finoguenov}, {Ghirardini},
  {Grandis}, {Kaefer}, {Sanders}, \& {Ramos-Ceja}}]{Eckert_2020}
{Eckert}, D., {Finoguenov}, A., {Ghirardini}, V., {et~al.} 2020, The Open
  Journal of Astrophysics, 3, 12

\bibitem[{{Eckert} {et~al.}(2011){Eckert}, {Molendi}, \&
  {Paltani}}]{Eckert_2011}
{Eckert}, D., {Molendi}, S., \& {Paltani}, S. 2011, \aap, 526, A79

\bibitem[{{Evrard} {et~al.}(1996){Evrard}, {Metzler}, \&
  {Navarro}}]{Evrard_1996}
{Evrard}, A.~E., {Metzler}, C.~A., \& {Navarro}, J.~F. 1996, \apj, 469, 494

\bibitem[{{Fabian}(2012)}]{Fabian_2012}
{Fabian}, A.~C. 2012, \araa, 50, 455

\bibitem[{{Finner} {et~al.}(2024){Finner}, {Jee}, {Cho}, {Hyeonghan}, {Lee},
  {van Weeren}, {Wittman}, \& {Yoon}}]{Finner_2024}
{Finner}, K., {Jee}, M.~J., {Cho}, H., {et~al.} 2024, arXiv e-prints,
  arXiv:2407.02557

\bibitem[{{Finner} {et~al.}(2017){Finner}, {Jee}, {Golovich}, {Wittman},
  {Dawson}, {Gruen}, {Koekemoer}, {Lemaux}, \& {Seitz}}]{Finner_2017}
{Finner}, K., {Jee}, M.~J., {Golovich}, N., {et~al.} 2017, \apj, 851, 46

\bibitem[{{Forman} {et~al.}(2020){Forman}, {Churazov}, {Heinz}, {Jones},
  {Nulsen}, {Kraft}, {Randall}, \& {Vikhlinin}}]{Forman_2020}
{Forman}, W., {Churazov}, E., {Heinz}, S., {et~al.} 2020, in IAU Symposium,
  Vol. 342, Perseus in Sicily: From Black Hole to Cluster Outskirts, ed.
  K.~{Asada}, E.~{de Gouveia Dal Pino}, M.~{Giroletti}, H.~{Nagai}, \&
  R.~{Nemmen}, 112--117

\bibitem[{{Fruscione} {et~al.}(2006){Fruscione}, {McDowell}, {Allen},
  {Brickhouse}, {Burke}, {Davis}, {Durham}, {Elvis}, {Galle}, {Harris},
  {Huenemoerder}, {Houck}, {Ishibashi}, {Karovska}, {Nicastro}, {Noble},
  {Nowak}, {Primini}, {Siemiginowska}, {Smith}, \& {Wise}}]{Fruscione_2006}
{Fruscione}, A., {McDowell}, J.~C., {Allen}, G.~E., {et~al.} 2006, in Society
  of Photo-Optical Instrumentation Engineers (SPIE) Conference Series, Vol.
  6270, Observatory Operations: Strategies, Processes, and Systems, ed. D.~R.
  {Silva} \& R.~E. {Doxsey}, 62701V

\bibitem[{{Gaspari} {et~al.}(2011{\natexlab{a}}){Gaspari}, {Brighenti},
  {D'Ercole}, \& {Melioli}}]{Gaspari_2011b}
{Gaspari}, M., {Brighenti}, F., {D'Ercole}, A., \& {Melioli}, C.
  2011{\natexlab{a}}, \mnras, 1003

\bibitem[{{Gaspari} {et~al.}(2011{\natexlab{b}}){Gaspari}, {Melioli},
  {Brighenti}, \& {D'Ercole}}]{Gaspari_2011a}
{Gaspari}, M., {Melioli}, C., {Brighenti}, F., \& {D'Ercole}, A.
  2011{\natexlab{b}}, \mnras, 411, 349

\bibitem[{{Ghizzardi} {et~al.}(2021){Ghizzardi}, {Molendi}, {van der Burg}, {De
  Grandi}, {Bartalucci}, {Gastaldello}, {Rossetti}, {Biffi}, {Borgani},
  {Eckert}, {Ettori}, {Gaspari}, {Ghirardini}, \& {Rasia}}]{Ghizzardi_2021}
{Ghizzardi}, S., {Molendi}, S., {van der Burg}, R., {et~al.} 2021, \aap, 646,
  A92

\bibitem[{{Gitti} {et~al.}(2012){Gitti}, {Brighenti}, \&
  {McNamara}}]{Gitti_2012}
{Gitti}, M., {Brighenti}, F., \& {McNamara}, B.~R. 2012, Advances in Astronomy,
  2012

\bibitem[{{Golovich} {et~al.}(2019){Golovich}, {Dawson}, {Wittman}, {van
  Weeren}, {Andrade-Santos}, {Jee}, {Benson}, {de Gasperin}, {Venturi},
  {Bonafede}, {Sobral}, {Ogrean}, {Lemaux}, {Brada{\v{c}}}, {Br{\"u}ggen}, \&
  {Peter}}]{Golovich_2019}
{Golovich}, N., {Dawson}, W.~A., {Wittman}, D.~M., {et~al.} 2019, \apj, 882, 69

\bibitem[{{Guo} \& {Mathews}(2010)}]{Guo_2010}
{Guo}, F. \& {Mathews}, W.~G. 2010, \apj, 717, 937

\bibitem[{{Harris} {et~al.}(2020){Harris}, {Millman}, {van der Walt},
  {Gommers}, {Virtanen}, {Cournapeau}, {Wieser}, {Taylor}, {Berg}, {Smith},
  {Kern}, {Picus}, {Hoyer}, {van Kerkwijk}, {Brett}, {Haldane}, {del R{\'\i}o},
  {Wiebe}, {Peterson}, {G{\'e}rard-Marchant}, {Sheppard}, {Reddy}, {Weckesser},
  {Abbasi}, {Gohlke}, \& {Oliphant}}]{Harris_2020}
{Harris}, C.~R., {Millman}, K.~J., {van der Walt}, S.~J., {et~al.} 2020, \nat,
  585, 357

\bibitem[{{Heinz} {et~al.}(2003){Heinz}, {Churazov}, {Forman}, {Jones}, \&
  {Briel}}]{Heinz_2003b}
{Heinz}, S., {Churazov}, E., {Forman}, W., {Jones}, C., \& {Briel}, U.~G. 2003,
  \mnras, 346, 13

\bibitem[{{HI4PI Collaboration} {et~al.}(2016){HI4PI Collaboration}, {Ben
  Bekhti}, {Fl{\"o}er}, {Keller}, {Kerp}, {Lenz}, {Winkel}, {Bailin},
  {Calabretta}, {Dedes}, {Ford}, {Gibson}, {Haud}, {Janowiecki}, {Kalberla},
  {Lockman}, {McClure-Griffiths}, {Murphy}, {Nakanishi}, {Pisano}, \&
  {Staveley-Smith}}]{HI4PI_2016}
{HI4PI Collaboration}, {Ben Bekhti}, N., {Fl{\"o}er}, L., {et~al.} 2016, \aap,
  594, A116

\bibitem[{{Kravtsov} \& {Borgani}(2012)}]{Kravtsov_2012}
{Kravtsov}, A.~V. \& {Borgani}, S. 2012, \araa, 50, 353

\bibitem[{{Landau} \& {Lifshitz}(1960)}]{Landau_1960}
{Landau}, L.~D. \& {Lifshitz}, E.~M. 1960, {Electrodynamics of continuous
  media}

\bibitem[{{Leccardi} {et~al.}(2010){Leccardi}, {Rossetti}, \&
  {Molendi}}]{Leccardi_2010}
{Leccardi}, A., {Rossetti}, M., \& {Molendi}, S. 2010, \aap, 510, A82

\bibitem[{{Liu} {et~al.}(2019){Liu}, {Sun}, {Nulsen}, {Clarke}, {Sarazin},
  {Forman}, {Gaspari}, {Giacintucci}, {Lal}, \& {Edge}}]{Liu_2019}
{Liu}, W., {Sun}, M., {Nulsen}, P., {et~al.} 2019, \mnras, 484, 3376

\bibitem[{{Lovisari} \& {Reiprich}(2019)}]{Lovisari_2019}
{Lovisari}, L. \& {Reiprich}, T.~H. 2019, \mnras, 483, 540

\bibitem[{{Markevitch} {et~al.}(2002){Markevitch}, {Gonzalez}, {David},
  {Vikhlinin}, {Murray}, {Forman}, {Jones}, \& {Tucker}}]{Markevitch_2002}
{Markevitch}, M., {Gonzalez}, A.~H., {David}, L., {et~al.} 2002, \apjl, 567,
  L27

\bibitem[{{Mathis} {et~al.}(2005){Mathis}, {Lavaux}, {Diego}, \&
  {Silk}}]{Mathis_2005}
{Mathis}, H., {Lavaux}, G., {Diego}, J.~M., \& {Silk}, J. 2005, \mnras, 357,
  801

\bibitem[{{McCarthy} {et~al.}(2008){McCarthy}, {Babul}, {Bower}, \&
  {Balogh}}]{McCarthy_2008}
{McCarthy}, I.~G., {Babul}, A., {Bower}, R.~G., \& {Balogh}, M.~L. 2008,
  \mnras, 386, 1309

\bibitem[{{McNamara} \& {Nulsen}(2007)}]{McNamara-Nulsen_2007}
{McNamara}, B.~R. \& {Nulsen}, P.~E.~J. 2007, \araa, 45, 117

\bibitem[{{Molendi} {et~al.}(2023){Molendi}, {De Grandi}, {Rossetti},
  {Bartalucci}, {Gastaldello}, {Ghizzardi}, \& {Gaspari}}]{Molendi_2023}
{Molendi}, S., {De Grandi}, S., {Rossetti}, M., {et~al.} 2023, \aap, 670, A104

\bibitem[{{Petrosian} \& {Bykov}(2008)}]{Petrosian_2008}
{Petrosian}, V. \& {Bykov}, A.~M. 2008, \ssr, 134, 207

\bibitem[{{Planck Collaboration XXVII} {et~al.}(2016){Planck Collaboration
  XXVII}, {Ade}, {Aghanim}, {Arnaud}, {Ashdown}, {Aumont}, {Baccigalupi},
  {Banday}, {Barreiro}, {Barrena}, {Bartlett}, {Bartolo}, {Battaner}, {Battye},
  {Benabed}, {Beno{\^\i}t}, {Benoit-L{\'e}vy}, {Bernard}, {Bersanelli},
  {Bielewicz}, {Bikmaev}, {B{\"o}hringer}, {Bonaldi}, {Bonavera}, {Bond},
  {Borrill}, {Bouchet}, {Bucher}, {Burenin}, {Burigana}, {Butler}, {Calabrese},
  {Cardoso}, {Carvalho}, {Catalano}, {Challinor}, {Chamballu}, {Chary},
  {Chiang}, {Chon}, {Christensen}, {Clements}, {Colombi}, {Colombo}, {Combet},
  {Comis}, {Couchot}, {Coulais}, {Crill}, {Curto}, {Cuttaia}, {Dahle},
  {Danese}, {Davies}, {Davis}, {de Bernardis}, {de Rosa}, {de Zotti},
  {Delabrouille}, {D{\'e}sert}, {Dickinson}, {Diego}, {Dolag}, {Dole},
  {Donzelli}, {Dor{\'e}}, {Douspis}, {Ducout}, {Dupac}, {Efstathiou},
  {Eisenhardt}, {Elsner}, {En{\ss}lin}, {Eriksen}, {Falgarone}, {Fergusson},
  {Feroz}, {Ferragamo}, {Finelli}, {Forni}, {Frailis}, {Fraisse}, {Franceschi},
  {Frejsel}, {Galeotta}, {Galli}, {Ganga}, {G{\'e}nova-Santos}, {Giard},
  {Giraud-H{\'e}raud}, {Gjerl{\o}w}, {Gonz{\'a}lez-Nuevo}, {G{\'o}rski},
  {Grainge}, {Gratton}, {Gregorio}, {Gruppuso}, {Gudmundsson}, {Hansen},
  {Hanson}, {Harrison}, {Hempel}, {Henrot-Versill{\'e}},
  {Hern{\'a}ndez-Monteagudo}, {Herranz}, {Hildebrandt}, {Hivon}, {Hobson},
  {Holmes}, {Hornstrup}, {Hovest}, {Huffenberger}, {Hurier}, {Jaffe}, {Jaffe},
  {Jin}, {Jones}, {Juvela}, {Keih{\"a}nen}, {Keskitalo}, {Khamitov}, {Kisner},
  {Kneissl}, {Knoche}, {Kunz}, {Kurki-Suonio}, {Lagache}, {Lamarre}, {Lasenby},
  {Lattanzi}, {Lawrence}, {Leonardi}, {Lesgourgues}, {Levrier}, {Liguori},
  {Lilje}, {Linden-V{\o}rnle}, {L{\'o}pez-Caniego}, {Lubin},
  {Mac{\'\i}as-P{\'e}rez}, {Maggio}, {Maino}, {Mak}, {Mandolesi}, {Mangilli},
  {Martin}, {Mart{\'\i}nez-Gonz{\'a}lez}, {Masi}, {Matarrese}, {Mazzotta},
  {McGehee}, {Mei}, {Melchiorri}, {Melin}, {Mendes}, {Mennella}, {Migliaccio},
  {Mitra}, {Miville-Desch{\^e}nes}, {Moneti}, {Montier}, {Morgante},
  {Mortlock}, {Moss}, {Munshi}, {Murphy}, {Naselsky}, {Nastasi}, {Nati},
  {Natoli}, {Netterfield}, {N{\o}rgaard-Nielsen}, {Noviello}, {Novikov},
  {Novikov}, {Olamaie}, {Oxborrow}, {Paci}, {Pagano}, {Pajot}, {Paoletti},
  {Pasian}, {Patanchon}, {Pearson}, {Perdereau}, {Perotto}, {Perrott},
  {Perrotta}, {Pettorino}, {Piacentini}, {Piat}, {Pierpaoli}, {Pietrobon},
  {Plaszczynski}, {Pointecouteau}, {Polenta}, {Pratt}, {Pr{\'e}zeau}, {Prunet},
  {Puget}, {Rachen}, {Reach}, {Rebolo}, {Reinecke}, {Remazeilles}, {Renault},
  {Renzi}, {Ristorcelli}, {Rocha}, {Rosset}, {Rossetti}, {Roudier}, {Rozo},
  {Rubi{\~n}o-Mart{\'\i}n}, {Rumsey}, {Rusholme}, {Rykoff}, {Sandri}, {Santos},
  {Saunders}, {Savelainen}, {Savini}, {Schammel}, {Scott}, {Seiffert},
  {Shellard}, {Shimwell}, {Spencer}, {Stanford}, {Stern}, {Stolyarov},
  {Stompor}, {Streblyanska}, {Sudiwala}, {Sunyaev}, {Sutton}, {Suur-Uski},
  {Sygnet}, {Tauber}, {Terenzi}, {Toffolatti}, {Tomasi}, {Tramonte},
  {Tristram}, {Tucci}, {Tuovinen}, {Umana}, {Valenziano}, {Valiviita}, {Van
  Tent}, {Vielva}, {Villa}, {Wade}, {Wandelt}, {Wehus}, {White}, {Wright},
  {Yvon}, {Zacchei}, \& {Zonca}}]{Planck_2016}
{Planck Collaboration XXVII}, {Ade}, P.~A.~R., {Aghanim}, N., {et~al.} 2016,
  \aap, 594, A27

\bibitem[{{Randall} {et~al.}(2011){Randall}, {Forman}, {Giacintucci}, {Nulsen},
  {Sun}, {Jones}, {Churazov}, {David}, {Kraft}, {Donahue}, {Blanton},
  {Simionescu}, \& {Werner}}]{Randall_2011}
{Randall}, S.~W., {Forman}, W.~R., {Giacintucci}, S., {et~al.} 2011, \apj, 726,
  86

\bibitem[{{Rasia} {et~al.}(2015){Rasia}, {Borgani}, {Murante}, {Planelles},
  {Beck}, {Biffi}, {Ragone-Figueroa}, {Granato}, {Steinborn}, \&
  {Dolag}}]{Rasia_2015}
{Rasia}, E., {Borgani}, S., {Murante}, G., {et~al.} 2015, \apjl, 813, L17

\bibitem[{{Ricker} \& {Sarazin}(2001)}]{Ricker_2001}
{Ricker}, P.~M. \& {Sarazin}, C.~L. 2001, \apj, 561, 621

\bibitem[{{Riva} {et~al.}(2024){Riva}, {Pratt}, {Rossetti}, {Bartalucci},
  {Kay}, {Rasia}, {Gavazzi}, {Umetsu}, {Arnaud}, {Balboni}, {Bonafede},
  {Bourdin}, {De Grandi}, {De Luca}, {Eckert}, {Ettori}, {Gaspari},
  {Gastaldello}, {Ghirardini}, {Ghizzardi}, {Gitti}, {Lovisari}, {Maughan},
  {Mazzotta}, {Molendi}, {Pointecouteau}, {Sayers}, {Sereno}, \&
  {Towler}}]{Riva_2024}
{Riva}, G., {Pratt}, G.~W., {Rossetti}, M., {et~al.} 2024, \aap, 691, A340

\bibitem[{{Robitaille} \& {Bressert}(2012)}]{aplpy_2012}
{Robitaille}, T. \& {Bressert}, E. 2012, {APLpy: Astronomical Plotting Library
  in Python}, Astrophysics Source Code Library, record ascl:1208.017

\bibitem[{{Rossetti} {et~al.}(2024){Rossetti}, {Eckert}, {Gastaldello},
  {Rasia}, {Pratt}, {Ettori}, {Molendi}, {Arnaud}, {Balboni}, {Bartalucci},
  {Batalha}, {Borgani}, {Bourdin}, {De Grandi}, {De Luca}, {De Petris},
  {Forman}, {Gaspari}, {Ghizzardi}, {Iqbal}, {Kay}, {Lovisari}, {Maughan},
  {Mazzotta}, {Pointecouteau}, {Riva}, {Sayers}, \& {Sereno}}]{Rossetti_2024}
{Rossetti}, M., {Eckert}, D., {Gastaldello}, F., {et~al.} 2024, \aap, 686, A68

\bibitem[{{Rossetti} {et~al.}(2016){Rossetti}, {Gastaldello}, {Ferioli},
  {Bersanelli}, {De Grandi}, {Eckert}, {Ghizzardi}, {Maino}, \&
  {Molendi}}]{Rossetti_2016}
{Rossetti}, M., {Gastaldello}, F., {Ferioli}, G., {et~al.} 2016, \mnras, 457,
  4515

\bibitem[{{Rossetti} \& {Molendi}(2010)}]{Rossetti_2010}
{Rossetti}, M. \& {Molendi}, S. 2010, \aap, 510, A83

\bibitem[{{Sanders}(2006)}]{Sanders_2006}
{Sanders}, J.~S. 2006, \mnras, 371, 829

\bibitem[{{Sarazin}(2002)}]{Sarazin_2001}
{Sarazin}, C.~L. 2002, in Astrophysics and Space Science Library, Vol. 272,
  Merging Processes in Galaxy Clusters, ed. L.~{Feretti}, I.~M. {Gioia}, \&
  G.~{Giovannini}, 1--38

\bibitem[{{Schellenberger} {et~al.}(2023){Schellenberger}, {O'Sullivan},
  {Giacintucci}, {Vrtilek}, {David}, {Combes}, {B{\^\i}rzan}, {Pan}, \&
  {Lin}}]{Schellenberger_2023}
{Schellenberger}, G., {O'Sullivan}, E., {Giacintucci}, S., {et~al.} 2023, \apj,
  948, 101

\bibitem[{{Stuardi} {et~al.}(2022){Stuardi}, {Bonafede}, {Rajpurohit},
  {Br{\"u}ggen}, {de Gasperin}, {Hoang}, {van Weeren}, \&
  {Vazza}}]{Stuardi_2022}
{Stuardi}, C., {Bonafede}, A., {Rajpurohit}, K., {et~al.} 2022, \aap, 666, A8

\bibitem[{{Ubertosi} {et~al.}(2023){Ubertosi}, {Gitti}, \&
  {Brighenti}}]{Ubertosi_2023}
{Ubertosi}, F., {Gitti}, M., \& {Brighenti}, F. 2023, \aap, 670, A23

\bibitem[{{van der Walt} {et~al.}(2011){van der Walt}, {Colbert}, \&
  {Varoquaux}}]{vanderWalt_2011}
{van der Walt}, S., {Colbert}, S.~C., \& {Varoquaux}, G. 2011, Computing in
  Science and Engineering, 13, 22

\bibitem[{{van Weeren} {et~al.}(2019){van Weeren}, {de Gasperin}, {Akamatsu},
  {Br{\"u}ggen}, {Feretti}, {Kang}, {Stroe}, \& {Zandanel}}]{vanWeeren_2019}
{van Weeren}, R.~J., {de Gasperin}, F., {Akamatsu}, H., {et~al.} 2019, \ssr,
  215, 16

\bibitem[{{Vikhlinin} {et~al.}(2001){Vikhlinin}, {Markevitch}, \&
  {Murray}}]{Vikhlinin_2001}
{Vikhlinin}, A., {Markevitch}, M., \& {Murray}, S.~S. 2001, \apj, 551, 160

\bibitem[{{Vikhlinin} {et~al.}(2005){Vikhlinin}, {Markevitch}, {Murray},
  {Jones}, {Forman}, \& {Van Speybroeck}}]{Vikhlinin_2005}
{Vikhlinin}, A., {Markevitch}, M., {Murray}, S.~S., {et~al.} 2005, \apj, 628,
  655

\bibitem[{{Vikhlinin} {et~al.}(2014){Vikhlinin}, {Kravtsov}, {Markevich},
  {Sunyaev}, \& {Churazov}}]{Vikhlinin_2014}
{Vikhlinin}, A.~A., {Kravtsov}, A.~V., {Markevich}, M.~L., {Sunyaev}, R.~A., \&
  {Churazov}, E.~M. 2014, Physics Uspekhi, 57, 317

\bibitem[{{Wallbank} {et~al.}(2022){Wallbank}, {Maughan}, {Gastaldello},
  {Potter}, \& {Wik}}]{Wallbank-2022}
{Wallbank}, A.~N., {Maughan}, B.~J., {Gastaldello}, F., {Potter}, C., \& {Wik},
  D.~R. 2022, \mnras, 517, 5594

\bibitem[{{Zhang} {et~al.}(2019){Zhang}, {Churazov}, {Forman}, \&
  {Jones}}]{Zhang_2019}
{Zhang}, C., {Churazov}, E., {Forman}, W.~R., \& {Jones}, C. 2019, \mnras, 482,
  20

\end{thebibliography}

\appendix

\section{Analysis of X-ray depressions}

\begin{figure*}
  \vspace{0.cm}
    \includegraphics[width=\textwidth,bb=36 281 577 511]{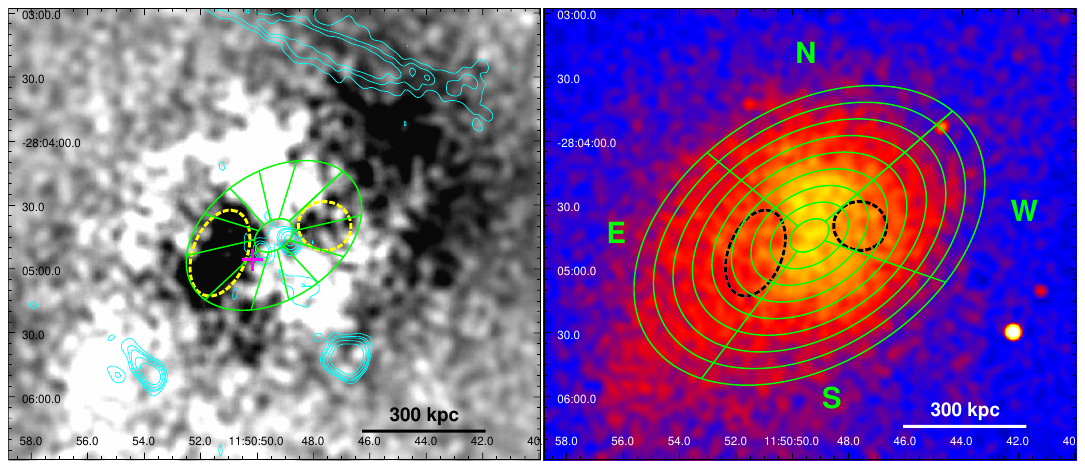}
%  \includegraphics[scale=0.69, angle=0]{figure/cavities-appendix.ps}
%\hspace{-0.2 cm}
%  \includegraphics[scale=0.3, angle=0]{figure/profile-ellipse-cavities.ps}
\vspace{0.0cm}
\caption{\label{cavity-appendix.fig}
Surface brightness regions used for the analysis of the depressions in the ICM of PLCKG287.
{\it Left panel:} Residual {\it
    Chandra} image of the central cluster region obtained by
  subtracting the best-fit $\beta$-model (same as in Figure
  {\color{black} \ref{residual-unsharp.fig}}, left panel) with superimposed the GMRT
  radio contours at 610 MHz (same as in Figure
  {\color{black} \ref{final-discussion.fig})}. The elliptical wedges used to evaluate the
  azimuthal SB profile around the center were chosen so as to mostly
  encompass the region of each depression (elliptical annulus between
  10$''$ and 45$''$ along the major axis, divided in 12 sectors of
  30$^\circ$ each) and are shown in green panda regions.  The magenta cross indicates the location
  of the main BCG.  {\it Right panel:} Background-subtracted, exposure
  corrected mosaic [0.5-7.0] keV {\it Chandra} image, smoothed with a
  kernel of 8 pixels, of the same central region of PLCKG287 as in the
  left panel. The elliptical wedges used to extract the radial SB
  profile in 5$''$-wide annular bin along four azimuthal sectors (N:
  8$^{\circ}$ - 18$^{\circ}$, E: 108$^{\circ}$ - 200$^{\circ}$, S:
  200$^{\circ}$ - 308$^{\circ}$, W: 308$^{\circ}$ - 368$^{\circ}$) are
  shown in green panda regions (to improve clarity, the annular bins
  shown in Figure are 10$''$-wide instead of 5$''$-wide).  The
  geometry is the same as that of the elliptical $\beta$-model (see
  Table \ref{beta.tab}).  In both panels, the dashed ellipses show the possible X-ray cavities 
  identified in Section \ref{morpho.sec} and discussed in Section \ref{feedback.sec}.}
\vspace{-0.1in}
\end{figure*}

We show in Figure \ref{cavity-appendix.fig} the sectors used to
investigate the significance of the X-ray depressions discussed in
Section \ref{morpho.sec}.

\section{Spectral analysis with $N_{\rm H}$ free}
\label{freenh.sec}

We present below the spectral analysis performed with the column
density left as a free parameter.

\subsection{Global properties inside $R_{500}$}

We first measured the global properties inside $R_{500}$ by
considering a circular region of radius = $294.7''$ (= 1541 kpc, see
green circle in Figure 3). We found that a free value of the galactic hydrogen
  column parameter, $N_{\rm H}$, was preferred by the fit (F-stat=68.85,
  F-prob=$4.52 \times 10^{-14}$) over the parameter fixed to the
tabulated value of $N_{\rm H}=6.93 \times 10^{20}$ cm$^{-2}$
  \citep{HI4PI_2016}. The best-fit parameters are: $N_{\rm H} = 1.47
  \pm 0.10 \times 10^{21}$ cm$^{-2}$, 
  temperature $kT = 10.24^{+0.38}_{-0.32}$ keV and abundance $Z =
  0.23^{+0.04}_{0.04}$ Z$_{\odot}$. With the model  {\ttfamily
    tbabs*clumin*apec} we estimated the unabsorbed luminosity in the soft,
  hard and bolometric bands to be $L_{0.3-2.4\,keV}= (6.23 \pm 0.03) \times 10^{44}$ erg
  s$^{-1}$, $L_{2-10\,keV}= (1.11 \pm 0.01) \times 10^{45}$ erg s$^{-1}$
  and $L_{\rm bol}= (2.36 \pm 0.01) \times 10^{45}$ erg s$^{-1}$, respectively.
We also verified that the addition 
of a power-law component
(model {\ttfamily tbabs*(apec+po)}), to account for the presence of
unresolved X-ray binaries, is not statistically significant (F-stat=2.24,
  F-prob=$0.107$) .
The spectral fitting results are reported in Table \ref{global-freeNH.tab}.

\begin{table*}
\caption{\label{global-freeNH.tab} Results of the spectral fits inside $R_{\rm 500}$. }
% title of Table
\centering                          % used for centering table
\begin{tabular}{c c c c c c c c}        % centered columns (4 columns)
  \hline
  \hline
  model & $N_{\rm H}$ & $kT$ & $Z$ & norm$_{\rm apec} $ & $\Gamma$ & norm$_{\rm po}$ & $\chi^2/$dof
  \\[+1mm]
  ~ & ($10^{22}$ cm$^{-2}$) & (keV) & (Z$_{\odot}$) & ($10^{-3}$) & ~ & ($10^{-3}$) & ~
 \\[+1mm]
  \hline
   {\ttfamily tbabs*apec} & 0.0693 fixed & $12.73^{+0.38}_{-0.40}$ & $0.26 \pm 0.05$ & $(3.380 \pm 0.003) $ & - & - & 513.96/441
  \\[+1mm]                                                             
    {\ttfamily tbabs*apec} & $0.147 \pm 0.010$  & $10.24^{+0.38}_{-0.32}$ & $0.23 \pm 0.04$ & $(3.586 \pm 0.004)$ & - & - & 451.52/440                                                                           
  \\[+1mm]
   {\ttfamily tbabs*(apec+po)} & $0.28^{+0.07}_{-0.10}$  & $10.86^{+1.12}_{-0.71}$ & $0.25^{+0.16}_{-0.05}$  & $3.45^{+0.13}_{-1.14}$ & $3.52^{+0.71}_{-2.08}$ & $0.16 \pm 0.08$ & 446.94/438
  \\[+1mm]
  \hline
  \hline
\end{tabular}
{\tablefoot{Column 1: Xspec model; column 2: galactic hydrogen
  column;
  column 3: temperature, $kT$; column 4: abundance, $Z$; column 5: normalization of the \ttfamily{apec} component; column 6: photon index, $\Gamma$; column 7: normalization of the \ttfamily{powerlaw} component; column 8: $\chi^2/{\rm dof}$.}}
\end{table*}  

%%%%%%%%%%%%%%%%%%%%%%%%%%%%%%%%%%%%%%%%%%%%%%%%%%%%%%%%%%%%%%%%%%%%%%%%%%%%%%%

\subsection{1D radial profiles of ICM thermodynamic properties}

In order to derive the azimuthally-averaged radial properties of the
ICM, we produced projected temperature and abundance profiles by
extracting spectra in circular annuli centered on the peak of the
X-ray emission. The annular regions were chosen in order to collect at
least 3,000-3,500 net counts (see Table \ref{annuli-freeNH.tab}). We verified
that freeing the column density parameter, linked among the different
annuli, provides a better fit ($\chi^2/{\rm dof}$= 2302/2325 = 0.99)
than that obtained by fixing it at the tabulated value
($\chi^2/{\rm dof}$= 2390/2326 = 1.03; F-stat= 89.2,
F-prob=$8.4 \times 10^{21}$).
% or allowing it to vary in each annulus (F-stat=, F-prob=).
We found a best-fit value of
$N_{\rm H} = 1.60 \pm 0.10 \times 10^{21}$ cm$^{-2}$.
The best-fit parameter obtained from the fits to the annular spectra
are summarized in Table \ref{annuli-freeNH.tab}.

\begin{table*}
\caption{\label{annuli-freeNH.tab} Results of the spectral fit to an absorbed apec model ({\ttfamily tbabs*apec}) performed in the [0.5-7.0] keV energy band in concentric annular regions extracted up to $R_{\rm 500}$.}
% title of Table
\centering                          % used for centering table
\begin{tabular}{c c c c c c c }        % centered columns (4 columns)
  \hline
  \hline
 $R_{\rm in}$ - $R_{\rm out}$ & $R_{\rm in}$ - $R_{\rm out}$ & Exposure & Net counts (\% total) & $kT$ & $Z$ & $\chi^2/$dof ($\chi^2_{\rm red})$  
\\[+1mm]
($''$) & (kpc) & (ksec) & ~ & (keV) & (Z$_{\odot}$) & ~ 
\\[+1mm]
  \hline
0-20 & 0-105 & 131.9 & 4333 (97.7) & $13.69_{-1.65}^{+1.40}$ & $0.47_{-0.19}^{+0.19}$ & 147.14/133 (1.11)
\\[+1mm]
20-30 & 105-157 & 131.9 & 4612 (97.4) & $11.42_{-0.96}^{+1.09}$ & $0.48_{-0.15}^{+0.15}$ & 134.03/140 (0.96)
\\[+1mm]
30-38 & 157-199 & 131.9 & 4039 (96.8) & $8.94_{-0.80}^{+1.05}$ & $0.38_{-0.13}^{+0.13}$ & 144.48/128 (1.13)
\\[+1mm]
38-45 & 199-235 & 131.9 & 3705 (96.2) & $9.94_{-0.99}^{+1.19}$ & $0.43_{-0.15}^{+0.15}$ & 105.30/117 (0.90)
\\[+1mm]
45-52 & 235-272 & 131.9 & 3402 (95.2) & $9.45_{-0.87}^{+0.95}$ & $0.25_{-0.13}^{+0.13}$ & 94.93/108 (0.88)
\\[+1mm]
52-59 & 272-309 & 131.9 & 3276 (94.5) & $10.54_{-1.11}^{+1.29}$ & $0.06_{-0.06}^{+0.15}$ & 129.00/106 (1.22)
\\[+1mm]
59-67 & 309-350 & 131.9 & 3373 (93.3) & $10.55_{-1.01}^{+1.13}$ & $0.04_{-0.04}^{+0.14}$ & 104.71/110 (0.95)
\\[+1mm]
67-76 & 350-398 & 131.9 & 3308 (91.4) & $10.64_{-1.08}^{+1.30}$ & $0.30_{-0.16}^{+0.16}$ & 110.22/113 (0.98)
\\[+1mm]
76-87 & 398-455 & 131.9 & 3489 (89.0) & $9.49_{-0.84}^{+1.00}$ & $0.32_{-0.14}^{+0.15}$ & 101.84/122 (0.84)
\\[+1mm]
87-100 & 455-523 & 175.1 & 4122 (87.6) & $10.71_{-1.00}^{+1.13}$ & $0.33_{-0.15}^{+0.15}$ & 124.89/143 (0.84)
\\[+1mm]
100-115 & 523-602 & 193.9 & 4165 (84.5) & $10.59_{-1.10}^{+1.28}$ & $0.38_{-0.16}^{+0.16}$ & 152.10/150 (1.01)
\\[+1mm]
115-135 & 602-706 & 244.3 & 4698 (80.2) & $9.70_{-0.91}^{+0.99}$ & $0.28_{-0.14}^{+0.14}$ & 152.54/170 (0.90)
\\[+1mm]
135-160 & 706-837 & 350.7 & 5067 (73.1) & $9.83_{-0.84}^{+1.07}$ & $0.04_{-0.04}^{+0.13}$ & 190.42/194 (0.98)
\\[+1mm]
160-192.2 & 837-1000 & 350.7 & 4556 (61.5) & $7.92_{-0.73}^{+1.87}$ & $0.14_{-0.13}^{+0.13}$ & 210.81/206 (1.02)
\\[+1mm]
191.2-294.7 & 1000-1541 & 350.7 & 7289 (39.1) & $7.28_{-0.60}^{+0.78}$ & $0.02_{-0.02}^{+0.12}$ & 399.58/385 (1.04)
\\[+1mm]
  \hline
  \hline
\end{tabular}
{\tablefoot{The temperature (in keV) and
metallicity \citep[in solar value,][]{Asplund_2009} were left as free
parameters. The absorbing column density was left as a free parameter but was linked among the different annuli.
The best-fit value is $N_{\rm H} = 1.60 \pm 0.10 \times 10^{21}$ cm$^{-2}$.
Column 1: inner and outer radius in arcsec; column 2: inner and outer radius in kpc ;
  column 3: combined exposure time; column 4: net counts (\% of total counts); column 5: temperature (keV); column 6: abundance (solar value); column 7:  $\chi^2/{\rm dof}$ . The global fit gives $\chi^2/{\rm dof}$= 2302/2325 = 0.99.}
  }
\end{table*}

%%%%%%%%%%%%%%%%%%%%%%%%%%%%%%%%%%%%%%%%%%%%%%%%%%%%%%%%%%%%%%%%%%%%%%%%%%%%%%%

%%%%%%%%%%%%%%%%%%%%%%%%%%%%%%%%%%%%%%%%%%%%%%%%%%%%%%%%%%%%%%%%%%%%%%%%%%%%%%%

\subsection{Study of the surface brightness discontinuities}

Given the geometry of the two nested fronts, which are aligned along
the same direction, we measured their spectral properties along a
common sector having an opening angle of $120^{\circ}$ (from
$0^{\circ}$ to $120^{\circ}$ from W and increasing counterclockwise). In particular, we
extracted the spectra of the ICM inside and outside each
discontinuity, with the addition of a central region and an outermost
region extending to $R_{\rm 500}$ to account for deprojection, for a
total of 5 spectral wedges.  Fitting the 0.5-7.0 keV band spectra with
a {\ttfamily projct*tbabs*apec} model (by keeping $N_{\rm H}$ as a
free parameter, linked among the different sectors) returned the deprojected
temperature and electron density, which were combined to derive the
pressure jump across each edge.  A summary of the deprojected
thermodynamic properties along the considered sector is reported in Table
\ref{spectra-front-freeNH.tab}.

Across the inner edge, located at $\sim 295$ kpc, we measure a
temperature jump $kT_{\rm in}/kT_{\rm out} = 0.60^{+0.37}_{-0.25}$ and
a pressure jump $p_{\rm in}/p_{\rm out} = 0.66^{+0.42}_{-0.28}$. These
are consistent with the interpretation of this edge as a cold front.  On the other hand, across
the outer edge located at $\sim 389$ kpc we measure a temperature jump
$kT_{\rm in}/kT_{\rm out} = 1.30^{+0.32}_{-0.23}$ and a pressure jump
$p_{\rm in}/p_{\rm out} = 5.08^{+1.25}_{-0.91}$. These are consistent
with the interpretation of this edge as a shock front\footnote{We verified that the temperature
  jump is detected also in the projected spectral fit:
  $kT_{\rm in}=10.13^{+1.53}_{-1.33}$ keV vs.
  $kT_{\rm out} = 7.41^{+0.67}_{-0.59}$ keV.}.

  \begin{table}[h]
    \caption{\label{spectra-front-freeNH.tab} Results of the deprojected spectral fit along a  $120^{\circ}$-wide sector enclosing the two fronts identified in Section \ref{fronts.sec}. }
  \begin{tabularx}{\linewidth}{*5{X}}
      \hline
      \hline
  Wedge no. (width) & Net counts & $kT_{\rm}$  (keV)  &  $n_{\rm e}$
  ($10^{-3}$ cm$^{-3}$) &  $p$  ($10^{-11}$ erg cm$^{-3}$)
\\[+1mm]
  \hline
  \#1 $(50'')$& 6459 & $9.92_{-1.12}^{+1.75}$ & $5.04_{-0.11}^{+0.11}$ & $14.65_{-1.68}^{+2.60} $
 \\[+1mm]                           
  \#2 $(8'')$& 1174 & $6.13_{-2.11}^{+3.69}$ & $3.70_{-0.35}^{+0.47}$ & $6.64_{-2.37}^{+4.09}$
  \\[+1mm]
    \end{tabularx}
    
    \nointerlineskip
    \begin{tabularx}{\linewidth}{ *1{X}}
 ---------------------  Cold front  ($R_{\rm cf} = 295 \pm 7$ kpc)    --------------------
 \\[+1mm]
   \end{tabularx}
    
    \nointerlineskip
    \begin{tabularx}{\linewidth}{ *5{X}}
   \#3 $(17'')$& 2013 & $10.25_{-1.36}^{+2.28}$ & $3.33_{-0.08}^{+0.08}$ & $10.01_{-1.35}^{+2.24}$ 
   \\[+1mm]
     \end{tabularx}
    
    \nointerlineskip
    \begin{tabularx}{\linewidth}{ *1{X}}
---------------------  Shock front  ($R_{\rm sh} = 389 \pm 6$ kpc)     --------------------
\\[+1mm]
  \end{tabularx}
    
    \nointerlineskip
    \begin{tabularx}{\linewidth}{ *5{X}}
   \#4 $(115'')$& 5649 & $7.89_{-0.80}^{+0.90}$ & $0.85_{-0.02}^{+0.02}$ & $1.97_{-0.20}^{+0.23}$
  \\[+1mm]
   \#5 $(100'')$& 1500 & $5.28_{-0.82}^{+1.14}$ & $0.37_{-0.01}^{+0.01}$ & $0.58_{-0.09}^{+0.13}$ 
  \\[+1mm]
  \hline
  \hline
    \end{tabularx}
    {\tablefoot{Column 1: wedge number (width in arcsec); column 2:
        net counts in the $[0.5-7.0]$ keV band; column 3: deprojected
        temperature; column 4: electron density; column 5:
        pressure. The best-fit absorbing column density, left as a
        free parameter but linked among the different annuli, is
        $N_{\rm H} = 1.83 \pm 0.03 \times 10^{21}$ cm$^{-2}$. The
        derived classification as cold front and shock front is
        indicated at the boundary of the second and third wedge,
        respectively.}}
\end{table}

\end{document}